\documentclass[useAMS,usenatbib,usegraphicx]{mn2e}
\usepackage{pslatex,color}
\usepackage{amsmath,amsfonts,amssymb,amsxtra,eufrak,url}

\def\idm#1{{\mbox{\scriptsize #1}}}
\newcommand{\au}{\mbox{au}}
\newcommand{\msun}{\mbox{m}_{\sun}}

\newcommand{\mE}{\mbox{m}_{\earth}}
\newcommand{\RE}{\mbox{R}_{\earth}}
\newcommand{\Mmean}{\mathcal{M}}
\newcommand{\mmegno}{\langle Y \rangle}
\def\mechanic{{\tt Mechanic}}
\def\v#1{{\pmb{#1}}}
\def\norm#1{|\,#1\,|}
\def\dott#1{{\frac{\mbox{d}\,#1}{\mbox{d\,}t}}}
\usepackage{color}
\definecolor{myred}{rgb}{0.7,0.1,0.1}
\definecolor{myblue}{rgb}{0.2,0.0,0.7}
\definecolor{mybrown}{rgb}{0.5,0.2,0.0}

\newcommand\corr[1]{{#1}}

\newcommand\hide[1]{}
\title[A dynamical analysis of the Kepler-11 planetary system]
{A dynamical analysis of the Kepler-11 planetary system}
\author[C. Migaszewski, M. S{\l}onina and K. Go\'zdziewski]
{Cezary Migaszewski$^{1}$\thanks{E-mail: c.migaszewski@astri.umk.pl}, 
Mariusz S{\l}onina$^{1}$\thanks{E-mail: m.slonina@astri.umk.pl} 
and Krzysztof Go\'zdziewski$^{1}$\footnotemark[1]\thanks{E-mail:
k.gozdziewski@astri.umk.pl}\\
$^{1}$Toru\'n Centre for Astronomy, Nicolaus Copernicus University, 
Gagarin Str. 11, 87-100 Toru\'n, Poland}
\begin{document}
%
\date{Accepted 2012 August 23.  Received 2012 August 20; in original form 2012 May 3}
\pagerange{\pageref{firstpage}--\pageref{lastpage}} \pubyear{2012}
\maketitle
\label{firstpage}
%
\begin{abstract}
The Kepler-11 star hosts at least six transiting super-Earth planets
detected through the precise photometric observations of the Kepler mission
(Lissauer et al.).  In this paper, we re-analyze the available Kepler data,
using the direct $N$-body approach rather than an indirect TTV method in the
discovery paper.  The orbital modeling in the realm of the direct approach
relies on the whole data set, not only on the mid--transits times. 
Most of the results in the original paper are confirmed and extended.  We
constrained the mass of the outermost planet~g to less than 30 Earth masses. 
The mutual inclinations between orbits~b and~c as well as between orbits~d
and e~are determined with a good precision, in the range of [1,5]~degrees. 
Having several solutions to four qualitative orbital models of the Kepler-11
system, we analyze its global dynamics with the help of dynamical maps. 
They reveal a sophisticated structure of the phase space, with narrow regions of
regular motion.  The dynamics are governed by a dense net of three-- and
four--body mean motion resonances, forming the Arnold web.  Overlapping of
these resonances is a main source of instability.  We found that the
Kepler-11 system may be long-term stable only in particular multiple
resonant configurations with small relative inclinations.  The mass-radius
data derived for all companions reveal a clear anti-correlation between the
mean density of the planets with their distance from the star.  This may
reflect the formation and early evolution history of the system.
\end{abstract}
%
%
%
\section{Introduction}
%
%
The Kepler space mission is a breakthrough in the field of searches for  the
Earth--like extrasolar  planets \citep{Borucki2010, Koch2010, Jenkins2010, 
Caldwell2010}. About of 150,000 solar dwarfs are monitored by 0.95---meter
Kepler telescope. The photometric data are publicly available from the MAST 
archive\footnote{\url{http://archive.stsci.edu/kepler}}.

To date, the mission identified more that 2,200 planetary candidates
\citep{Batalha2012}. Among  them, many multi-planet systems are found. For instance,
planets were  confirmed in two--planet configurations, i.e., Kepler-10
\citep{Batalha2011,Fressin2011},  Kepler-25, 26, 27, 28 \citep{Steffen2012},
Kepler-29, 31, 32 \citep {Fabrycky2012b}, Kepler-23, 24 \citep{Ford2012};
in three-planet systems  Kepler-9 \citep{Holman2010}, Kepler-30 \citep{Fabrycky2012b},
Kepler-18  \citep{Cochran2011}; in four--planet systems \citep{Borucki2011}, as well as in five--planet configurations Kepler-20 
\citep{Gautier2011, Fressin2011b}, Kepler-33 \citep{Lissauer2012}. 
The Kepler-11 hosts six planetary  companions
\citep{Lissauer2011}. The transiting planet candidates can be  confirmed through
determining their masses with the help of the so-called  Transit Timing Variations
method \citep[TTV,][]{Holman2005,Agol2005}. In this approach, the (O-C) variations 
between observed mid-transit times and their ephemeris are the observables, which
can be fitted by an appropriate orbital model. 
\corr{In recent papers, also additional 
observables are analysed, like the so-called Transits Duration Variations (TDVs)
\citep[see, e.g.,][]{Nesvorny2012}.}

In this paper, we re-analyse the photometric data of Kepler-11 with a  modified,
direct approach providing an alternate estimation of masses and  orbital elements.
To describe this method further in the paper, we recall shortly  the main conclusions
in \citep{Lissauer2011}. Using the TTV method and  an assumption of strictly
coplanar model of the system, they determined  masses of five inner planets in the
range of a few Earth masses. The  outermost planet interacts weakly with the inner
companions, and its mass  could be roughly constrained as smaller than the Jupiter
mass. It has been not confirmed as a planet, although the probability of blending 
is very small, $\sim 0.001$. Orbital eccentricities in the Kepler-11 system
were determined only for the five inner objects. Due to the assumption of
coplanarity,  a~determination of mutual inclinations between the orbits was not
possible.  \cite{Lissauer2011} argue that these inclinations should remain in  the
range of [0,2]~degrees. The dynamical analysis have revealed that the system is not
involved in the mean motion resonances (MMRs),  however a pair of planet~b and
planet~c is close to 5:4~MMR. 

The determined masses and radii of the planets imply constrains on their  chemical
composition. Planets~d, e~and f~might have similar internal compositions to 
those of Uranus
or Neptune, while planets~b and~c  are rather ice~rich, with a smaller amount of
H$_2$/He mixture than these planets in the Solar system.

In this paper, we focus mostly on the global dynamics of the system and a few
aspects which were not addressed in the discovery paper.

First of all, we model the available Kepler data  through a direct algorithm that
relies on the self-consistent $N$-body fitting  of the light-curves, instead of the
TTV method applied in the discovery  work. The TTV algorithm makes use of the
transit times {\em a~posteriori}, after they are determined from the light curves. 
Through extensive numerical experiments,  we found that the direct approach brings
more information  than the TTV method. For instance, we could constrain the mass of
the outermost planet  to less than $\sim 30$~Earth masses. We also found 
significant bounds for
the mutual inclinations to less than $5^{\circ}$ for planets~b
and~c as well as for planets~d and~e.

\corr{
The direct model, also called the dynamical-photometric model, already was
used in a few papers.  For instance, it was applied to analyse the light
curve of the triple-star system KOI-126 \citep{Carter2011}, and to estimate
masses of two planets transiting Kepler-36 \citep{Carter2012}.  This
algorithm also verified the Kepler-9 model, which was found first with
the help of the TTV algorithm \citep{Holman2010}.
}

A number of initial conditions found with the direct approach makes it  possible to
investigate the dynamics of the system. We focus on  the  short--term time scale,
governed by the mean motion resonances. We study  the multi-dimensional structure of
the phase space with the help of  dynamical maps. In the vicinity a few qualitative
transit  models considered in this work, the dynamics are governed by a dense net
of  3--body and 4--body mean motion resonances. This net may be identified with  the
Arnold web, which is a feature of close to integrable Hamiltonian  systems. The
Kepler-11  appears as strongly resonant extrasolar system, and this  feature may
reflect its trapping into MMRs at the early stages of the  formation and evolution.

Using a new determination of the masses and radii, we found a  curious mass-radius
relation implying a clear anti-correlation between the mean  density of the planets
and their distances from the star. Their densities  exhibit a sequence of planet~b
which is denser than Neptune, through the  Neptune-like planet c, Uranus-like
planet~d, Jupiter-like planets e~and f,  and planet~g which is likely 
Jupiter/Saturn-like. 

The paper is structured as follows. In Sect.~\ref{sec:modeling}, we shortly 
describe the photometric data of Kepler-11 available in the MAST archive. We also  
refine the
observational TTV model. In Sect.~\ref {sec:results}, we present the
results derived through intensive  computations with the bootstrap algorithm.
Furthermore, we discuss a possible  composition of the planets
(Sect.~\ref{sec:composition}). Section~\ref {sec:dynamics} is devoted to the
dynamical analysis of the 
Kepler-11 system.  Conclusions and prospects for a future work are
given at the end of this paper.
%
%
\section{Transits in a multi-planet system}
\label{sec:modeling}
The photometric data of Kepler-11 were taken from the MAST archive. At the time of
writing this paper,  the publicly available light-curves span about of $500$~days in six
parts. These data were  binned on $\sim 30$-minute intervals. We analysed 
a ``de-trended'' data set derived through a smoothing 
\corr{procedure.  At first, we isolate all transits from the light curve. 
Then the moving average with a time-step of $0.5$~days
provides the mean level of the flux.  Next, we
construct an interpolated, reference light curve with the cubic spline 
on these nodes.  Finally, we divide the raw flux, with
all transits data, by its values of the reference, mean level flux curve.  
}

The
de-trended data available in the  MAST database exhibit a growth of the flux shortly
before and after a  particular transit.  In some parts of the available
light-curves, spanning  approximately $\corr{300}$~days, the measurements appear in the raw
form.  We did not use these data, aiming to analyze a possibly uniform set of 
observations.
%
%
\subsection{Modeling the stellar flux}
%
A common model of photometric observations of a star transited by  planetary
companions consists of two major parts. The first part concerns  the flux deficit
due to small, dark objects passing in front of the star.  At first, the average
orbital periods are determined. Then   transit depths and duration times are
parametrized  on the basis of phase--folded light-curves. Single mid-transit times
are also determined. At the second level, we can estimate the planetary masses and
orbital elements fitting a model of motion of mutually interacting planets.

We focus on the first level of the photometric analysis.  To compute the flux
deficit, we use the quadratic limb darkening model \citep{Mandel2002}, recalling
that the Kepler-11 light-curves are relatively noisy and sampled with a low
frequency,
\begin{equation}
\Delta I(r) = 1 - \gamma_1 \left( 1 - \cos\theta \right) - 
              \gamma_2 \left( 1 - \cos\theta \right)^2,
\end{equation}
where $r$ is the normalized radial coordinate w.r.t. the centre of the stellar disk,
$\theta$  is the angle between the direction to the observer and the normal to the 
stellar surface. The two limb-darkening coefficients $\gamma_1$ and $\gamma_2$  must be
positive and $\gamma_1 + \gamma_2 < 1$ \citep[see a study  of the limb darkening
coefficients for a few target stars of the Kepler  mission,][]{Howarth2011}. For
small ratio $p \equiv R_p/R_s$ of planet radius $R_p$ to the stellar radius $R_s$,
\cite{Mandel2002} found  an analytic approximation of the flux deficit, $\Delta F =
\Delta F(z; p,  \gamma_1, \gamma_2)$ which depends on the normalized distance~$z$
between  the centers of stellar and planetary disks, projected onto the sky plane 
(see Eq.~8 in the cited paper), as well as on $p$ and $\gamma_{1,2}$.

If more than one planet transits the star at the same time, the total flux  deficit
can be computed as the sum of the deficits caused by particular  planets.
Obviously,  $\gamma_1, \gamma_2$ are the same for all planets,  while $p$ and $z$
are different for each object. If transiting planets are small,  we can use a simple
model of independent transits rather than  more  general treatment
\citep[e.g.,][]{Pal2011}. Because we model the  photometric measurements directly, 
by {\em reconstructing the whole light-curve},  we are  not restricted to single
transits and mid-transit times. Also   multiple transits can be covered. In
light of relatively narrow observational window, multiple transits are very helpful
to constrain orbital elements of the transit model.
\begin{figure*}
\hbox{\includegraphics[width=0.33\textwidth]{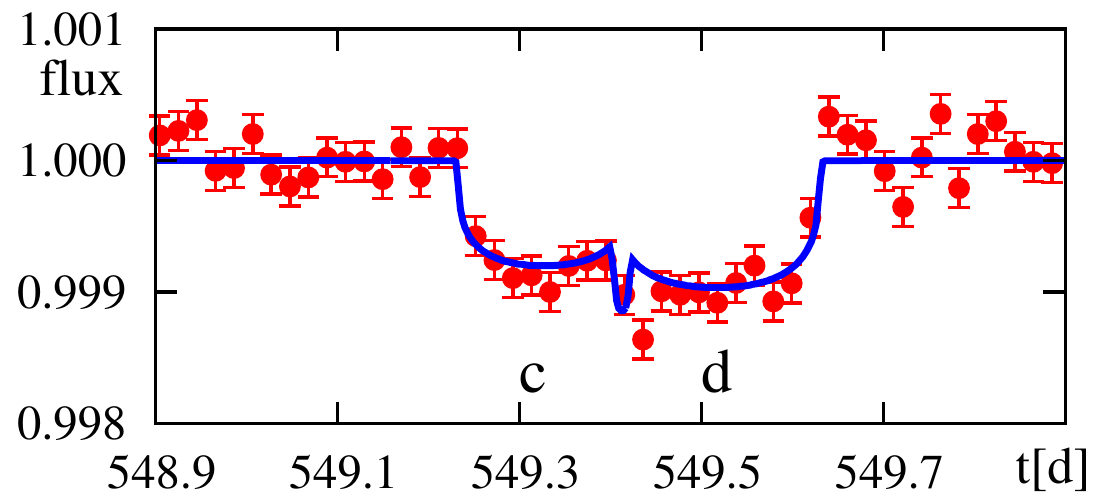}
\includegraphics[width=0.33\textwidth]{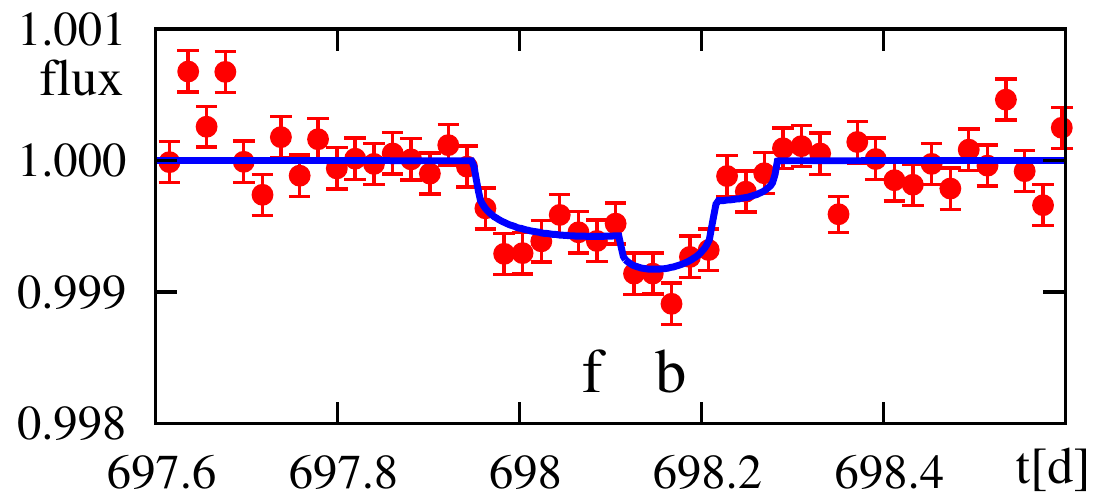}
\includegraphics[width=0.33\textwidth]{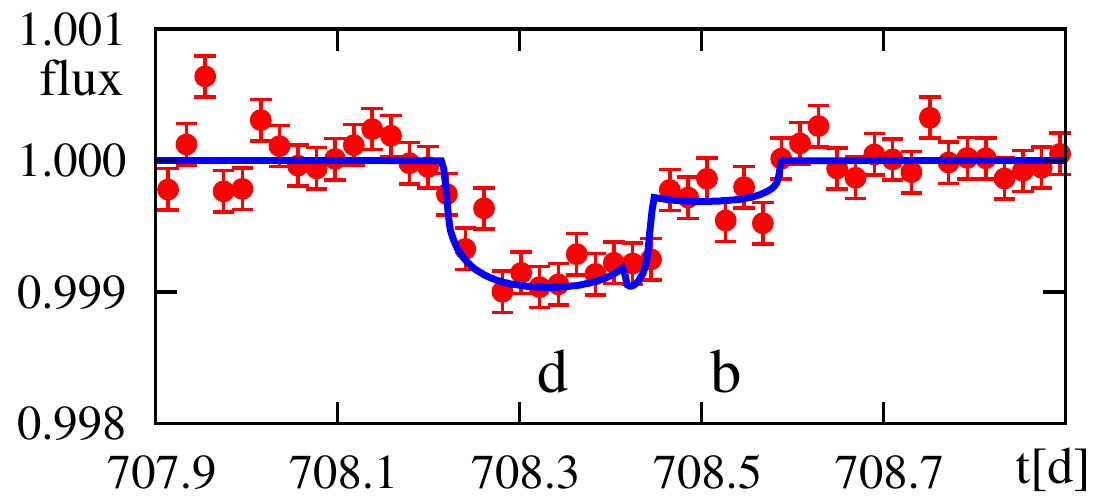}}
\hbox{\includegraphics[width=0.33\textwidth]{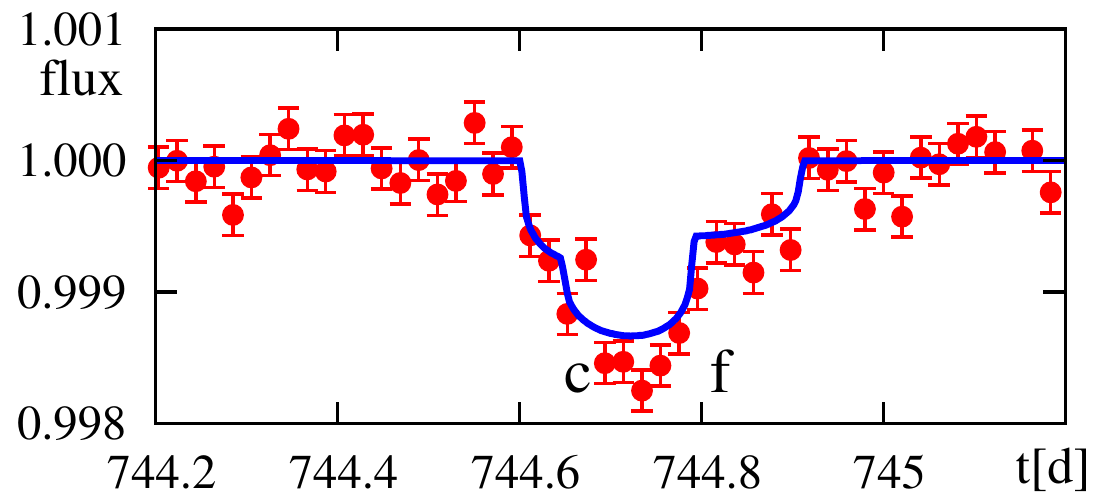}
\includegraphics[width=0.33\textwidth]{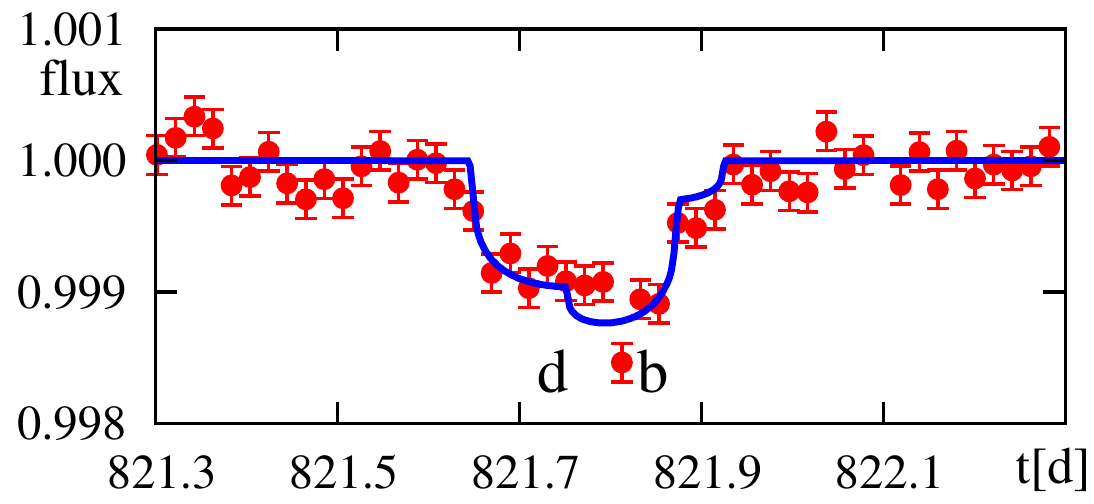}
\includegraphics[width=0.33\textwidth]{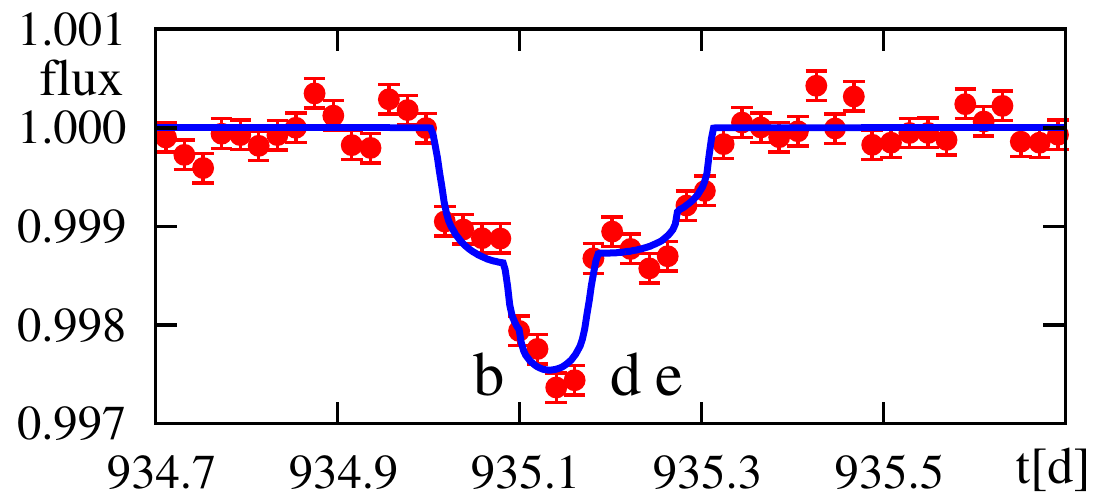}}
\caption{
Sample synthetic light curves over-plotted on photometric measurements of
Kepler-11.  Red dots are for the observational data,  blue solid curve is for
synthetic light-curve derived in this work.  The reference epoch is
JD~2,455,500.
}
\label{fig:light_curves2}
\end{figure*}

Figure~\ref{fig:light_curves2} displays a few selected fragments of the  data set
marked with red dots and error bars which are over-plotted on  the synthetic curve
best--fitting the data (blue curve). The fitting  procedure will be described in
more detail in Sect.~2.3. The last panel shows  transits of three planets (b, d
and~e). 
%
%
\subsection{The model of orbital motion}
%
The orbital motion of multiple planetary system is described in terms of the full
$N$-body problem in the Poincar{\'e} reference frame
\citep[e.g.,][]{Morbidelli2001}.  In this frame, the Cartesian coordinates of  the
planets are astrocentric, while their velocities are barycentric. The  equations of
motion are integrated with the second order symplectic integrator SABA$_2$ 
\citep{Wisdom1991, Laskar2001}. It provides 2-3 times  better CPU performance than
other algorithms, which we tested (like the  Bulirsh-Stoer-Gragg scheme, BGS)
constrained with the same time--step accuracy.  To speed-up the computations even
more, we did not  integrate the system at all measurements moments. This would
force  $\sim 30$-minute step-size of the integrator. Instead, we fixed this 
step-size to $\Delta t \sim 1/20$ of the innermost period of planet b,  i.e.,
$\Delta t \approx 0.5$~day. Furthermore, the flux function $F(t)$  is computed only
close to the mid-transits. Ingress and egress times   of particular events are
tabulated. When a transit takes  place, the coordinates of particular planet at time
$t$ required to evaluate the  flux deficit are determined through the polynomial
interpolation on five  nodes around $t$. Through a comparison with the direct,
full-accuracy  integrations with the BGS algorithm, we found that the selected  time
time-step and the number of interpolation nodes provide 
a sufficient precision and acceptable CPU overhead. 
\corr{We examined this method by changing the number of nodes
in the polynomial interpolation, 
as well as the time step-size.  The flux
level, interpolated on five nodes and with $\Delta~t \approx 0.5$~day, 
differs from its exact value by less than $10^{-9}$.}
%
%
\subsection{Optimization algorithm and error estimation}
%
We searched for the best--fit model of the transits by a common minimization of the
$\chi_{\nu}^2$ function. This function is defined as follows:
\begin{equation}
\chi_{\nu}^2 = { \frac{1}{ N_{\idm{obs}}-N_{\idm{p}}-1} 
\sum_{j=1}^{N_{\idm{obs}}} \frac{1}{\sigma_j^2} \Big[ F_j - F(t_j) \Big]^2},
\label{eq:chi_def}
\end{equation}
where $N_{\idm{obs}}$ is the number of observations,  $N_{\idm{p}}$ is the number of
free parameters, $\nu=N_{\idm{obs}}-N_{\idm{p}}-1$ is the number of  the degrees of
freedom, $\sigma_j$ is the error of the $j$-th observation  $F_j$, and $F(t_j)$ is a
model function evaluated at time $t_j$. This form  of the $\chi_{\nu}^2$--function
is correct if the uncertainties are uncorrelated \citep[see, e.g.,][]{Baluev2009}.
\corr{
To verify whether the available photometric data fulfill this assumption,
one has to use a more general statistical model incorporating the red--noise
effect.  However, under particular settings of our 
$N$-body photometric model, this would require an
enormous CPU overhead.  Hence,  we use equation~\ref{eq:chi_def} as
a reasonable first order approximation.
}

The best--fit parameters of the transits model are searched through a two--step 
optimization method. In the first step, we apply a robust and well tested
quasi-global Genetic Algorithm  \citep[GA,
see,][]{Charbonneau1995,Deb2004}\footnote{We use publicly  available implementation
by Kalyanmoy Deb, see \url {http://www.iitk.ac.in/kangal/pub.htm}} which makes it
possible to find  promising solutions. A local, fast  gradient method (here, the 
Levenberg-Marquardt algorithm) is then used to refine the solutions found  in the GA
step. Such an approach is called the hybrid optimization \citep [see][and references
therein]{Gozdziewski2008}. Let us note that the  parameter space is huge as it has
dimension of 50. Some of these parameters  can be determined very well, like the orbital
periods of the transiting  planets. Unfortunately, due to the relatively short
observational time window,  many parameters which are critical for the stability
(relative  inclinations, masses, nodal lines) cannot be well constrained. It makes
the  fitting process a challenging problem.

The parameter errors are estimated through the bootstrap algorithm \citep[see, 
e.g.,][]{Press1992}. The bootstrap is CPU--demanding, but it is straightforward
method to estimate standard errors in  high-dimensional problems  and for large
number of data. The light--curves which we analyzed have $\sim$22,000 points.  The
bootstrap algorithm requires to find the best--fit solutions to a large number of 
synthetic sets derived through random sampling with replacement from the original
measurements. To obtain reliable error estimates  of the best--fit parameters, one
needs at least $\sim 10^3$--$10^4$  synthetic solutions. When such a large set of
the best--fit models  is gathered, we constructed normalized histograms for each
free parameter.  These histograms reflect the parameter distribution in response to
the errors of the measurements, and may be  smoothly approximated by an asymmetric
Gaussian function. This makes it possible to determine the standard uncertainties.
To perform the bootstrap  procedure, at first one needs to find reliable best--fit
parameters for the  nominal data set. This step was done through an intensive
quasi-global search  with the help of the hybrid algorithm. The bootstrap
computations are CPU-time  consuming and were performed on the \verb+reef+{}
CPU-cluster  of the Pozna{\'n}{} Supercomputing Centre.
%
%
\subsection{Numerical setup of the dynamical analysis}
%
%
In spite of small eccentricities and apparently co-planar orbits, the  Kepler-11
systems is orbitally very active.  It appears as dynamically packed planetary system
\citep[the definition is given in][]{Barnes2008},  with only narrow stable zones in
the phase space. For this reason, we used the  best-fit model solutions  gathered in
the bootstrap search as the input  data to extensive dynamical study of this system.
As we will discuss later,  a study of the stability is a challenging problem. Due
to  relatively short observational window ($\sim 500$~days), weak transits  having
depths comparable with the measurements errors and 
\corr{a small number of data points covering particular transits (typically
$10-15$)}, the derived initial conditions may be 
shifted away from the real configurations. 

To investigate the dynamics of the Kepler-11 system in a global manner, we applied
an approach in our previous  papers which is well established in the literature. It
relies on reconstructing the structure of the phase space with  the fast indicator
MEGNO \citep[][]{Cincotta2000,Cincotta2003}.  This dynamical characteristic makes it
possible to distinguish between regular (stable) and irregular (chaotic, unstable)
trajectories in the phase space by computing relatively short numerical orbits.
Having  representative solutions selected in the bootstrap  statistics, we study
their neighborhood on the dynamical maps.  Constructing a dynamical map relies on
two model parameters, e.g., the semi-major  axes of a pair of planets. The selected
parameters are varied in the given  range at a discrete grid. The remaining
components of the initial parameter vector are fixed at their nominal values. If it
is necessary, they are  altered to preserve the observational constraints. Then we
calculate MEGNO at  each point of the grid. Dynamical maps are informative and
become a standard  numerical tool helpful to understand the global dynamics of
multiple systems.

To compute the MEGNO indicator, we must solve the variational equations to the
equations  of motion of the planetary $N$-body problem. The Kepler-11 system
architecture with  low--eccentric orbit and small masses is an ideal target for an
efficient symplectic algorithm  described in
\citep{Gozdziewski2003a,Gozdziewski2008a}.  The general-purpose integrators, like
the Runge-Kutta or  Bulirsh-Stoer-Gragg schemes are  not efficient  nor accurate
enough in this  case. These methods introduce a systematic drift of the energy and
other  integrals. To avoid such errors, and to solve the variational equations, we
apply the tangent map  introduced by \cite{Mikkola1999}. As the very basic step, it
requires to differentiate the ``drift'' and  ``kick'' maps of the standard
leap--frog algorithm.  The variations may be  then propagated within the same
symplectic scheme, as the equations of motion.  Having the variational vector
$\v{\delta}$ computed at  discrete times, we  find temporal $y$ and mean $Y$ of the
MEGNO at the $j$-th  integrator step $j=1,2,\ldots$,
\citep{Cincotta2003,Gozdziewski2008a}:
\begin{eqnarray*}
Y(j) &=& \frac{(j-1) Y(j-1) + y(j)}{j}, \\
y(j) &=& \frac{j-1}{j} y(j-1) + 2\ln \left(\frac{\delta_j}{\delta_{j-1}}\right)
\end{eqnarray*}
with initial conditions $y(0)=0$, $Y(0)=0$, $\delta = \norm{\v{\delta}}$. The MEGNO
maps tend  asymptotically to
\[
 Y(j) = a h j + b,
\]
where $a=0,b \sim 2$ for quasi-periodic orbits, $a=b=0$ for stable, periodic  orbit,
and $a=(1/2)\sigma, b=0$ for chaotic orbit with the maximal Lyapunov  exponent
$\sigma$.  The  tangent MEGNO map is linear, hence the variational vector can be
normalized, if its value  grows too large for chaotic orbits. In practice, we stop
the integration if  the MEGNO indicator reaches a given limit (usually, $Y=5$). 

The symplectic maps were propagated with the 4-th order SABA$_4$ scheme in
\citep{Laskar2001}.  A choice of the fixed step-size must be carefully controlled.
We   did this, checking whether the relative energy error  is ``flat'' across the
dynamical map \citep{Gozdziewski2008a} and sufficiently small. Indeed, the
step--size $\sim 0.5$~day preserved this error at a level of  $10^{-11}$ over the
total integration times up to $T \sim 40,0000$~yr  ($\sim 100,000$ periods of the
outermost planet). This time scale is long enough to detect the most significant
2-body and 3--body MMRs  though even such integration period may be insufficient to
detect all ``dangerous'' unstable resonances. Weakly chaotic motions due to
multi-body MMRs still may lead to catastrophic  events after much longer time
\citep{Gozdziewski2008a}.

The dynamical maps in this paper have typical resolution up to $512\times512$ 
pixels. This requires an enormous CPU-time. It is basically not possible to  perform
such intensive computations on a single workstation. Therefore, we  used our new
Message Passing Interface (MPI) based  environment \mechanic{} \citep{Slonina2012}
to perform the computations in a reasonable time in CPU-clusters\footnote{Informations
on this project may found at \url {http://git.astri.umk.pl/projects/mechanic}.}. 
They were performed on the {\tt reef} cluster at the Pozna\'n{}  Supercomputing
Centre. A \mechanic{} run of a typical dynamical map occupied up to 1200 CPU cores
for $\sim 16$~hours.
%
%
\subsection{Free parameters of the transit model}
%
The free parameters of the transit model are the stellar radius $R_0$, the limb
darkening  coefficients $\gamma_1, \gamma_2$; the mass $m_i$, radius $R_i$ and 
orbital elements of each planet in the system, where $i=$b,c,d,e,f,g.  Planetary
orbits are described through the Poincar\'e{} geometric, osculating elements at the
epoch of the  first observation JD~2455964.51128: a tuple $(a_i, e_i, I_i, \Omega_i,
\omega_i, \Mmean_i)$ is for the semi-major axis,  eccentricity, inclination to the
plane of the sky, the longitude of ascending  nodes, the argument of pericenter, and
the mean anomaly, respectively. The  orbital node of the first planet,
$\Omega_{\idm{b}}=0^{\circ}$ due to invariance of the model with respect to a
rotation of the whole  system. The inclinations are obviously close to $90^{\circ}$.
A deviation  from $90^{\circ}$ is irrelevant only for single-planet systems. 

In a multi-planet system, some orbits may be inclined to the sky plane  by angle
$\neq 90^{\circ}$, which implies different {\em relative}  inclinations between
orbits of particular planets, even for the same  longitudes of nodes. Due to the
invariance of transits with respect to the  direction of the total angular momentum
of the system, a combination of  $(I_i \le 90^{\circ},\Omega_i)$ means the same
geometry as  $([180^{\circ}-I_i] \ge 90^{\circ},-\Omega_i)$. Thus, when necessary,
for a  given planet~p we can fix the range of $I_{\idm{p}} \leq 90^{\circ}$ and 
$(I_i,\Omega_i)$ are corrected for remaining companions, in accord  with the
invariance relation.

Orbital elements $(a_i, e_i, \omega_i, \Mmean_i)$ are not fully  suitable for
transiting systems with small relative inclinations and small  eccentricities. To
avoid singularities and weakly constrained elements, like $\omega_i$ when $e_i=0$
(circular or weakly eccentric orbits), we use the Poincar\'e{} modified elements 
$(X_i \equiv e_i \cos\omega_i$, $Y_i  \equiv e_i \sin\omega_i)$ instead of $(e_i,
\omega_i$).

Similarly, the orbital period $P_i$ is more suitable for model fitting than $a_i$ 
since the semi--major axis depends on the planetary mass $m_i$ (a free parameter) 
and on the stellar mass $m_0$, which is fixed to  $0.95\,\msun$, but it can be also
fitted. Hence, we define $P_i$ as one of  osculating elements related to $a_i$
through the IIIrd Keplerian law. 

The mean anomaly $\Mmean_i$ strongly depends on $\omega_i$. It  determines the
relative orbital phase (the mean longitude) but is also  related to $e_i$. This may
be avoided by choosing the time of the first  transit  $T_i$ as a free  parameter
instead of $\Mmean_i$, because it is one of the directly determined observables from the
light-curves. Simple relations between $T_i$ and Poincar\'e{}  canonical elements
may be derived easily.
%
\subsection{{\em Direct} and {\em indirect} transit model parameters}
%
The {\em direct} parameters of the transit model are determined from the basic 
observables: it is the mean period of transits $P^*_i$, the depths, duration  times
of the transits, and shapes of the light-curves (through the limb--darkening
coefficients). These data are  usually derived from the period--phased light--curves
of particular planets.  The depths and durations of transit determine the ratio of
planetary and stellar  radii, $R_i/R_0$. If the stellar mass $m_0$ is fixed then
$R_i$ and  $R_0$ may be resolved. We can also determine $I_i$ up to the angular 
momentum direction invariance, and $T_i$. In general, the mean period of  transit
events $P^*_i$ is different from the osculating orbital period at  the epoch of the
first observation, $P_i$. A shape  of the event--period phased light--curves make it
possible to fit the limb  darkening coefficients, $\gamma_1, \gamma_2$. 

These parameters of the transit model are independent on the  the $N$-planet
dynamics. Hence the remaining are {\em indirect} parameters.  To resolve them, a
dynamical model of the orbital evolution  is required. The indirect parameters
consist of planetary masses $m_i$ as well  as orbital elements, $e_i$, $\Omega_i$,
$\omega_i$ and $P_i$ (instead of  $P^*_i$). Knowing $m_i$ and $P_i$, we may fix the
osculating semi-major axis  $a_i$ at the date of the first (or prescribed)
observation.

\corr{
We would like to note, that the above distinction for two types of model
parameters is somehow arbitrary in our photometric model.  In our algorithm
both the direct and indirect parameters are fitted simultaneously, unlike,
for instance, the TTV algorithm, in which the direct parameters are fitted
at the first stage, and the indirect parameters are fitted in the next step.
}

Usually, the direct parameters can be estimated much more reliably  than the
indirect parameters. Even a potential derivation of the  indirect parameters depend
on the particular model of motion. i.e., kinematic ---  Keplerian, or dynamic ---
Newtonian, and on the used method of modelling the  observations.  In the Keplerian 
(kinematic) model \citep[see, e.g.,][]{Agol2005}, mid-transits of a given  planet
are governed by {\em geometric} reflex motion of the star  around the center of mass
in a sub-system composed of the star and all  inner planets.  For instance, transits
of planet~d are affected by  planets~b and~c, but any outer planet does not affect
transits of its  inner companions. Hence, in accord with the Keplerian model, the
indirect  parameters  $(m_{\idm{g}}, a_{\idm{g}}, e_{\idm{g}}, \Omega_{\idm{g}},
\omega_{\idm{g}})$  of the outermost planet~g in the Kepler-11 system  cannot be
determined at all.

In a given pair of planets, the outer companion affects the transits times  of the
inner planet only through gravitational mutual perturbations  which lead to changes
of osculating orbital elements. To account for the mutual  interactions,  one has to
apply the self-consistent $N$-body model of  motion of the system.

Usually, to resolve the indirect parameters from photometric observations, the well
known TTV  method is used~\citep{Agol2005}. It has two steps. At first, we
determine  the mean periods, the mid-transits, and then the (O-C) residua,  i.e.,
differences between the measured and ephemeris transit times.  The  (O-C) variations
are observables in the second step during which  we search for masses,
eccentricities, and arguments of pericenters of  planetary companions.  The TTV
method in this form has a limitation, because  it does not make any use of transit
depths nor their duration times. If the  individual inclinations of planets are
different, the planets transit the  parent star usually at different attitudes.
Hence the transit depths as well as  duration times may vary, like the (O-C) of
mid-transits. This  information can be used to better constrain the transit model.

The mutual inclinations depend on the longitudes of ascending nodes in accord with
\[
 \cos \Delta I_{i,j} = \cos I_i \cos I_j - \sin I_i \sin I_j \cos( \Omega_i - \Omega_j). 
\]
Because the inclinations of transiting planets, ($I_i$, $I_j$) must be close to 
$90^{\circ}$ then $\Delta I_{i,j} \approx |\Omega_i - \Omega_j|$. Within  this
approximation, the TTV method is apparently not sensitive for  individual
$\Omega_i$. In fact, different values of $\Omega_i$ imply  different mutual
inclinations affecting the dynamics and (O-C). However,  the dynamical variability
of (O-C) due to mutual interactions is weaker  than the geometric variability due to
changes of transit depths and  duration times reflecting the motion of the star
around the mass center of  the system.

Overall, by direct modeling of the light-curves (photometric measurements),  rather
than the mid-transit times, we can resolve the  (O-C) with an improved precision.
Modeling the light-curves in terms of the  $N$-body model is CPU-demanding, but it
makes it possible to  estimate individual longitudes of nodes and mutual
inclinations. In  particular, as will be shown later, the direct method helped  us
to derive accurate relative inclinations between planets~b and~c, as  well as
between~d and~e $\sim 2^{\circ} \pm 2^{\circ}$.
%
%
\section{Results of the bootstrap analysis}
%
\label{sec:results}
We performed the direct bootstrap TTV analysis of a few different orbital  models of
the Kepler-11 system. In the most general case~(I), all  parameters discussed in the
previous section are the free parameters of the fit  model. Some of them are poorly
constrained by the observations, in  particular, the eccentricity of planet~g and
particular longitudes of  nodes. Therefore, we also studied less general models, in
which some of  weakly constrained parameters are fixed. In the second model~(II), 
$X_{\idm{g}}=0$, $Y_{\idm{g}}=0$, i.e., $e_{\idm{g}} = 0$. In the third 
model~(III), also $\Omega_{\idm{g}}=0^{\circ}$, while in the last  model~(IV),
$\Omega_{\idm{b}}, \Omega_{\idm{c}}, \Omega_{\idm{d}},  \Omega_{\idm{e}},
\Omega_{\idm{f}}$ are all fixed at $0^{\circ}$.  Because  inclinations $I_i$ are not
exactly equal to $90^{\circ}$, also $\Delta  I_{i,j} \geq 0^{\circ}$.

For each of these four transit models, we applied the bootstrap algorithm and we
gathered sets of $\sim 1500$ solutions for each instance of the transit model.
%
%
\subsection{Model I: systems with $\pmb{e_g \neq 0}$}
%
%
\begin{figure*}
\hbox{\includegraphics[width=0.33\textwidth]{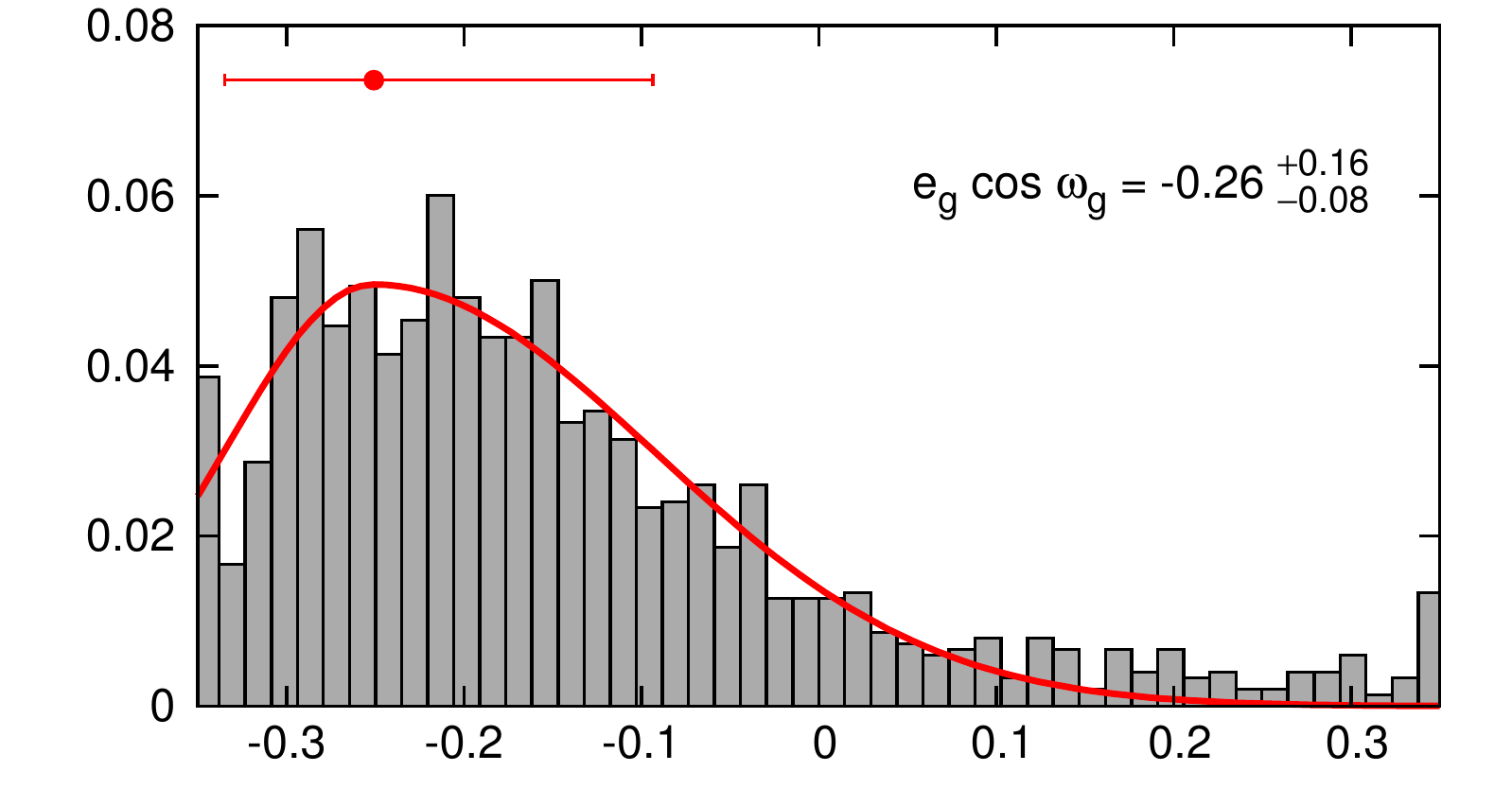}
\includegraphics[width=0.33\textwidth]{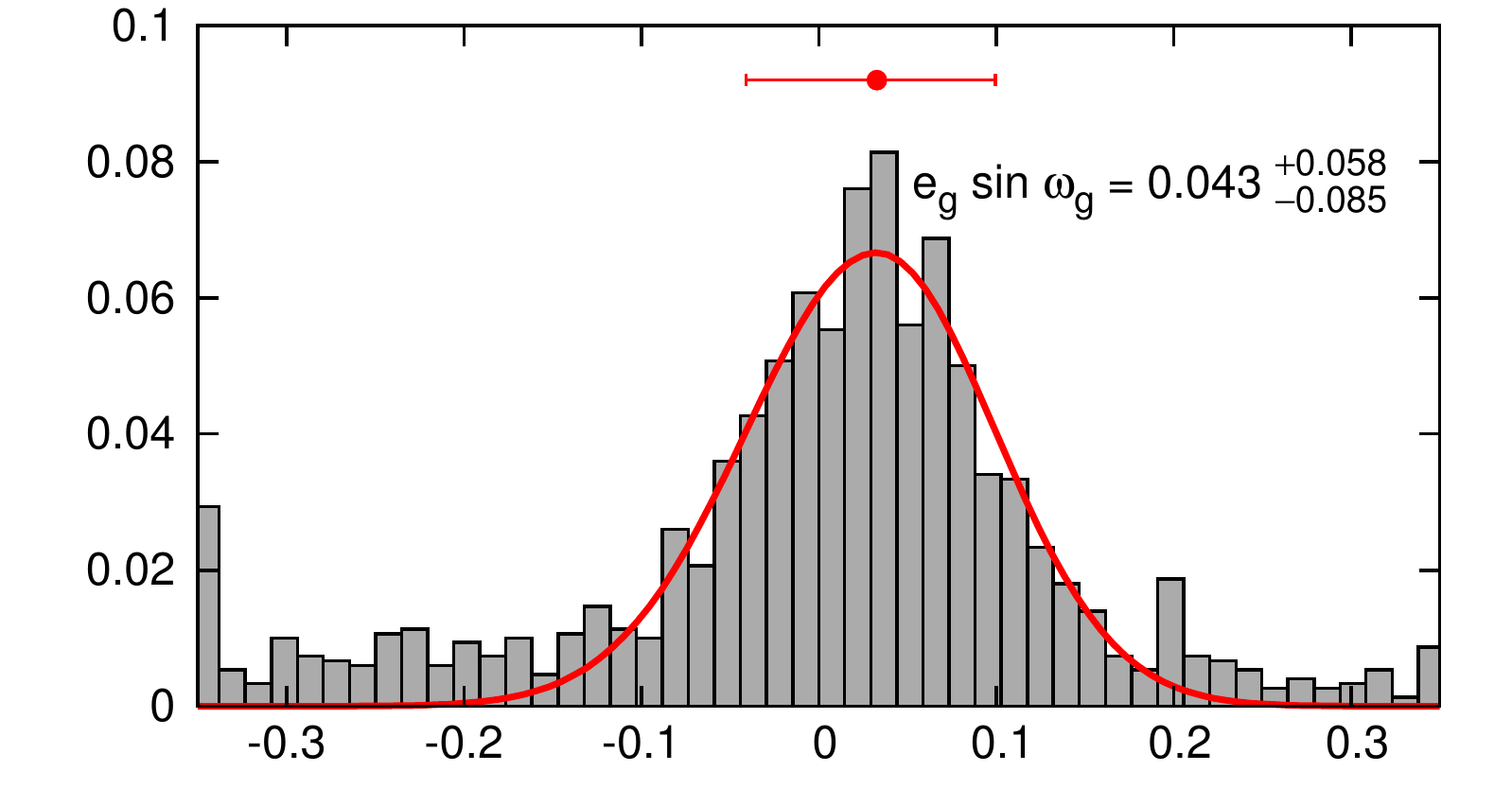}
\includegraphics[width=0.33\textwidth]{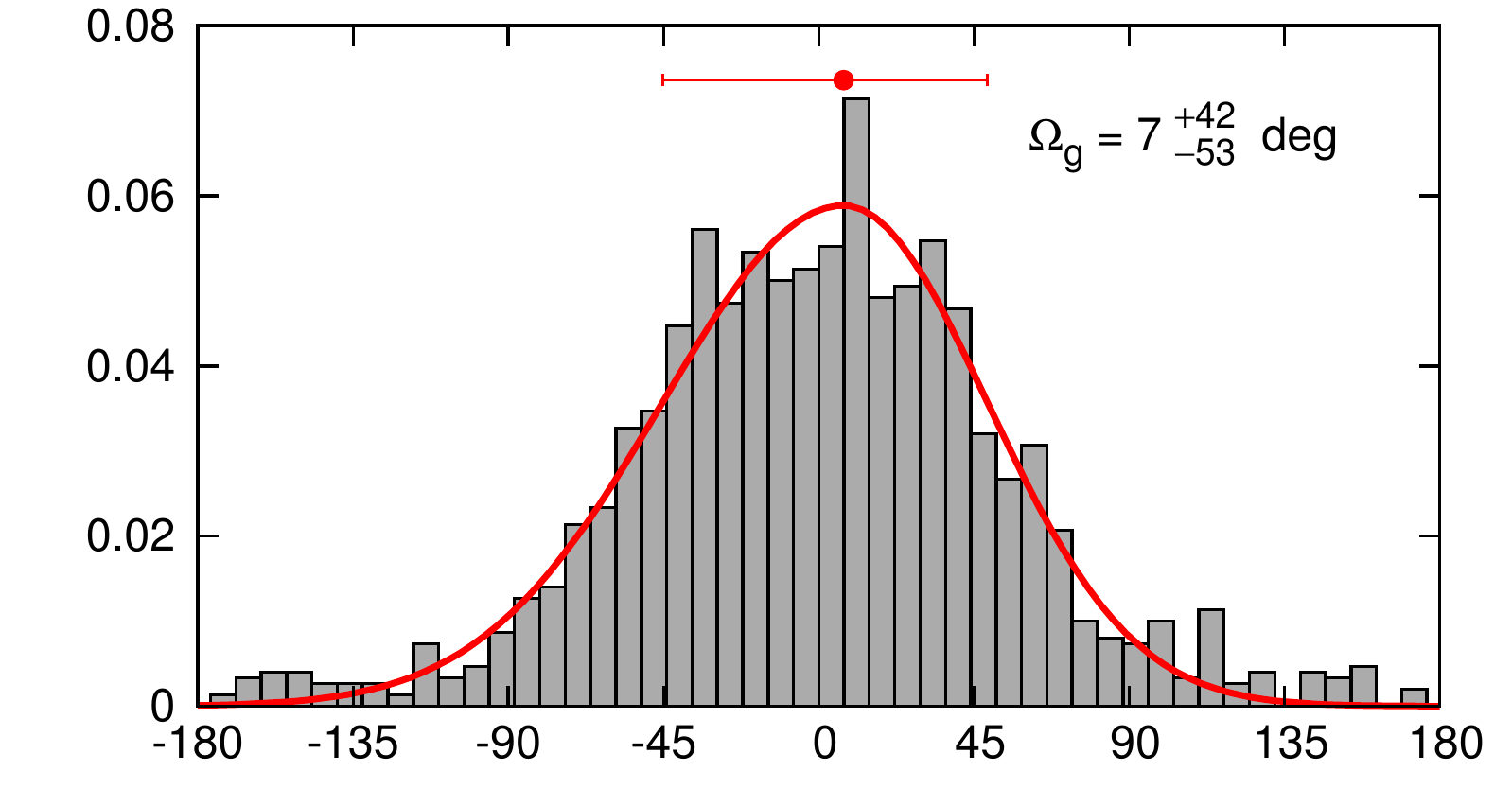}}
\caption{
Bootstrap histograms for $X_{\idm{g}}, Y_{\idm{g}}, \Omega_{\idm{g}}$, transit
model~I. See the text for more detail.
}
\label{fig:modelI}
\end{figure*}
Figure~\ref{fig:modelI} shows an outcome of the bootstrap algorithm in the  form of
normalized histograms  constructed for $X_{\idm{g}}, Y_{\idm{g}}$ and $\Omega_g$,
and  depicted from the left to the right panel, respectively.  The red solid curves
illustrate the best fit asymmetric Gauss function to  the histogram bins. The formal
$1\sigma$ errors are marked with red bars  displayed above the histograms. The best
fit parameters corresponding to  the maximum of the Gaussian distribution are
written in the respective  panels, and they may be compared with the nominal
solutions given in Table~ \ref{tab:bootstrapI}. The uncertainties of the
eccentricity and longitude  of node of planet~g are relatively large. 

Because the nominal system is dynamically unstable, we examined the  whole set of
$\sim 1500$ bootstrap solutions by calculating their MEGNO  indicator $\mmegno$ on
the time interval of $\sim 8000$~yr. It  corresponds to $\sim 25,000$ periods of the
most distant companion. Such a characteristic time scale should be long enough to
detect unstable  solutions due to low--order 2--body and 3--body mean motion
resonances  \citep[][and references therein]{Gozdziewski2008a}. Unfortunately, all 
initial configurations exhibit large values of $\mmegno$, indicating that  the
system is strongly chaotic.  The main source of  instability are crossing orbits in
the system, that lead to disruptive  events
\corr{, i.e., one or more of the planets were ejected from the system or
collided with the parent star}.  None of the tested solutions passed the
direct integration over 10~Myr.

The parameter space of the Kepler-11 system is $\sim 50$-dimensional, and the 
dimension of the phase space of the $N$-body model is $36$-dimensional. The 
$\mmegno$ experiments indicate that this system can be locally chaotic and its phase
space is filled with mostly unstable solutions. Then  only small regions of stable
MMRs may be present. In the light of a large  dimension of the phase space, the
gathered statistics of best--fit configurations  is still very poor. We conclude
that due to short data span of only $\sim 500$ ~days, and unconstrained elements of
the most general model, we cannot find reliably stable solutions assuming the most
general transit model~I. Unfortunately, in this high--dimensional problem an
alternate GAMP algorithm  that relies on the optimization with imposed stability
constrains  \citep{Gozdziewski2008} would be CPU-time expensive.

\begin{table*}
\caption{
Bootstrap results for transit model~I. Mass of the star is $0.95\,\msun$ (fixed). 
The best-fitting stellar parameters of this model are $R_0 =
1.140^{+0.030}_{-0.027}$,  $\corr{\gamma_1 = 0.33^{+0.47}_{-0.30}}$, $\corr{\gamma_2 =
0.41^{+0.24}_{-0.34}}$,  $\gamma_1 + \gamma_2 = 0.74^{+0.23}_{-0.23}$.  Osculating
Poincar\'e{} elements are given at the epoch of the first observation 
JD~2455964.51128.
}
\begin{tabular}{c c c c c c c c}
\hline
\hline
parameter/planet & b & c & d & e & f & g \\
\hline
$m \, [\mE]$ & $4.2^{+2.4}_{-3.0}$ & $9.2^{+3.8}_{-6.9}$ & $8.9^{+3.5}_{-2.7}$ & $10.7^{+2.4}_{-2.1}$ & $3.6^{+5.4}_{-2.6}$ & $18^{+24}_{-15}$  \\
$R \, [\RE]$ & $2.04^{+0.18}_{-0.10}$ & $3.25^{+0.13}_{-0.14}$ & $3.58^{+0.17}_{-0.14}$ & $4.71^{+0.20}_{-0.18}$ & $2.82^{+0.19}_{-0.14}$ & $3.80^{+0.15}_{-0.14}$ \\
$\bar{\rho} \, [\bar{\rho}_{\earth}]$ & $0.50^{+0.38}_{-0.29}$ & $0.27^{+0.14}_{-0.17}$ & $0.19^{+0.07}_{-0.07}$ & $0.10^{+0.04}_{-0.02}$ & $0.16^{+0.22}_{-0.14}$ & $0.33^{+0.47}_{-0.23}$ \\
$a \, [\au]$ & $0.091089\left(^{+13}_{-11}\right)$ & $0.106522\left(^{+7}_{-12}\right)$ & $0.154241\left(^{+19}_{-10}\right)$ & $0.193937\left(^{+15}_{-21}\right)$ & $0.249489\left(^{+39}_{-25}\right)$ & $0.463918\left(^{+59}_{-28}\right)$  \\
$e \, \cos\omega$ & $0.010^{+0.017}_{-0.021}$ & $0.005^{+0.017}_{-0.018}$ & $-0.013^{+0.008}_{-0.022}$ & $-0.020^{+0.008}_{-0.022}$ & $-0.006^{+0.011}_{-0.018}$ & $-0.26^{+0.16}_{-0.08}$   \\
$e \, \sin\omega$ & $-0.011^{+0.031}_{-0.025}$ & $-0.004^{+0.028}_{-0.020}$ & $-0.009^{+0.006}_{-0.015}$ & $-0.016^{+0.007}_{-0.011}$ & $-0.017^{+0.016}_{-0.021}$ & $0.008^{+0.058}_{-0.085}$   \\
$I^* \,$~[deg] & $88.40^{+0.76}_{-0.42}$ & $91.17^{+0.40}_{-0.20}$ & $89.18^{+0.22}_{-0.17}$ & $88.743^{+0.062}_{-0.060}$ & $89.30^{+0.12}_{-0.09}$ & $90.23^{+0.16}_{-0.11}$  \\
$\Omega \,$~[deg] & $0$ (fixed) & $3.2^{+4.2}_{-2.9}$ & $-33^{+13}_{-11}$ & $-31^{+12}_{-11}$ & $-32^{+30}_{-27}$ & $-65^{+51}_{-46}$  \\
$\mathcal{M} + \omega \,$~[deg] & $204.5^{+2.2}_{-2.5}$ & $265.3^{+2.0}_{-2.0}$ & $182.8^{+2.3}_{-1.0}$ & $197.4^{+1.8}_{-1.4}$ & $89.5^{+1.5}_{-1.8}$ & $6^{+7}_{-19}$  \\
$P \, [\mbox{d}]$ & $10.3023\left(^{+24}_{-18}\right)$ & $13.0284\left(^{+12}_{-20}\right)$ & $22.7002\left(^{+41}_{-24}\right)$ & $32.0051\left(^{+42}_{-46}\right)$ & $46.700\left(^{+11}_{-6}\right)$ & $118.410\left(^{+16}_{-10}\right)$  \\
$T_0 \, [\mbox{JD}]$ & $471.505\left(^{+20}_{-7}\right)$ & $471.175\left(^{+20}_{-4}\right)$ & $481.455\left(^{+16}_{-6}\right)$ & $487.178\left(^{+22}_{-8}\right)$ & $464.673\left(^{+15}_{-8}\right)$ & $501.916\left(^{+42}_{-21}\right)$  \\
\hline
\hline
\end{tabular}
\label{tab:bootstrapI}
\end{table*}

%
%
\subsection{Transit model II: systems with $\pmb{e_g = 0}$}
\begin{figure*}
\hbox{\includegraphics[width=0.33\textwidth]{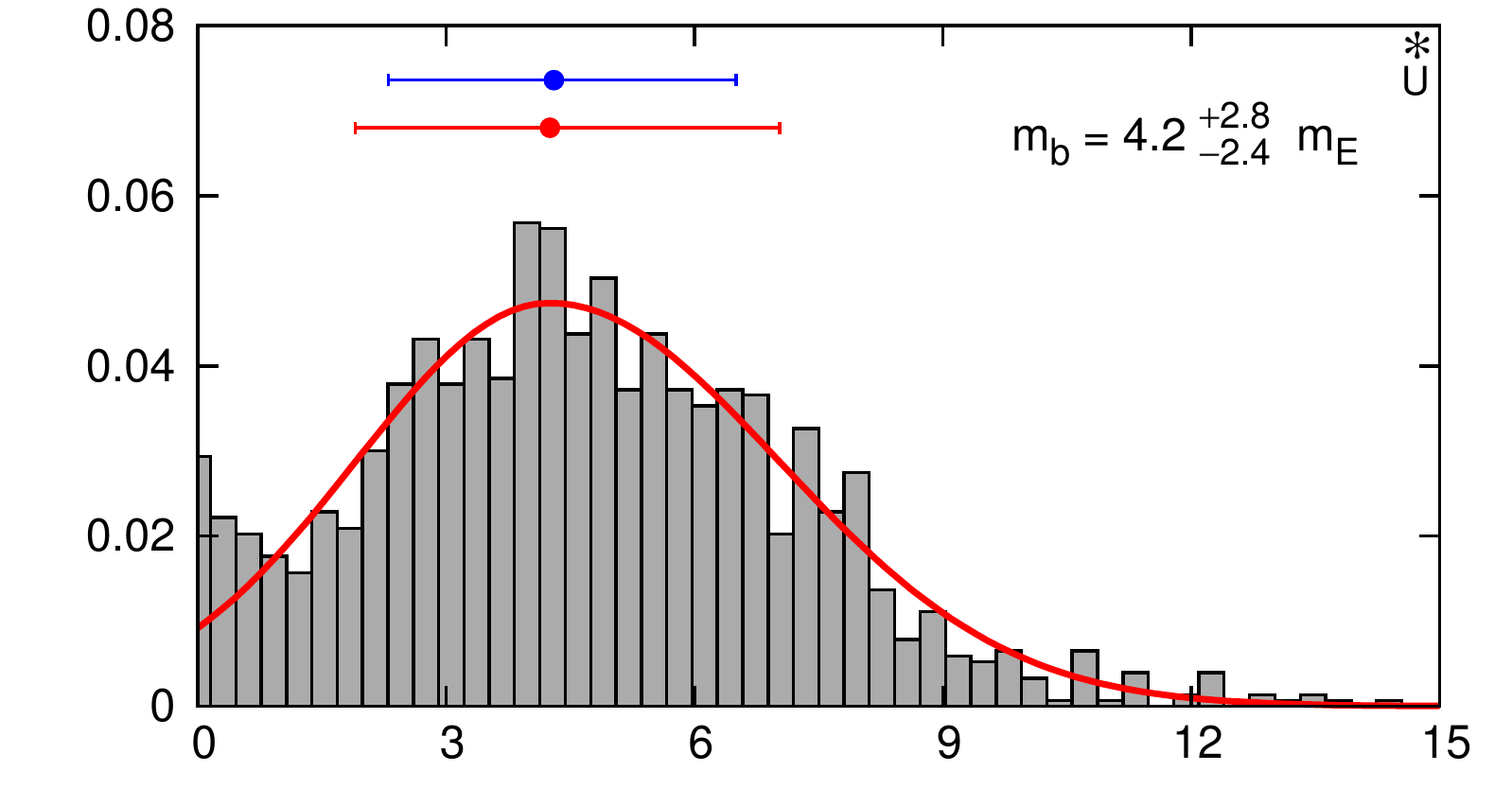}
\includegraphics[width=0.33\textwidth]{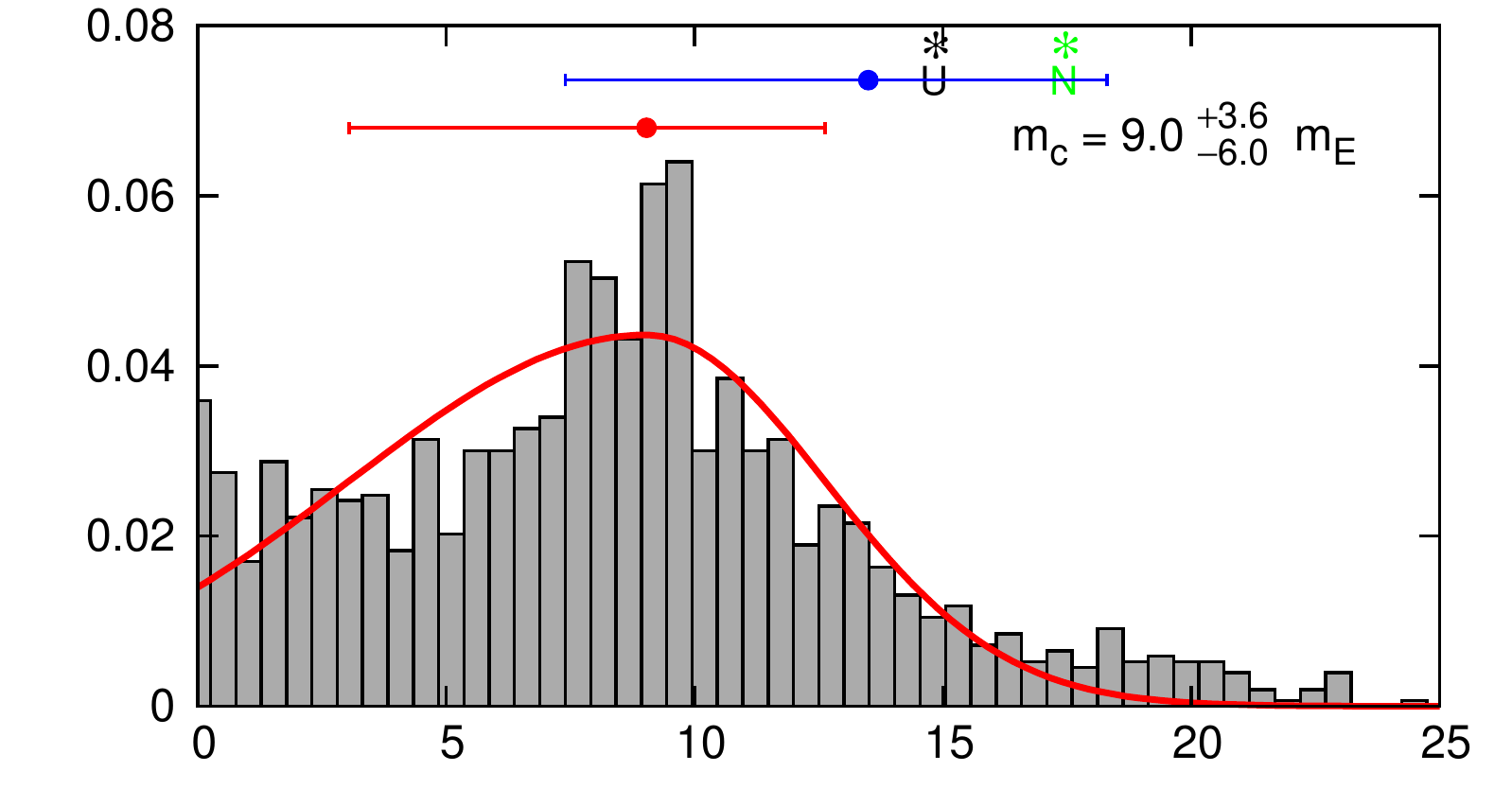}
\includegraphics[width=0.33\textwidth]{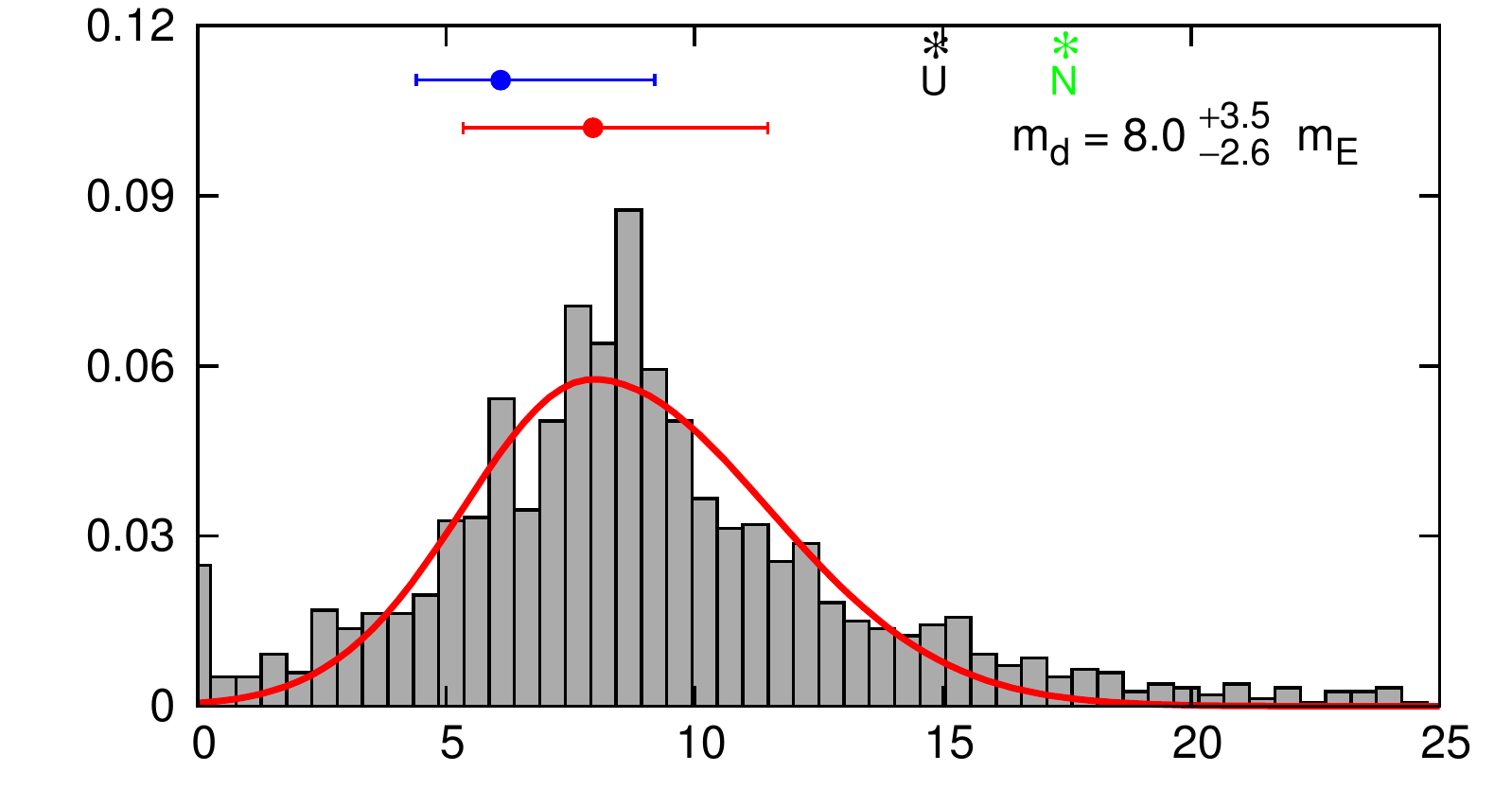}}
\hbox{\includegraphics[width=0.33\textwidth]{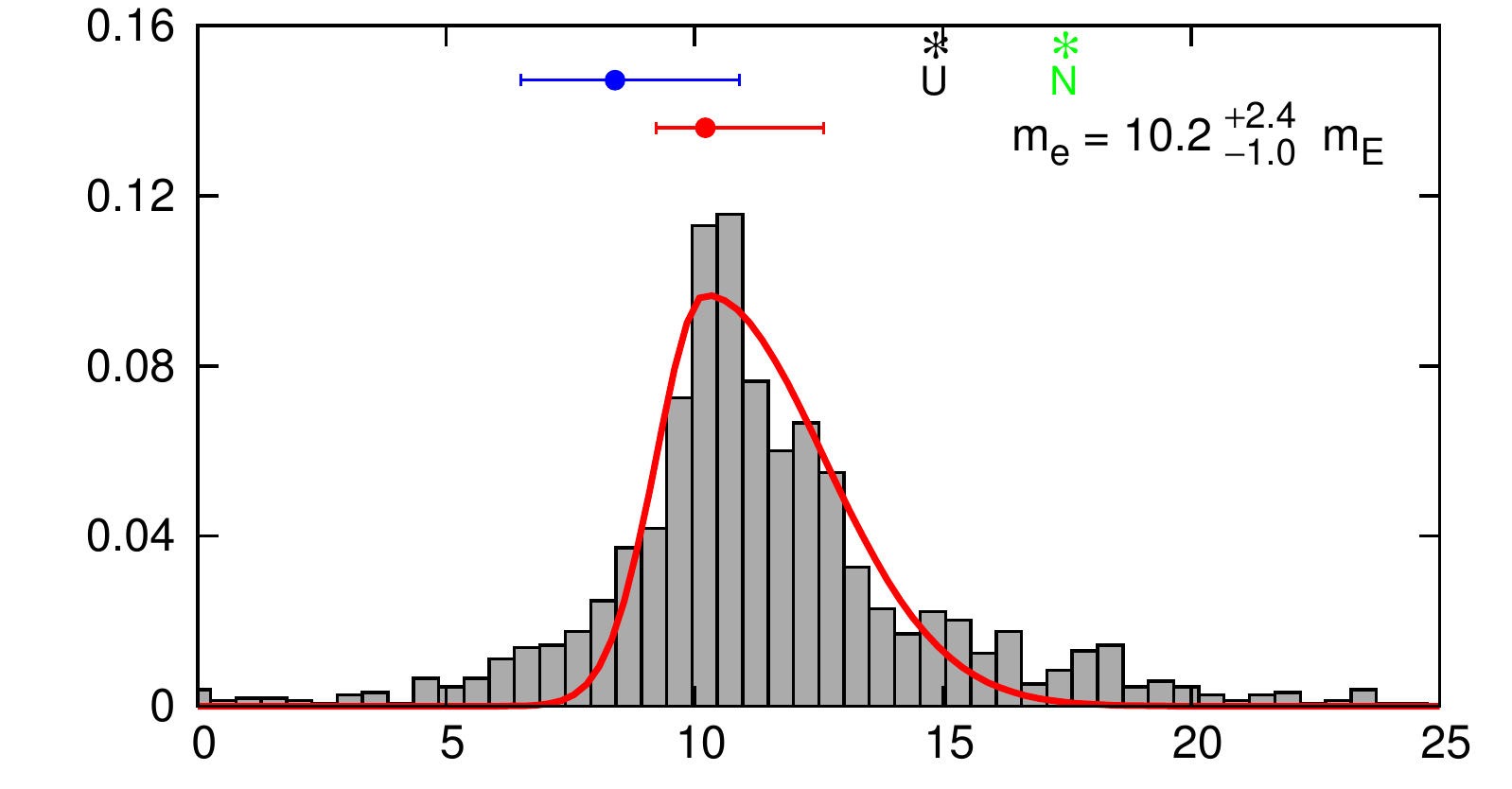}
\includegraphics[width=0.33\textwidth]{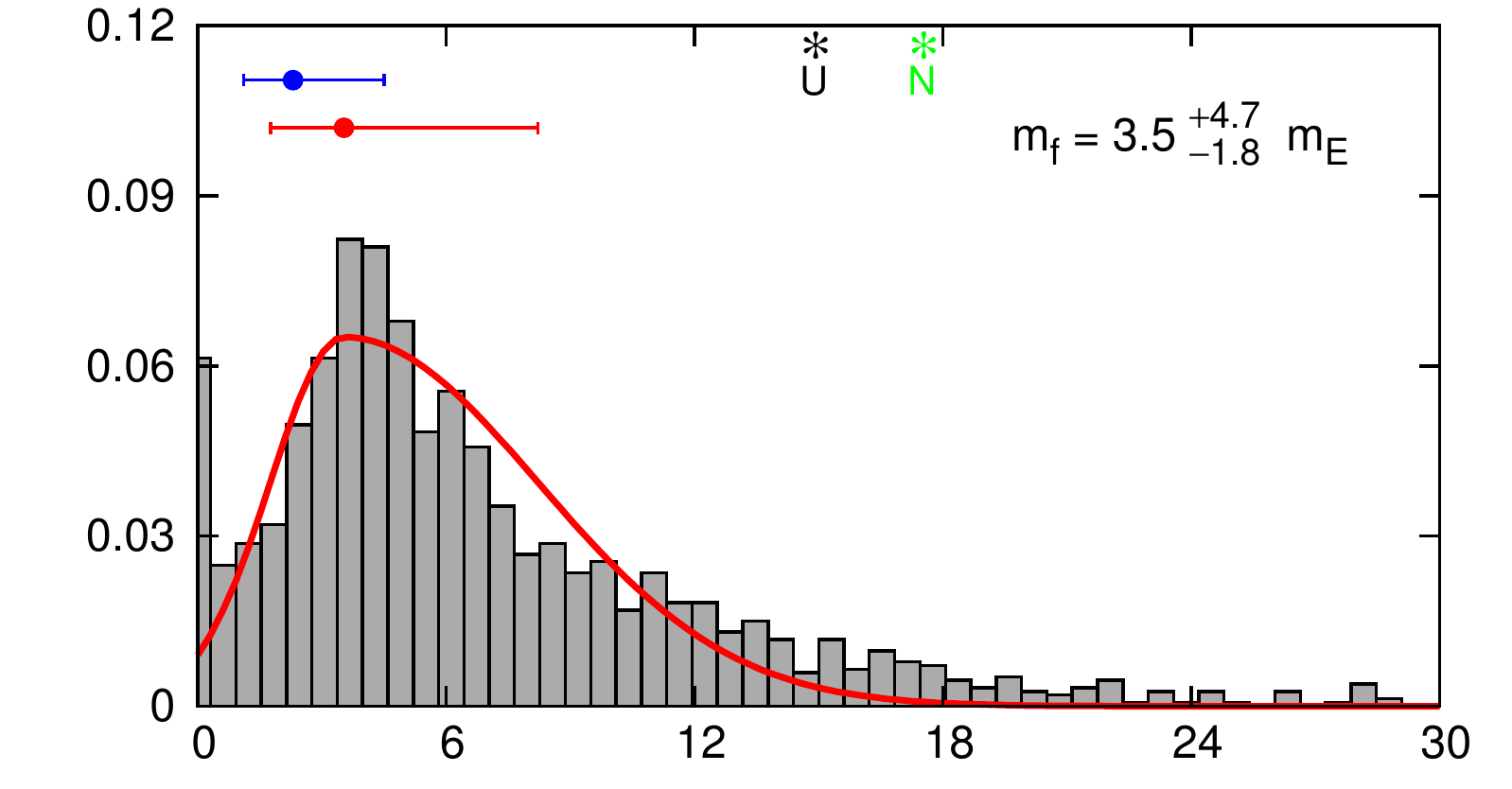}
\includegraphics[width=0.33\textwidth]{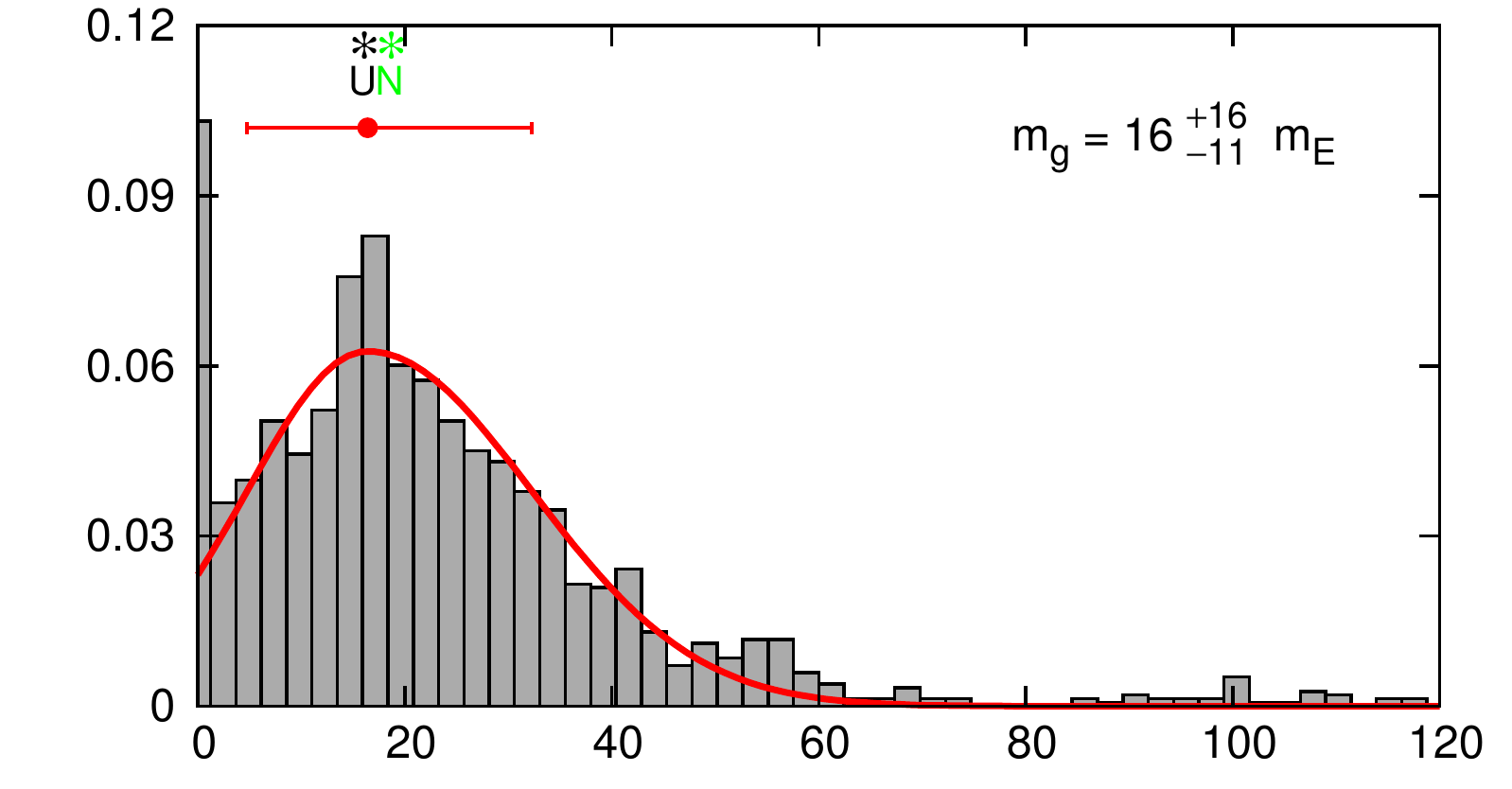}}
\caption{Bootstrap histograms for planetary masses, transit model II.}
\label{fig:masses}
\end{figure*}

\begin{figure*}
\hbox{\includegraphics[width=0.33\textwidth]{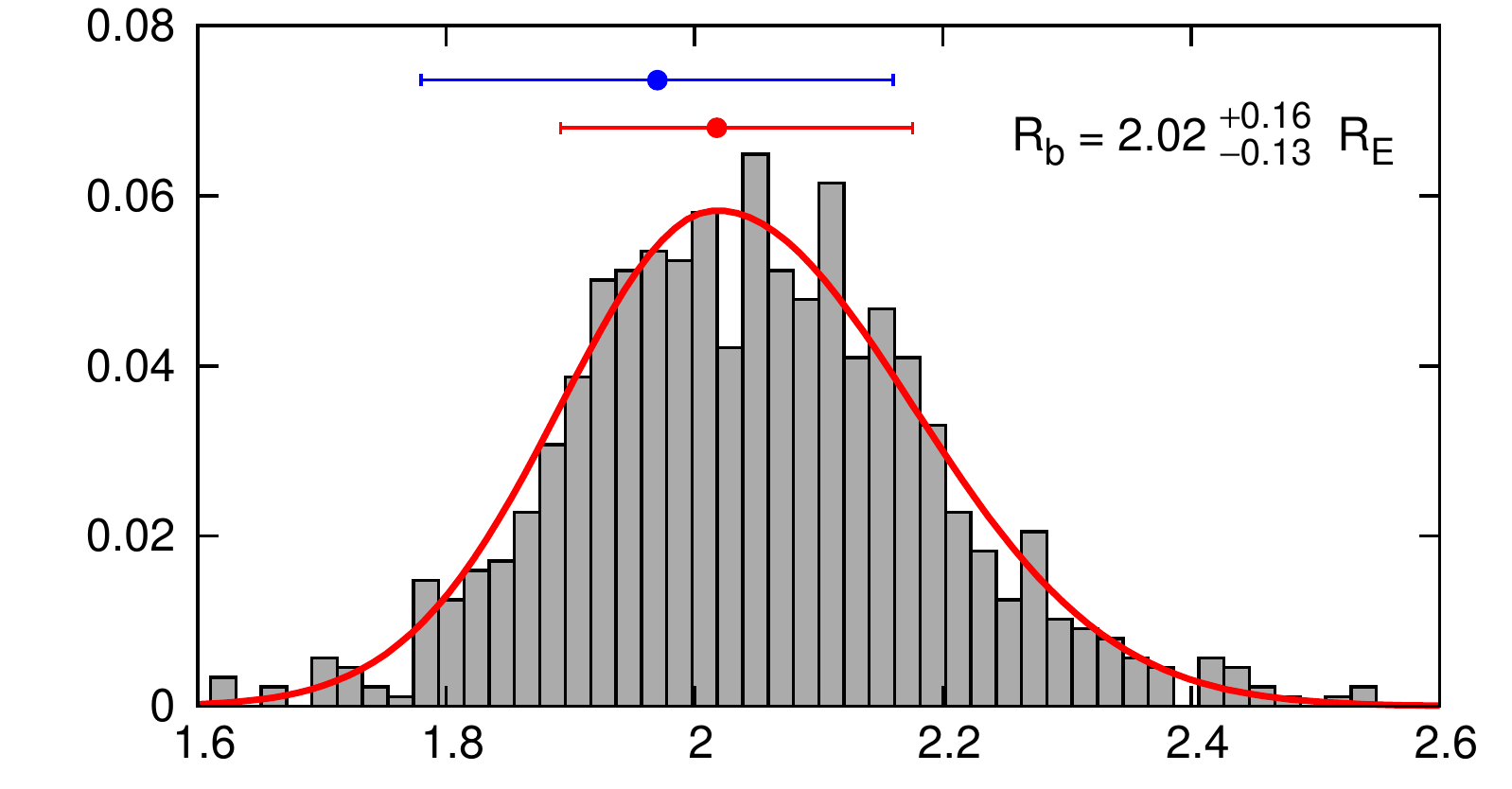}
\includegraphics[width=0.33\textwidth]{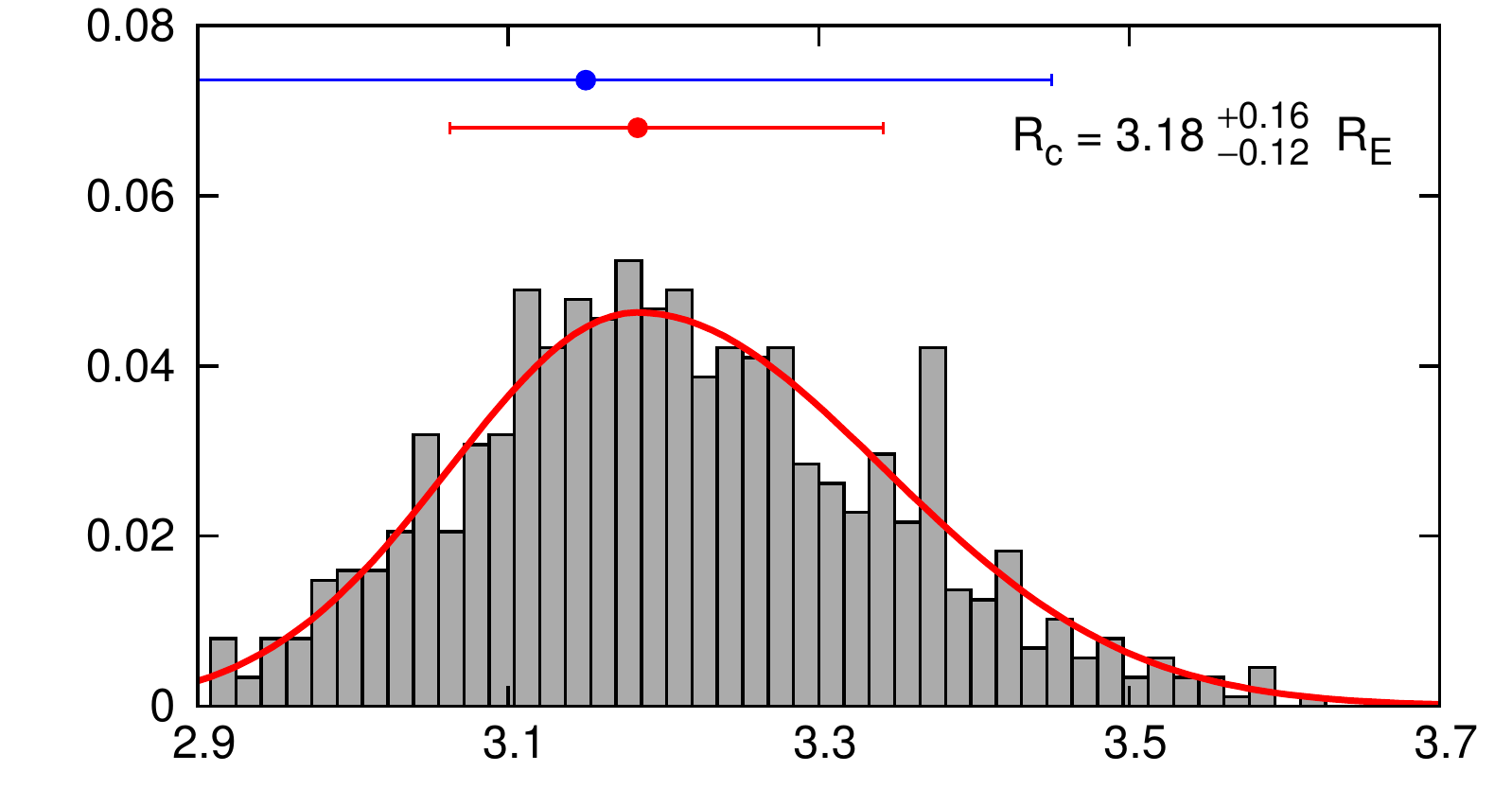}
\includegraphics[width=0.33\textwidth]{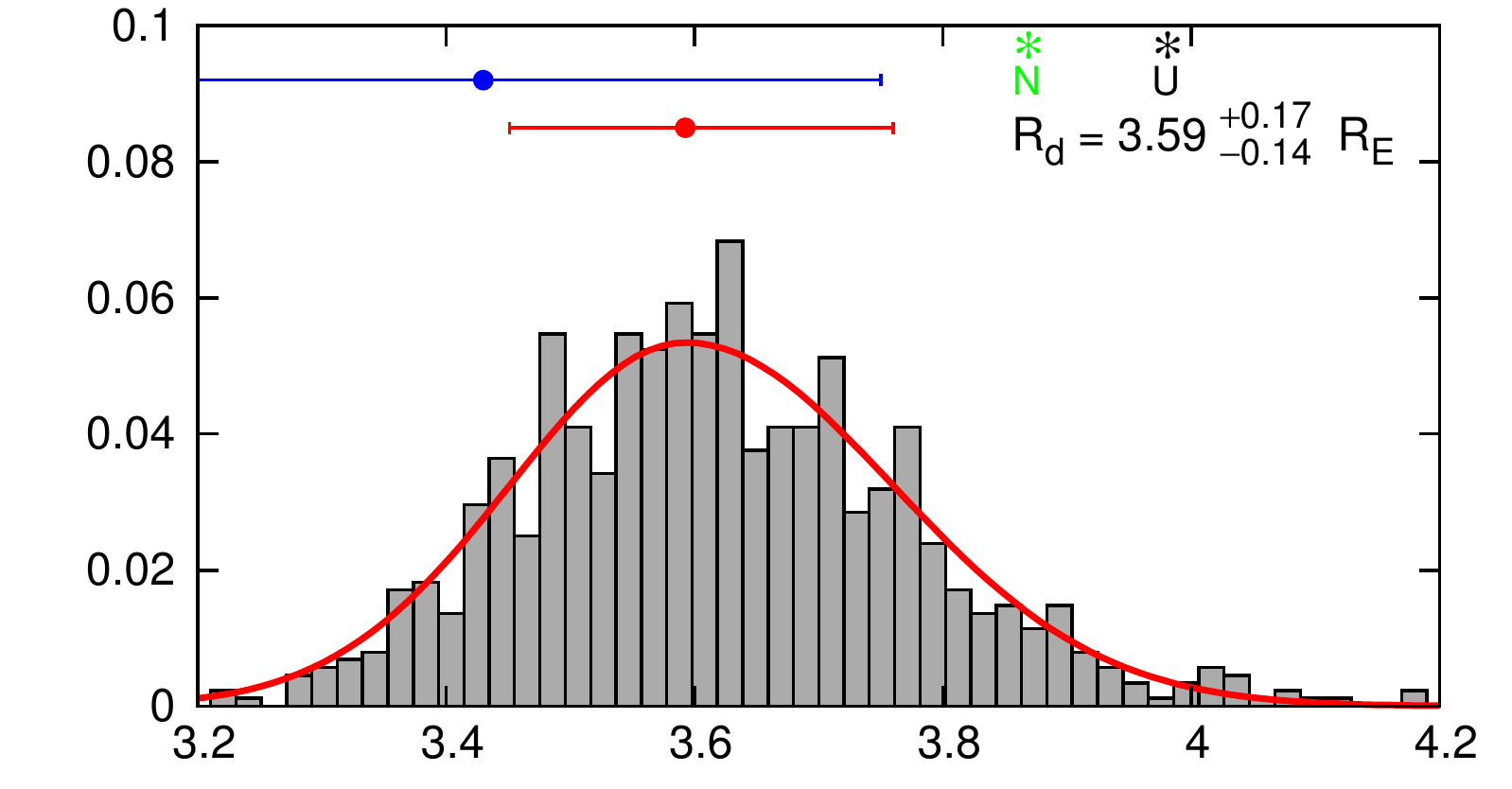}}
\hbox{\includegraphics[width=0.33\textwidth]{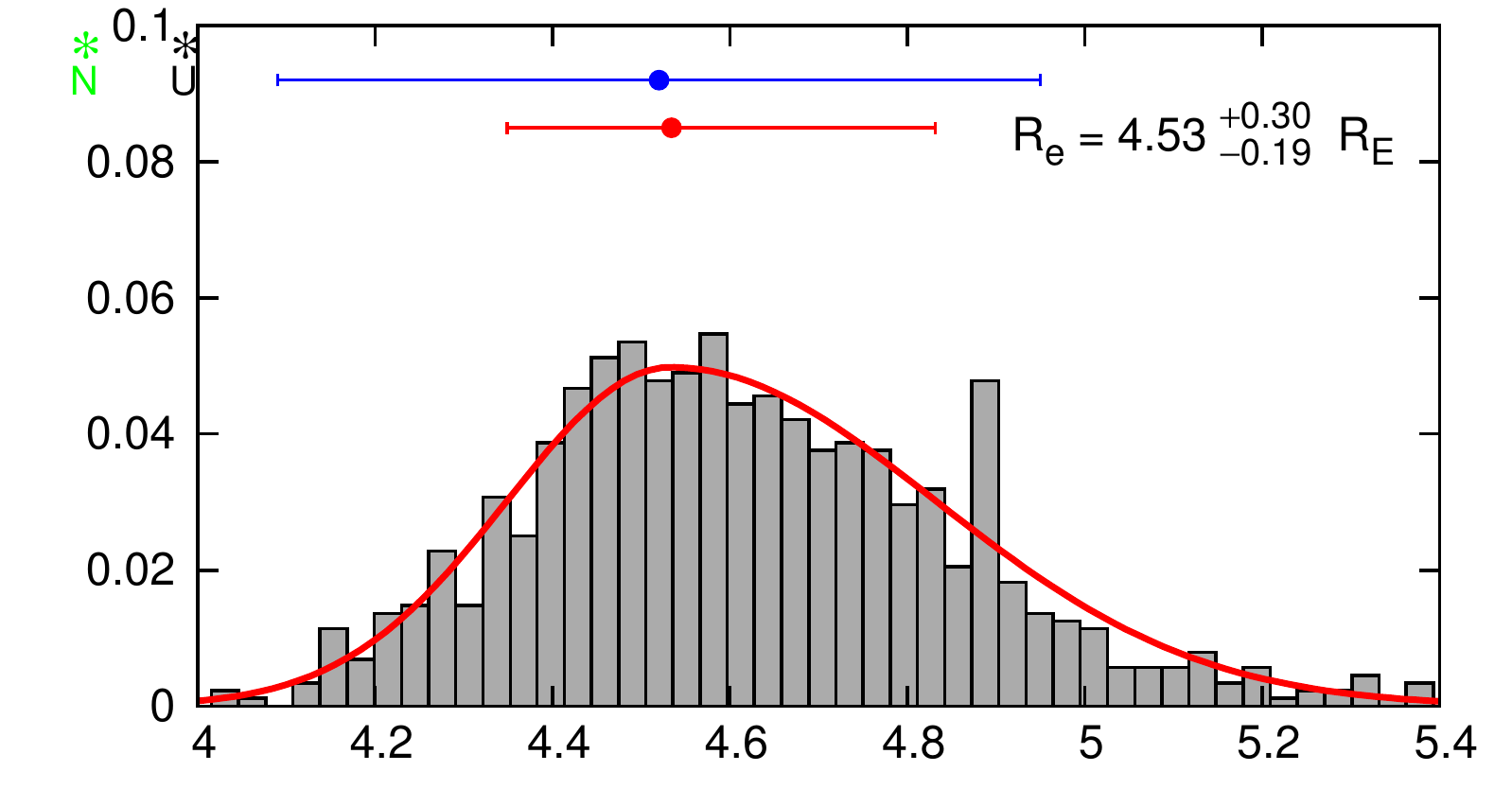}
\includegraphics[width=0.33\textwidth]{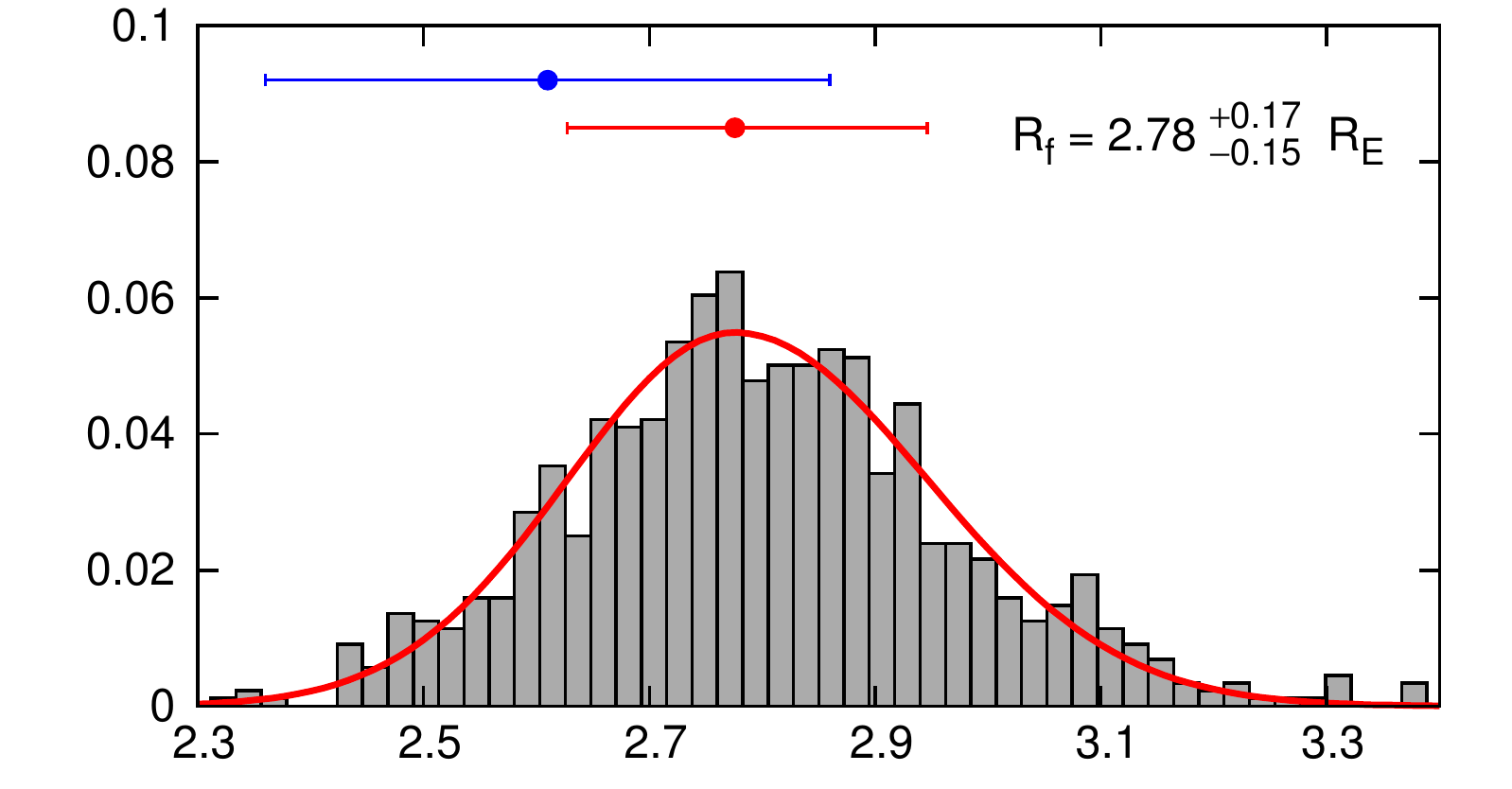}
\includegraphics[width=0.33\textwidth]{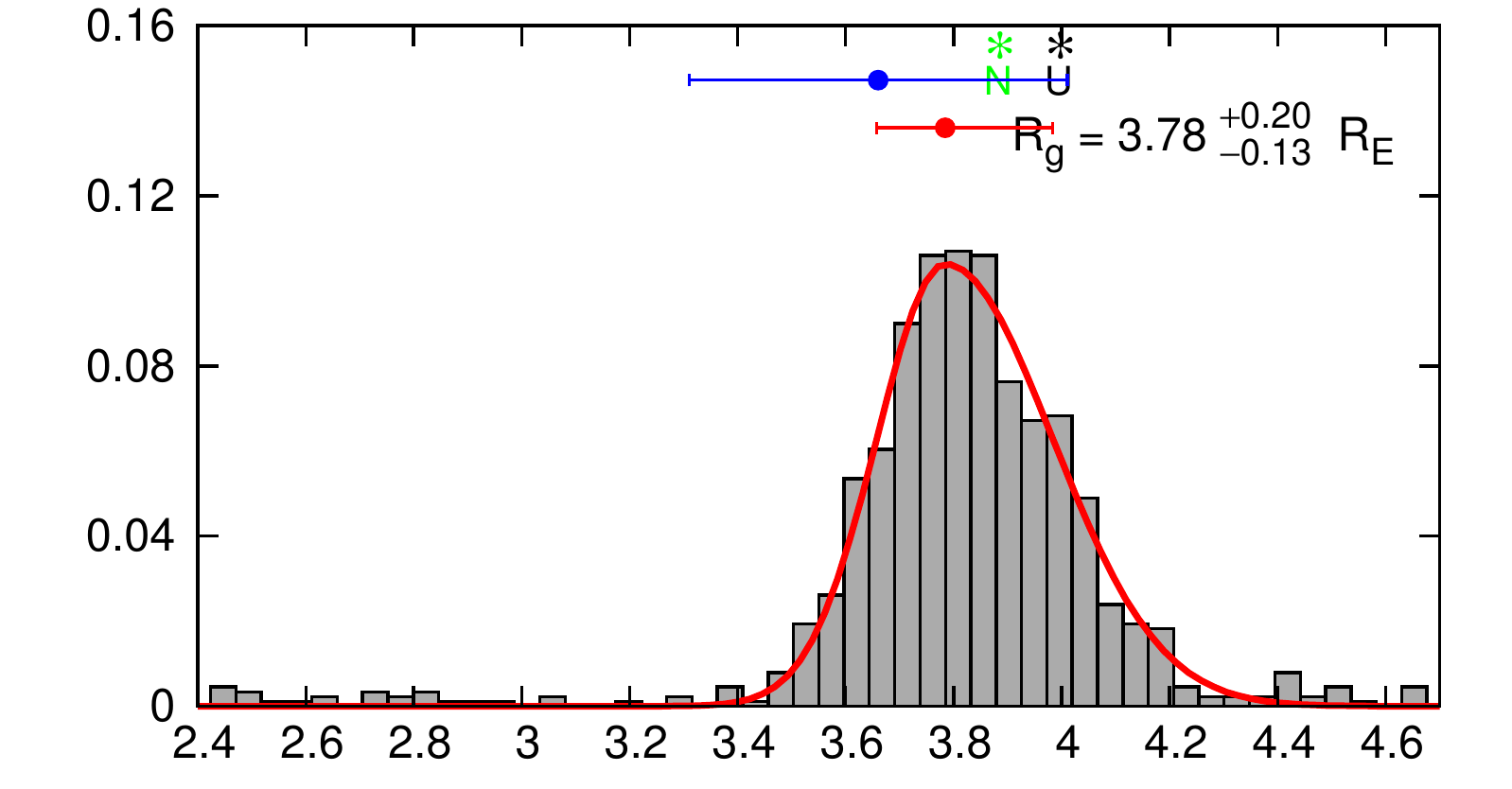}}
\caption{Bootstrap histograms for the planetary radii, transit model II.}
\label{fig:radii}
\end{figure*}

\begin{figure*}
\hbox{\includegraphics[width=0.33\textwidth]{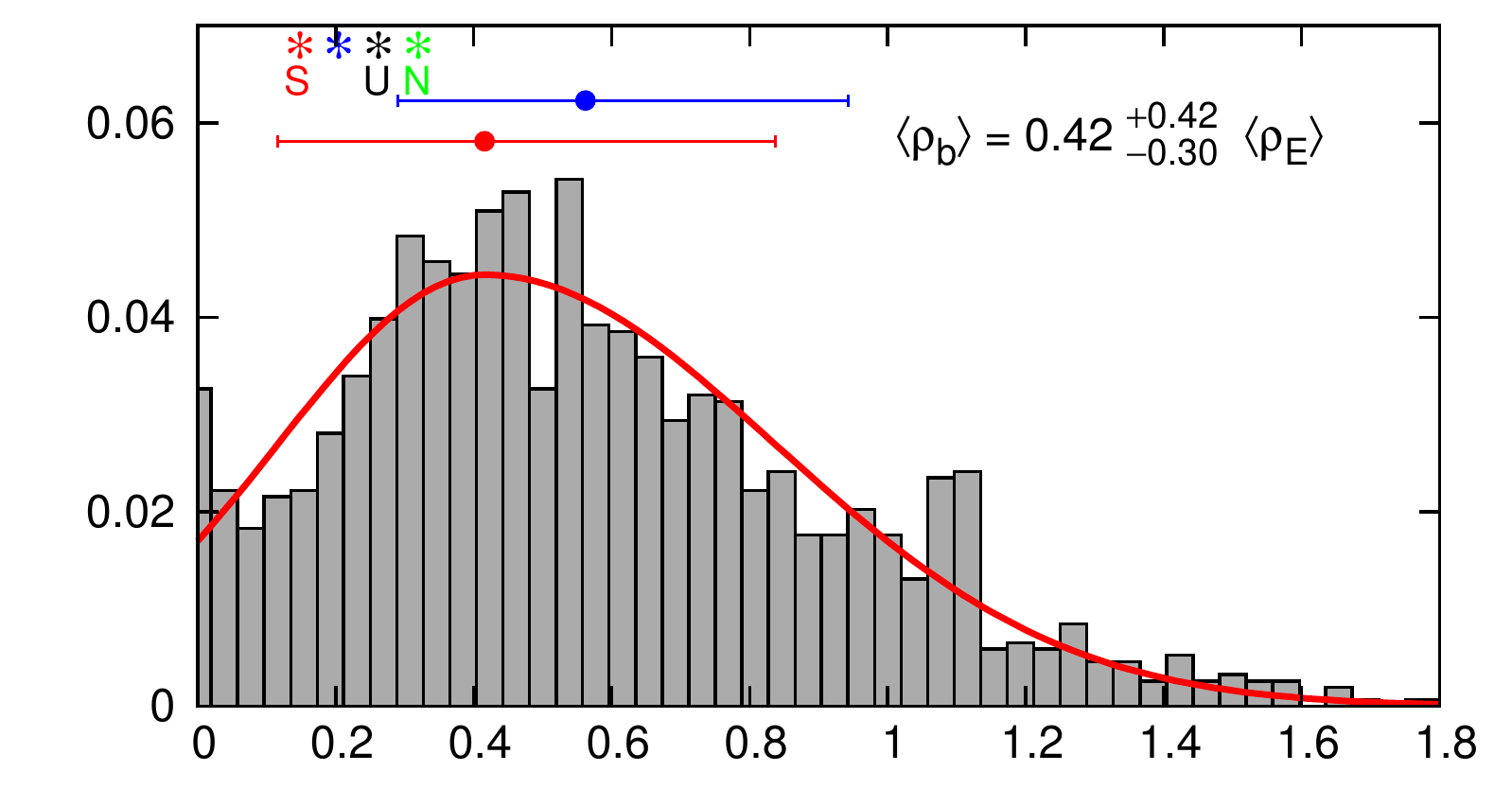}
\includegraphics[width=0.33\textwidth]{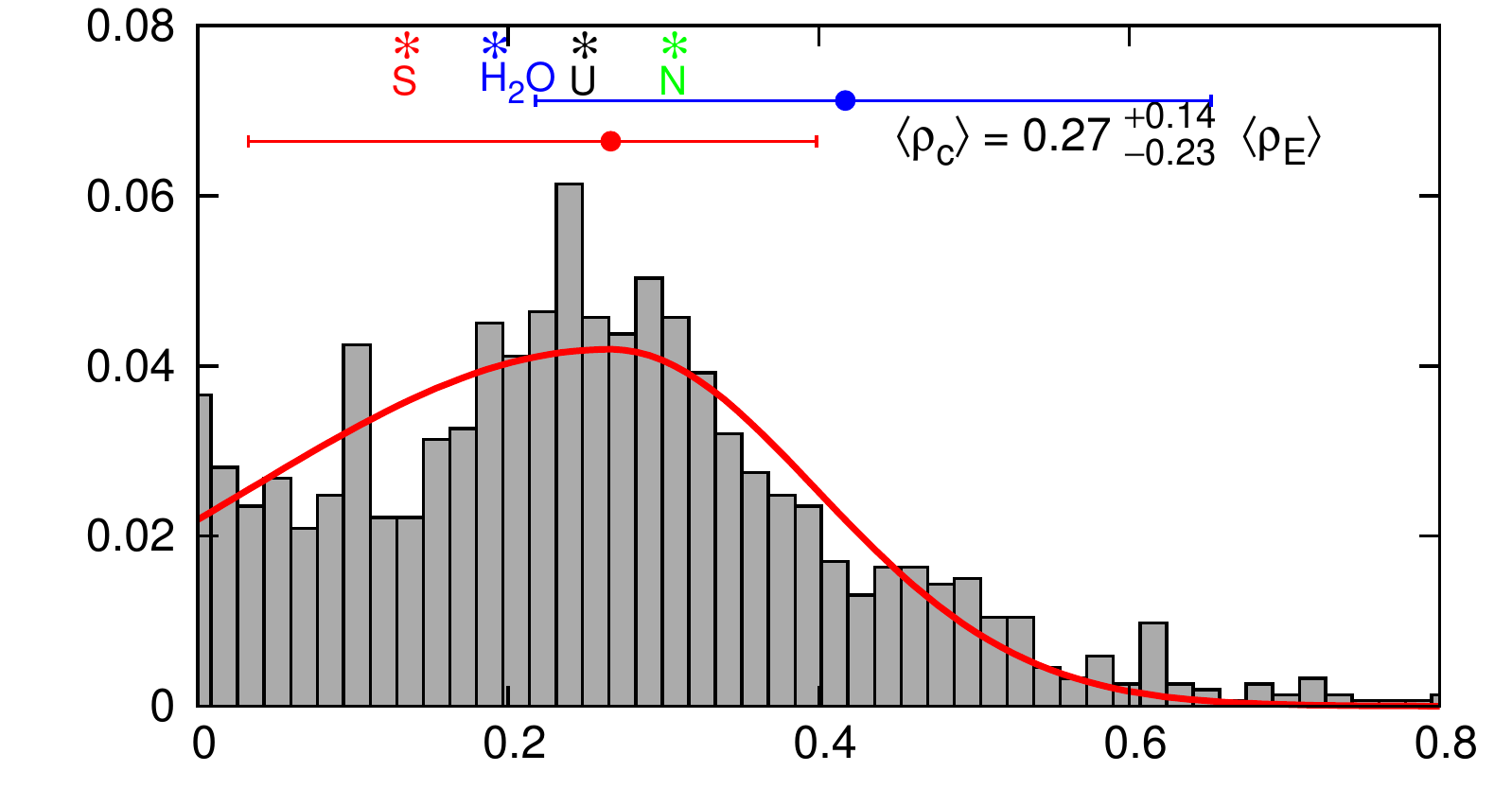}
\includegraphics[width=0.33\textwidth]{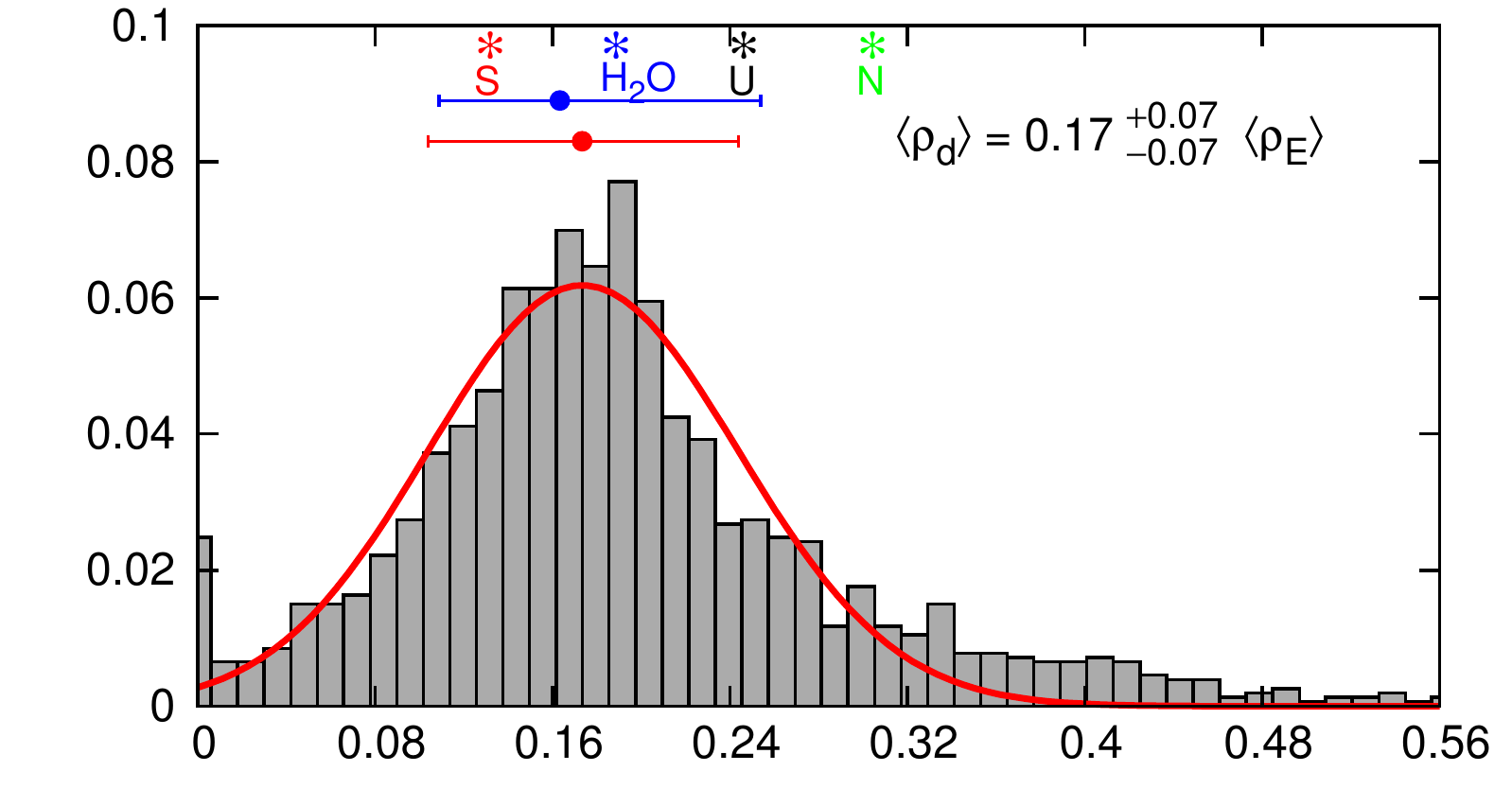}}
\hbox{\includegraphics[width=0.33\textwidth]{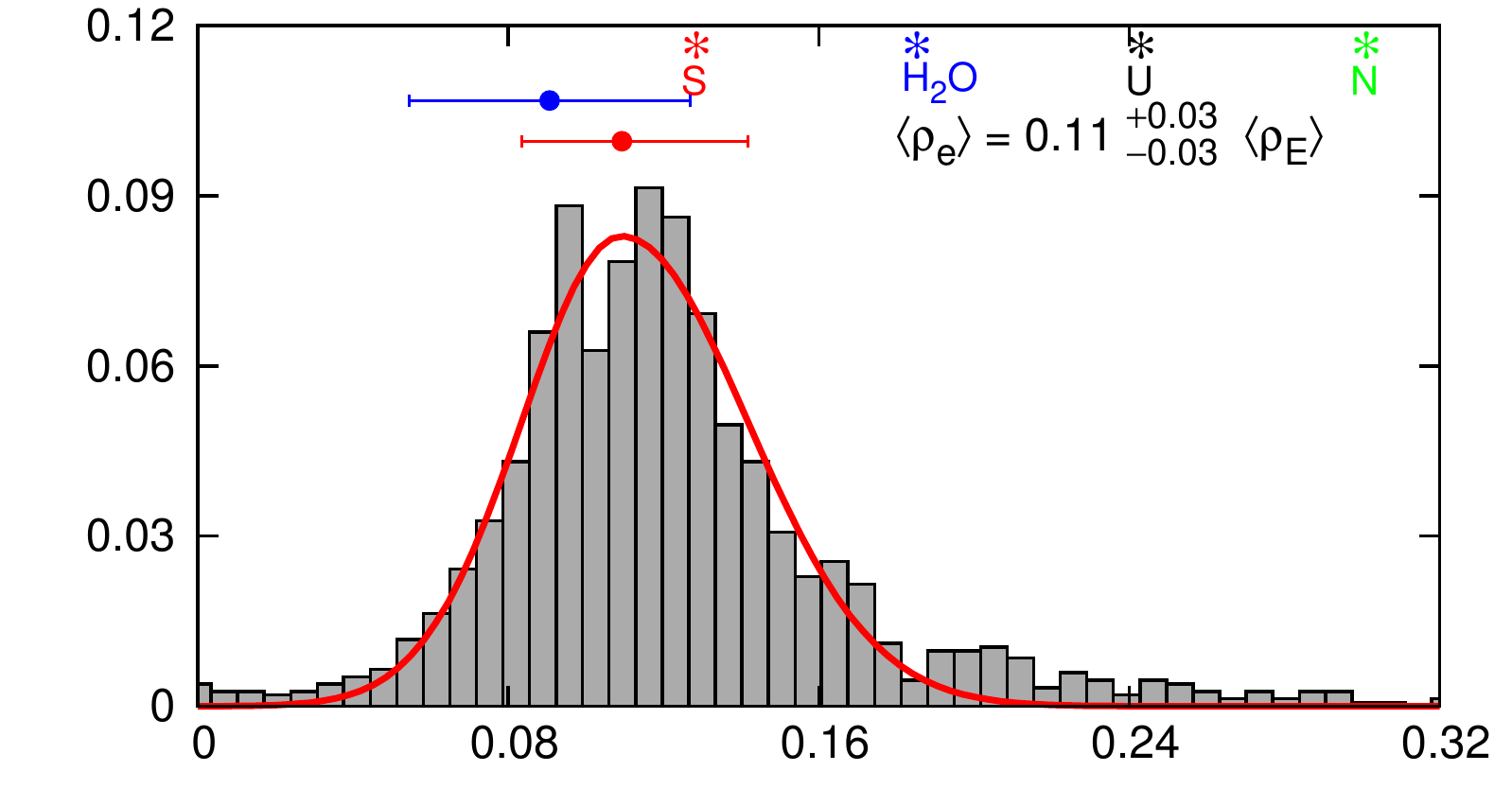}
\includegraphics[width=0.33\textwidth]{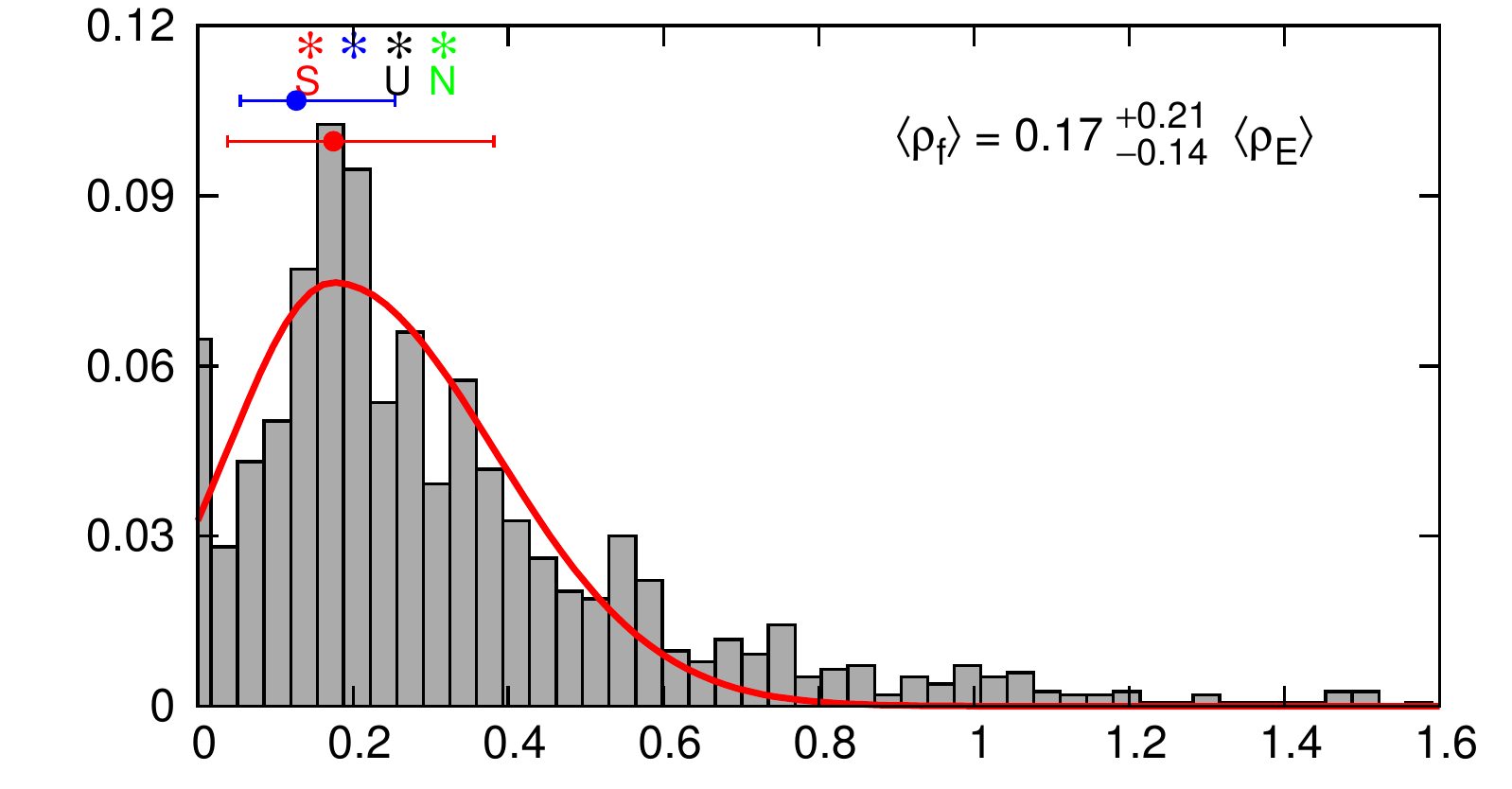}
\includegraphics[width=0.33\textwidth]{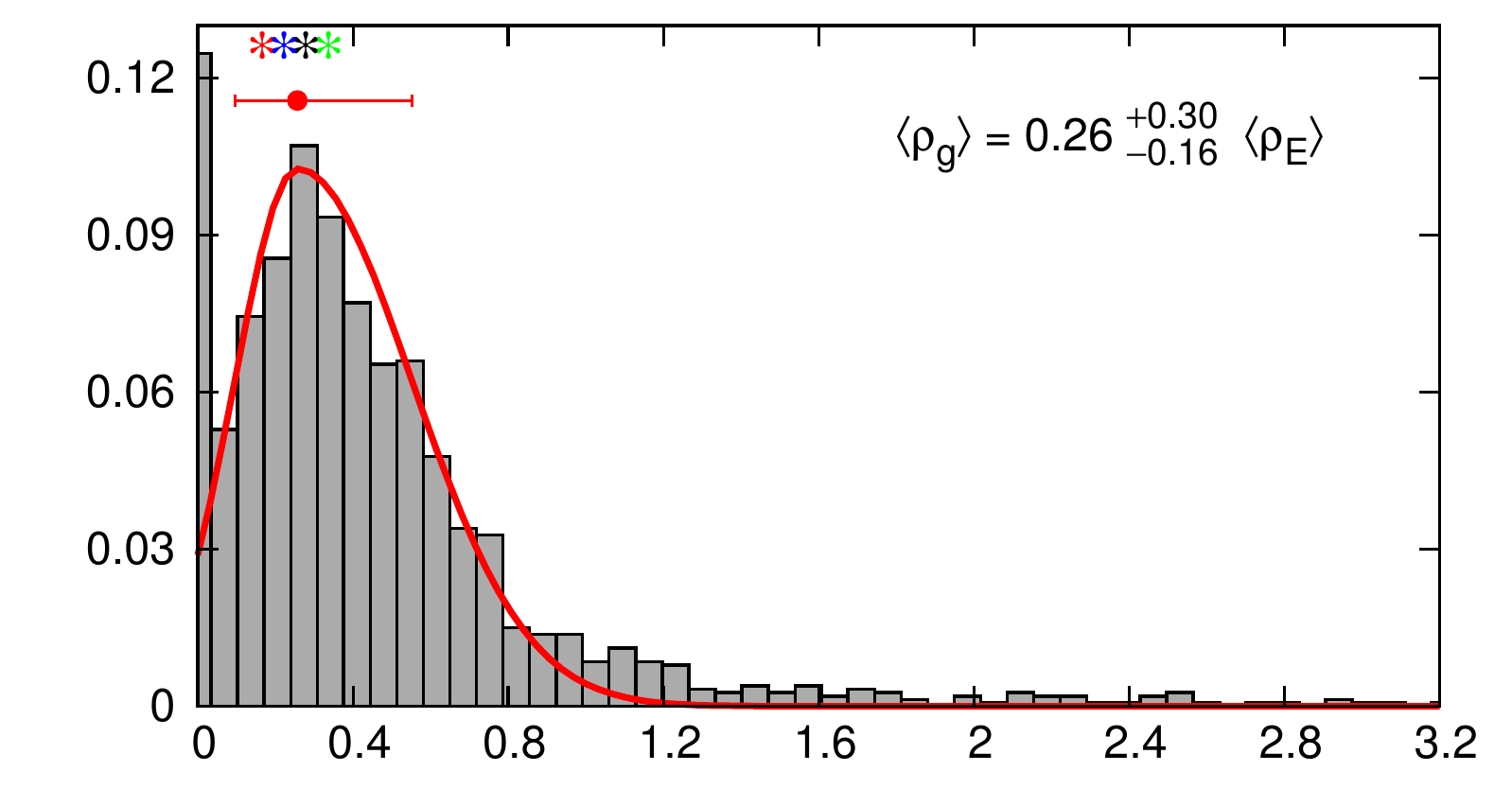}}
\caption{Bootstrap histograms for the mean densities, transit model II.}
\label{fig:densities}
\end{figure*}

\begin{figure*}
\hbox{\includegraphics[width=0.33\textwidth]{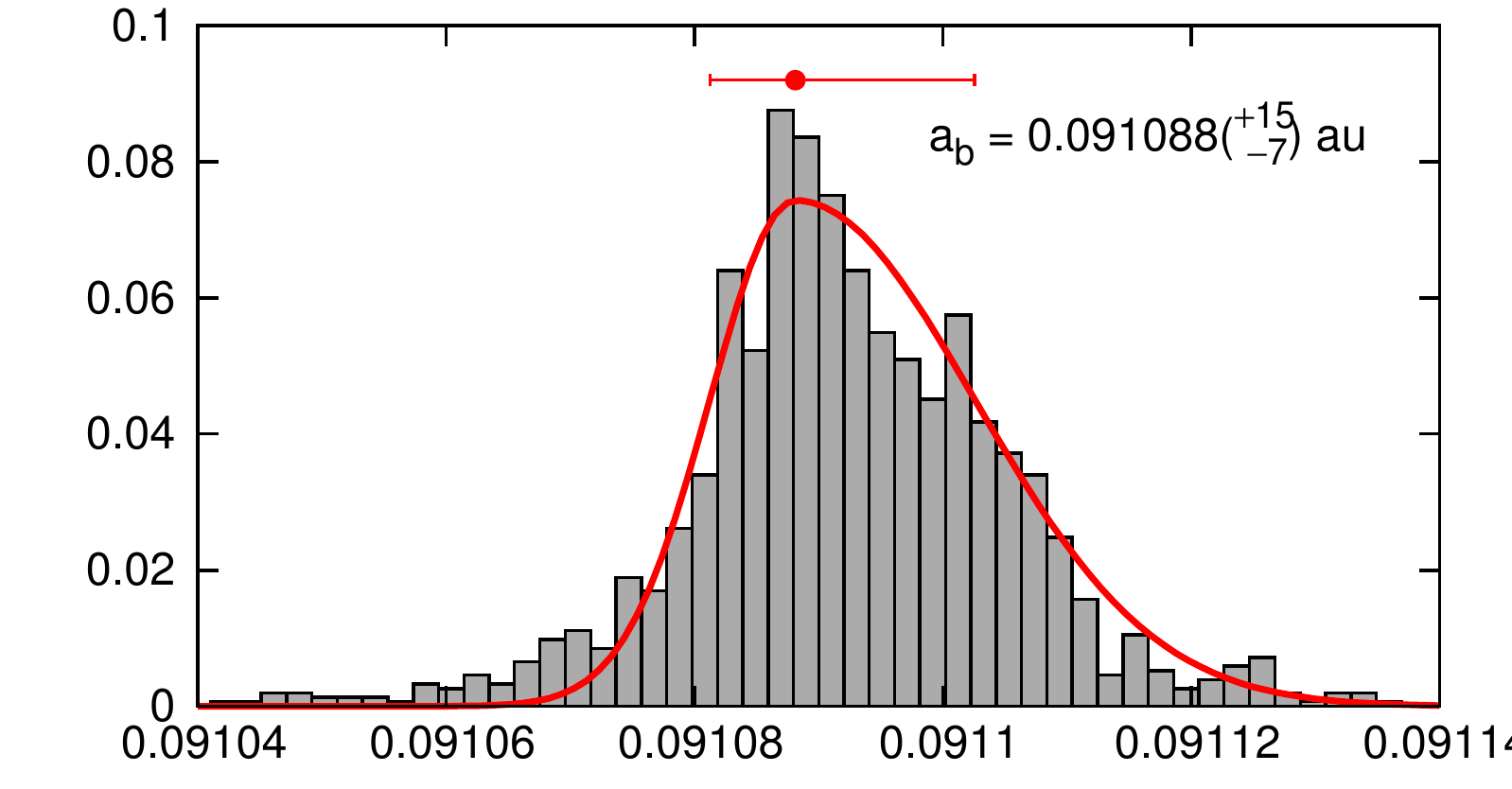}
\includegraphics[width=0.33\textwidth]{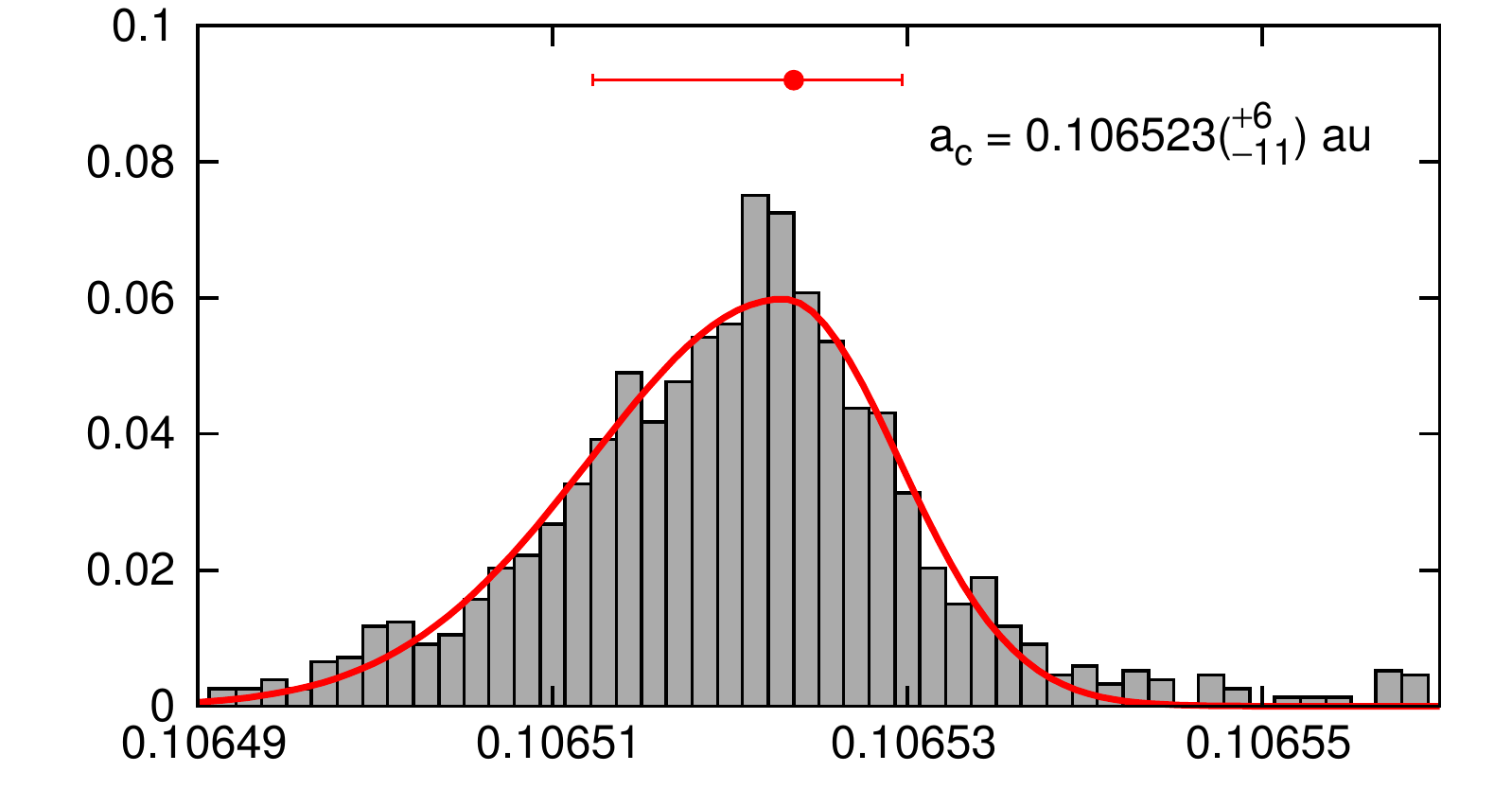}
\includegraphics[width=0.33\textwidth]{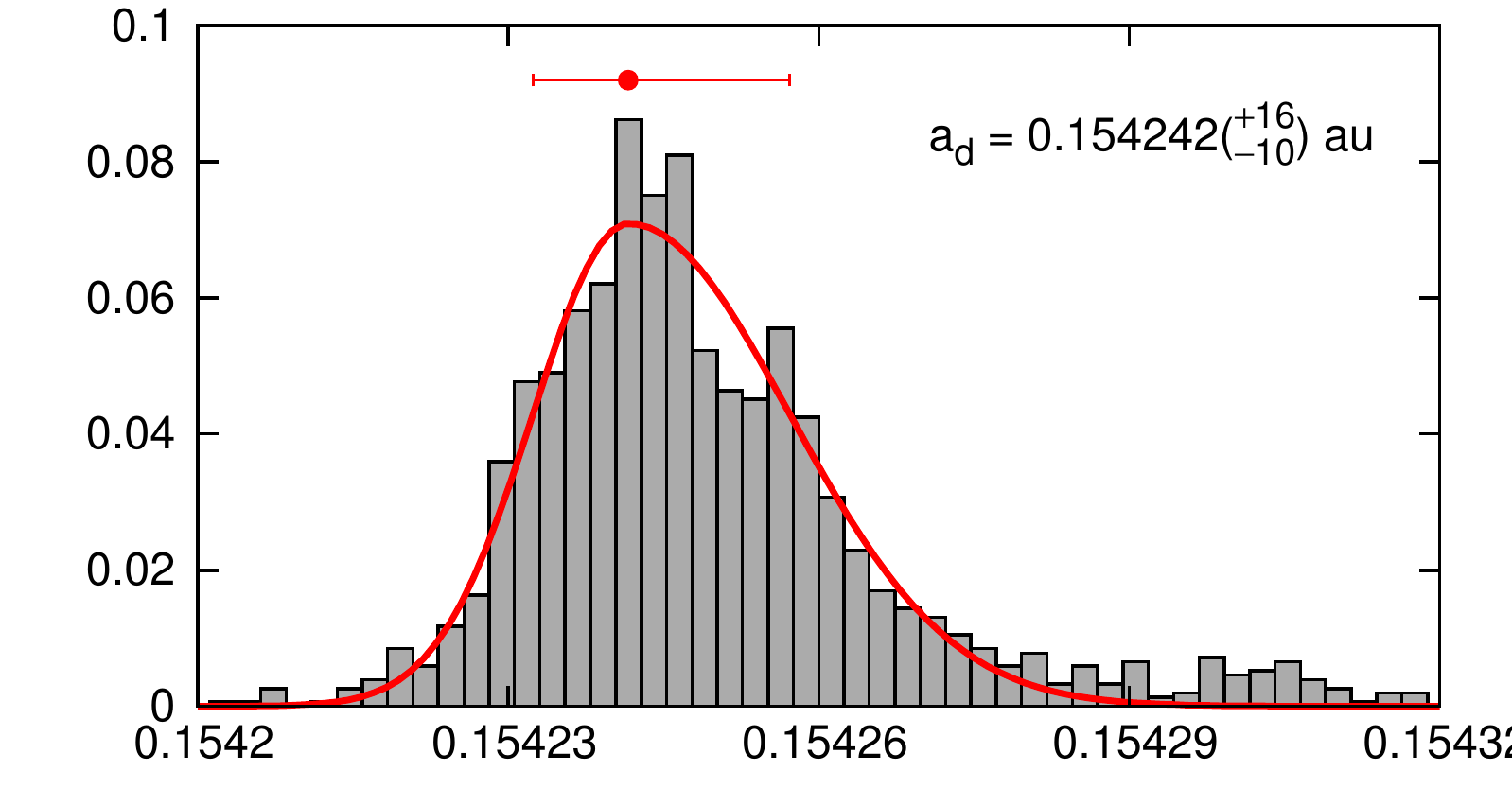}}
\hbox{\includegraphics[width=0.33\textwidth]{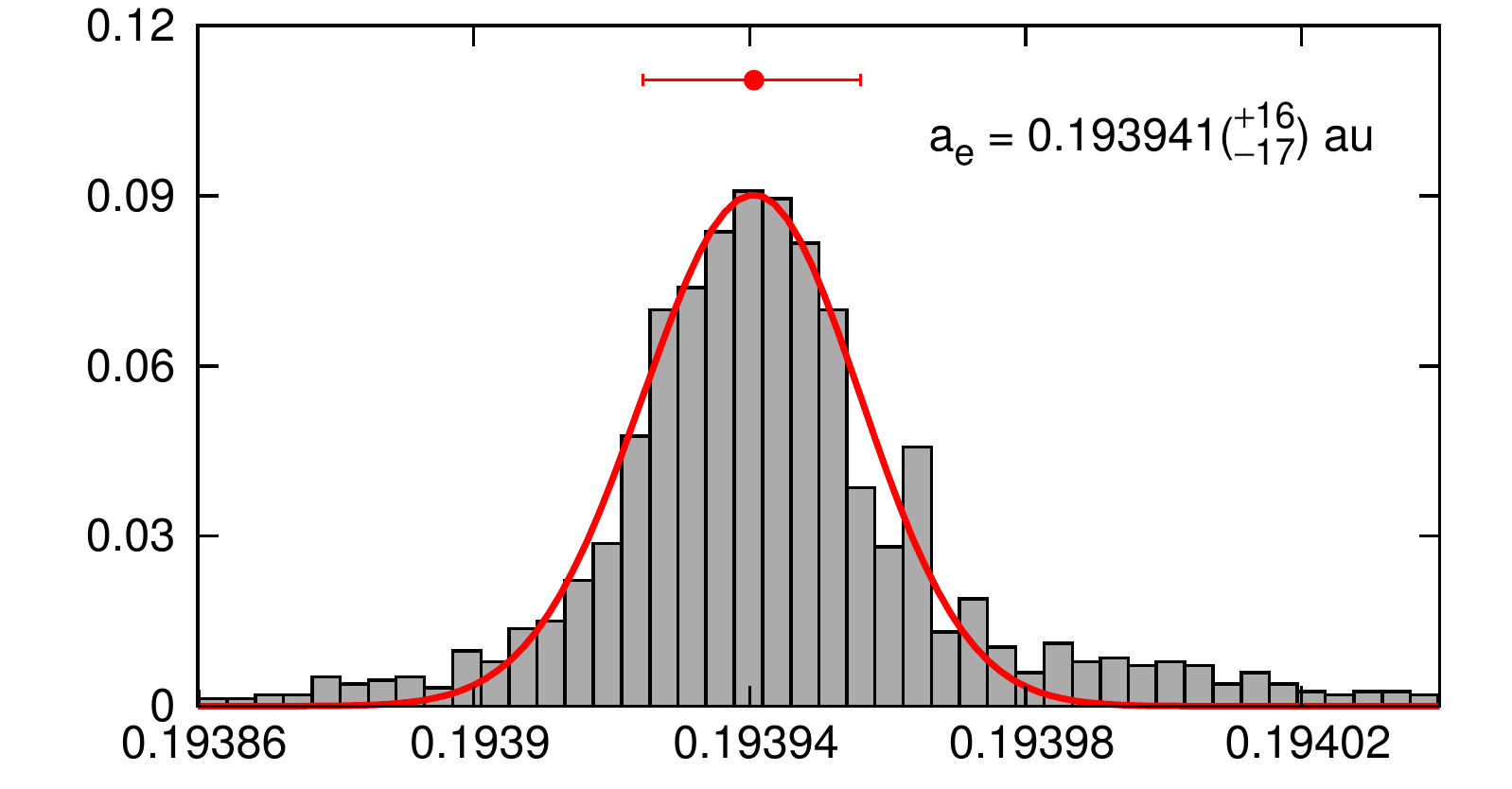}
\includegraphics[width=0.33\textwidth]{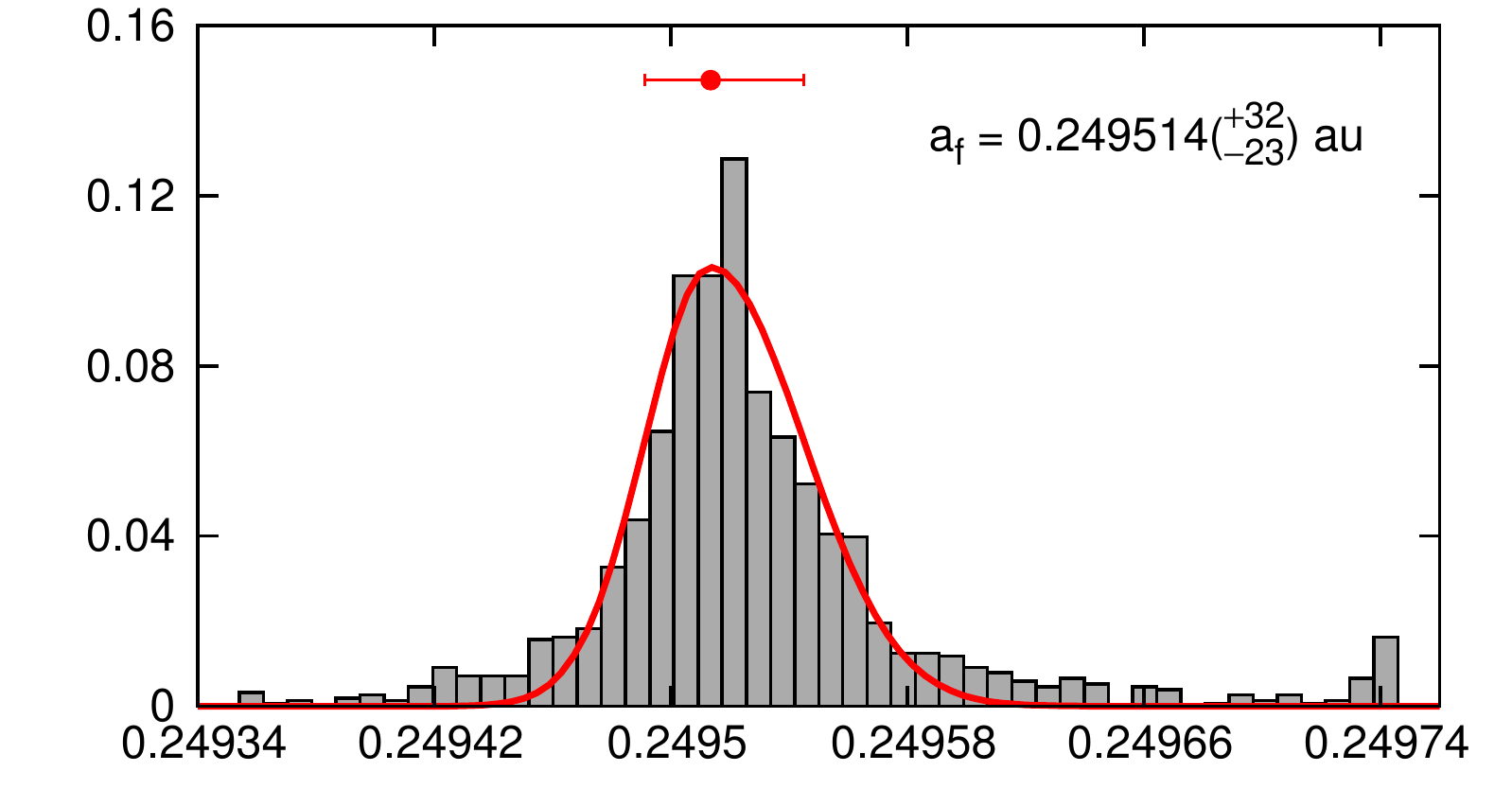}
\includegraphics[width=0.33\textwidth]{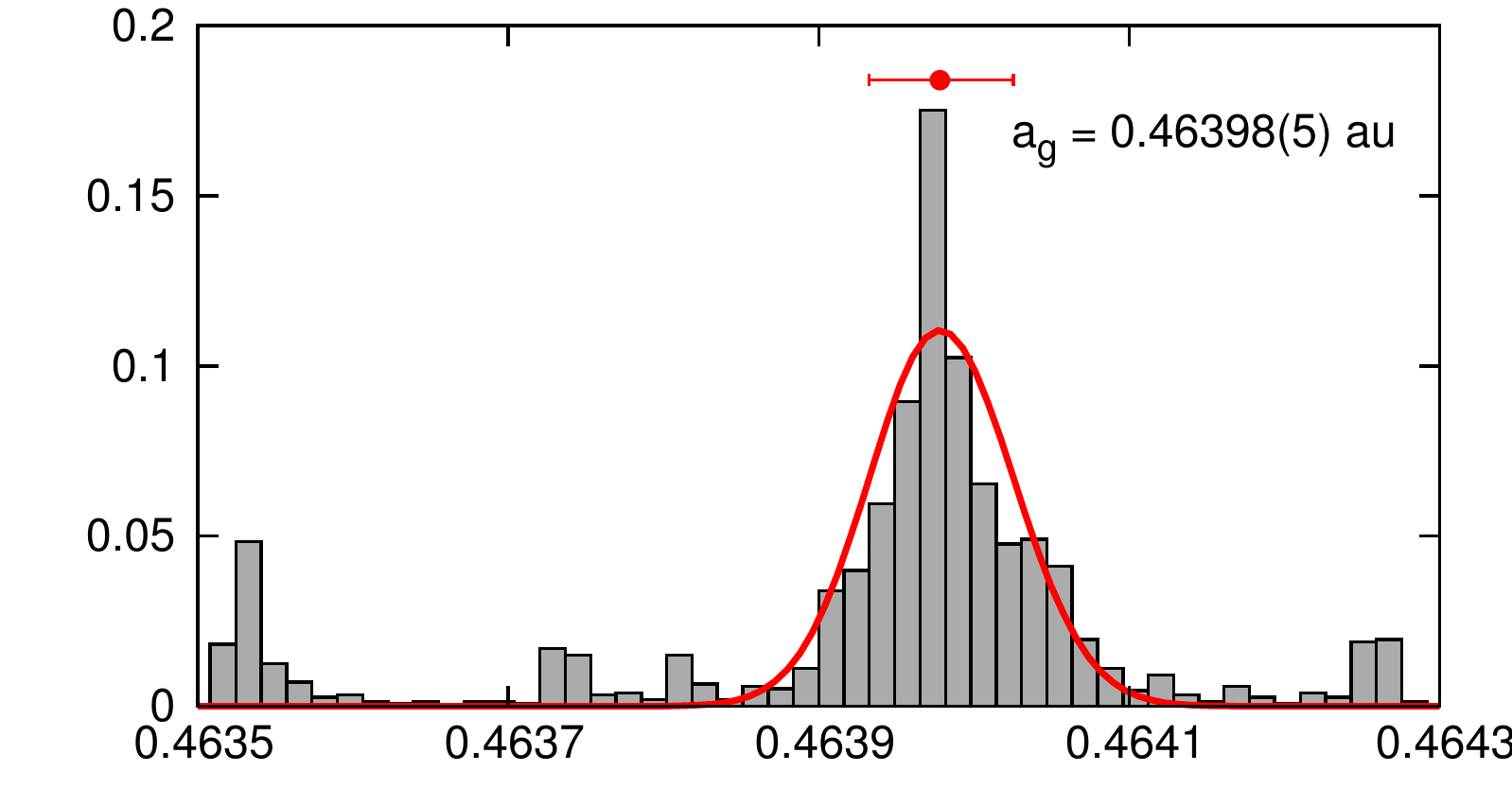}}
\caption{Bootstrap histograms for the semi-major axes, transit model~II.}
\label{fig:semi-major-axes}
\end{figure*}

\begin{figure*}
\hbox{\includegraphics[width=0.33\textwidth]{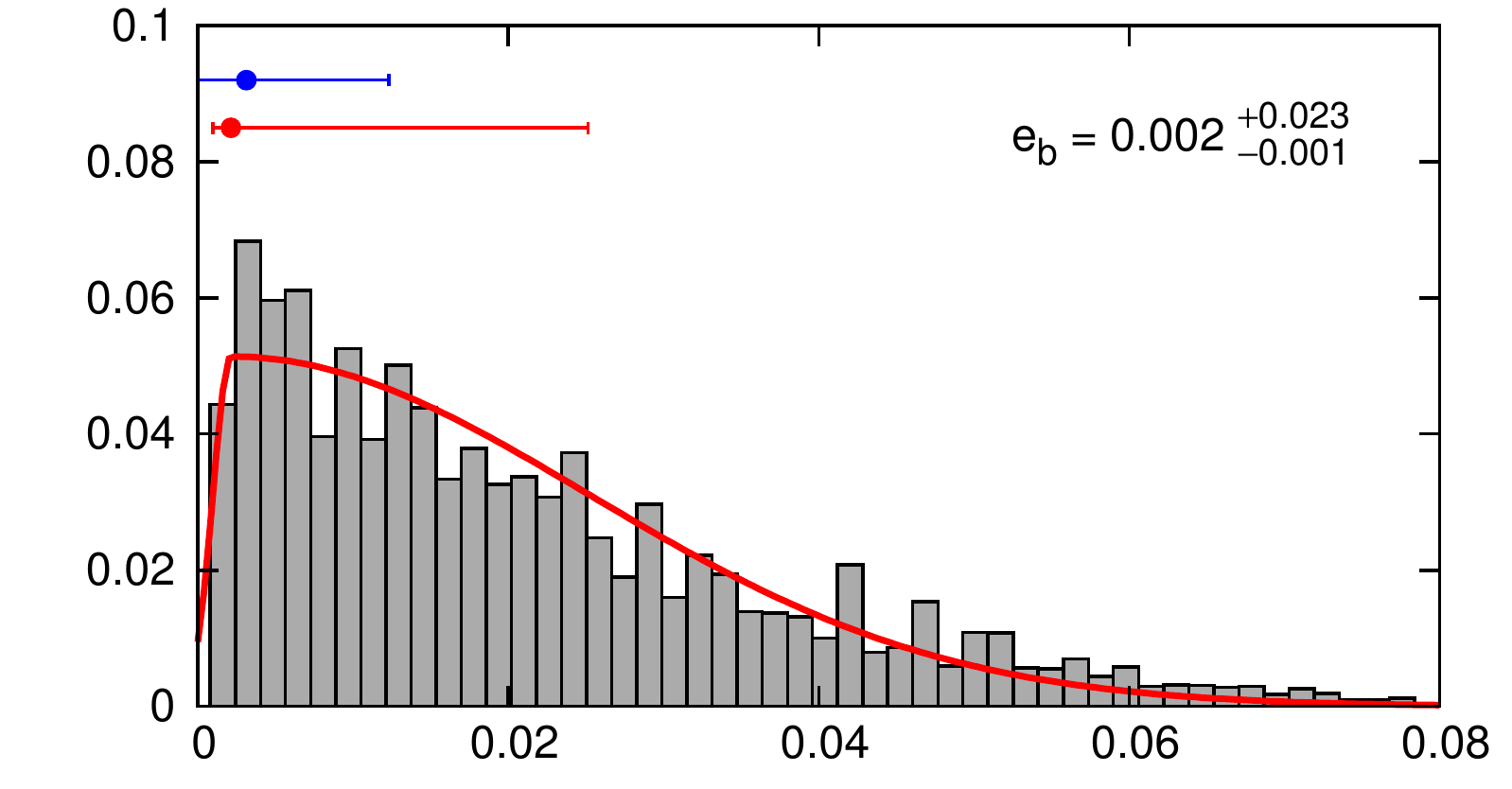}
\includegraphics[width=0.33\textwidth]{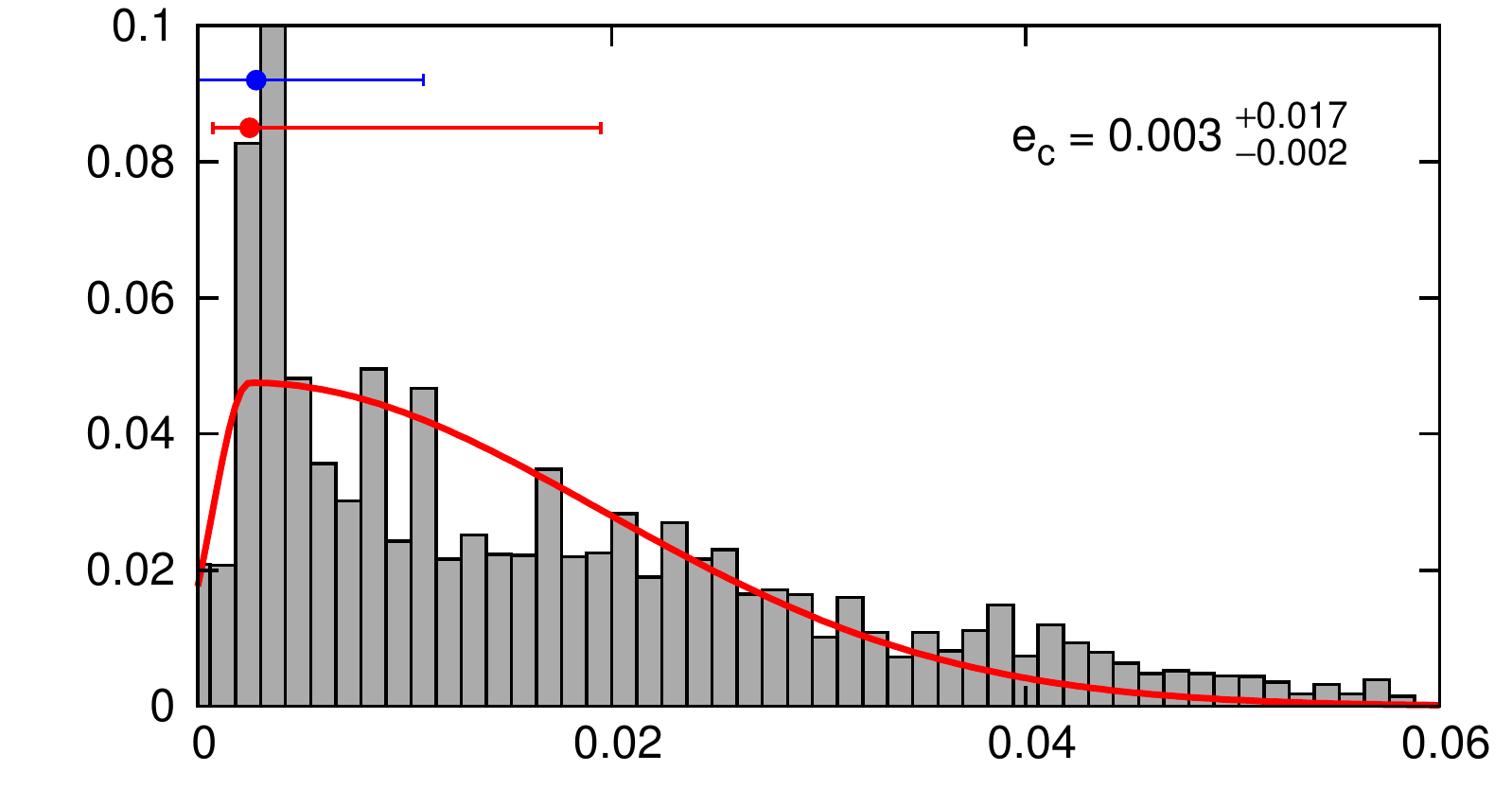}
\includegraphics[width=0.33\textwidth]{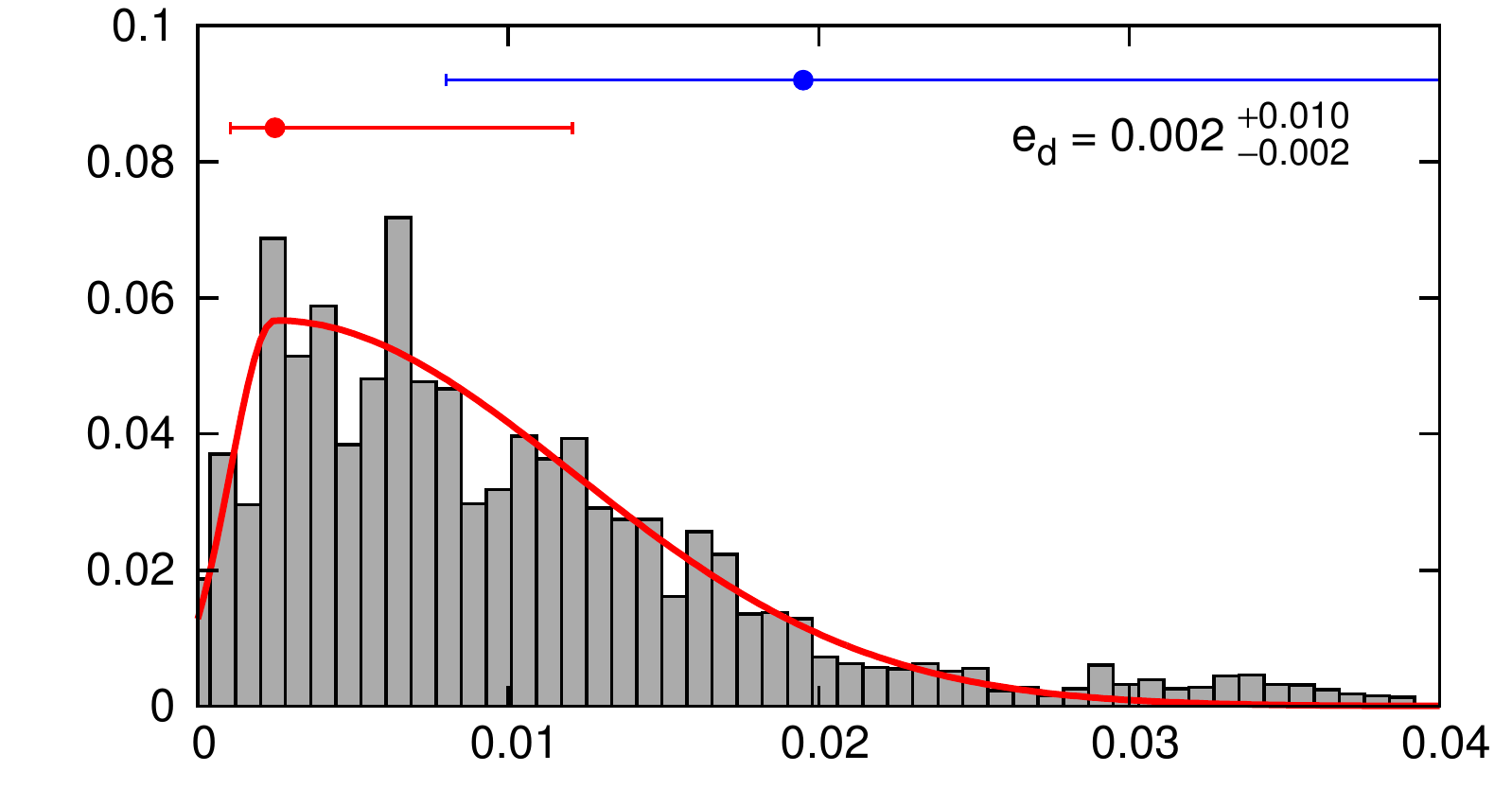}}
\hbox{\includegraphics[width=0.33\textwidth]{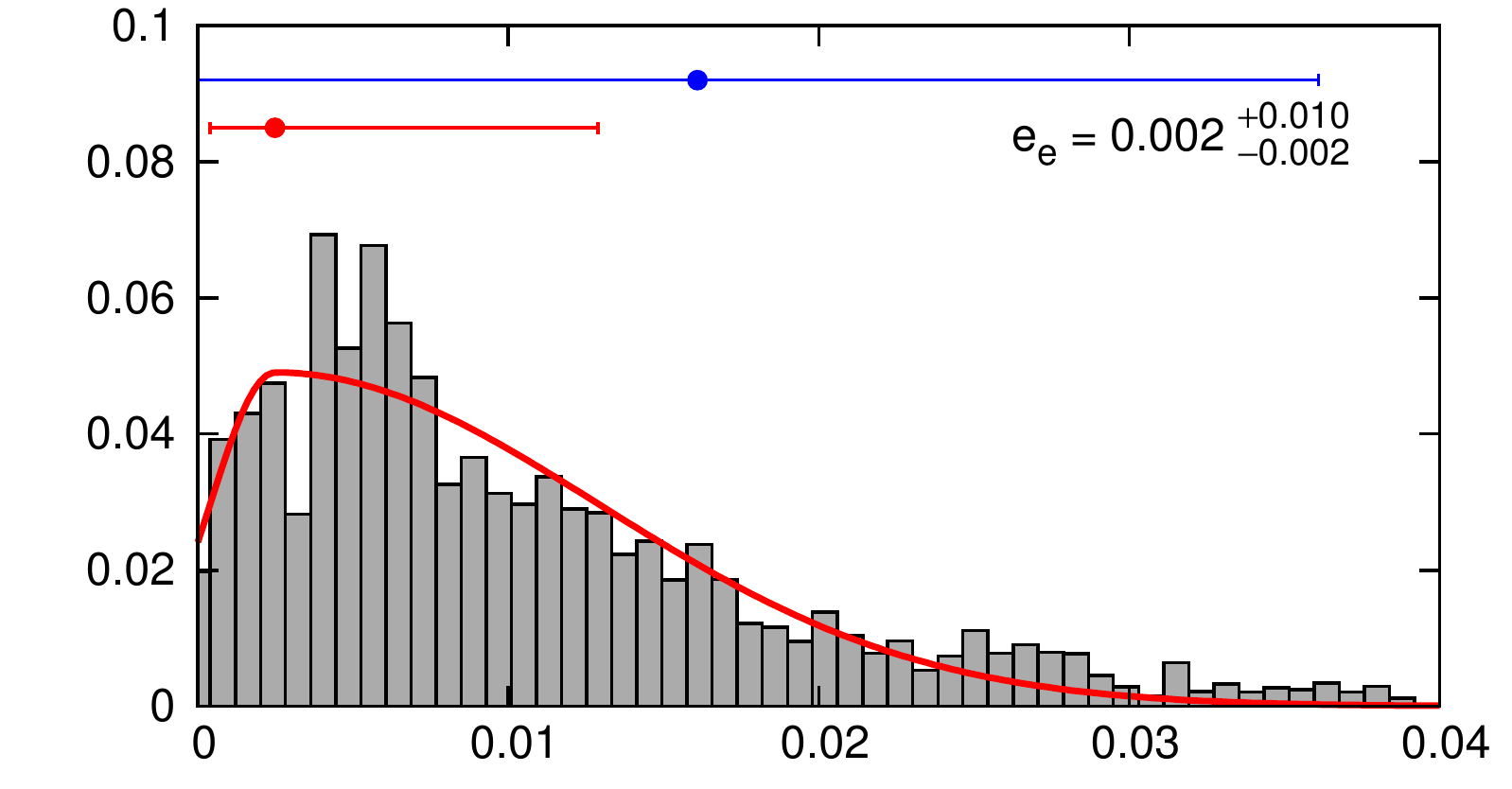}
\includegraphics[width=0.33\textwidth]{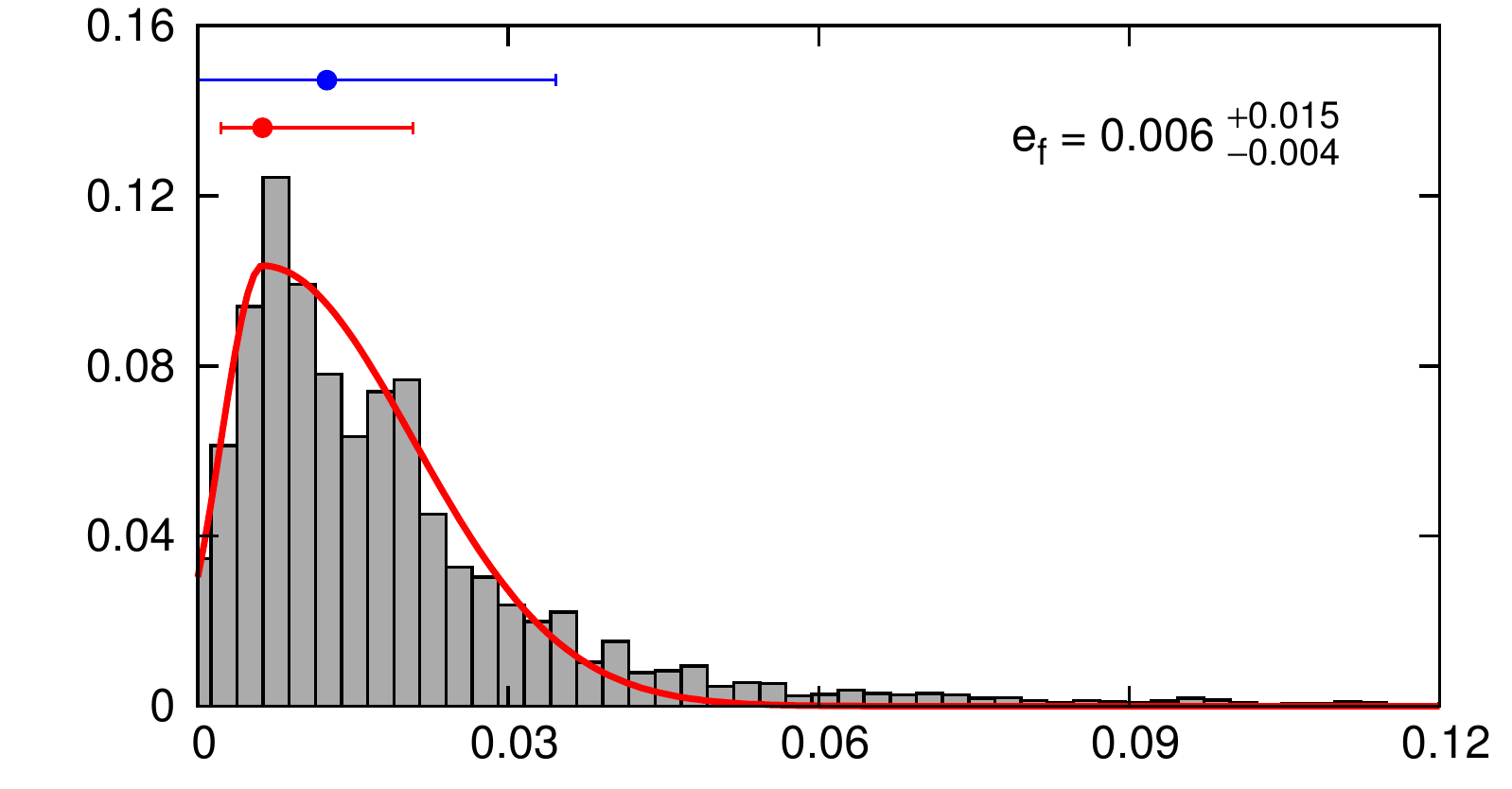}
\includegraphics[width=0.33\textwidth]{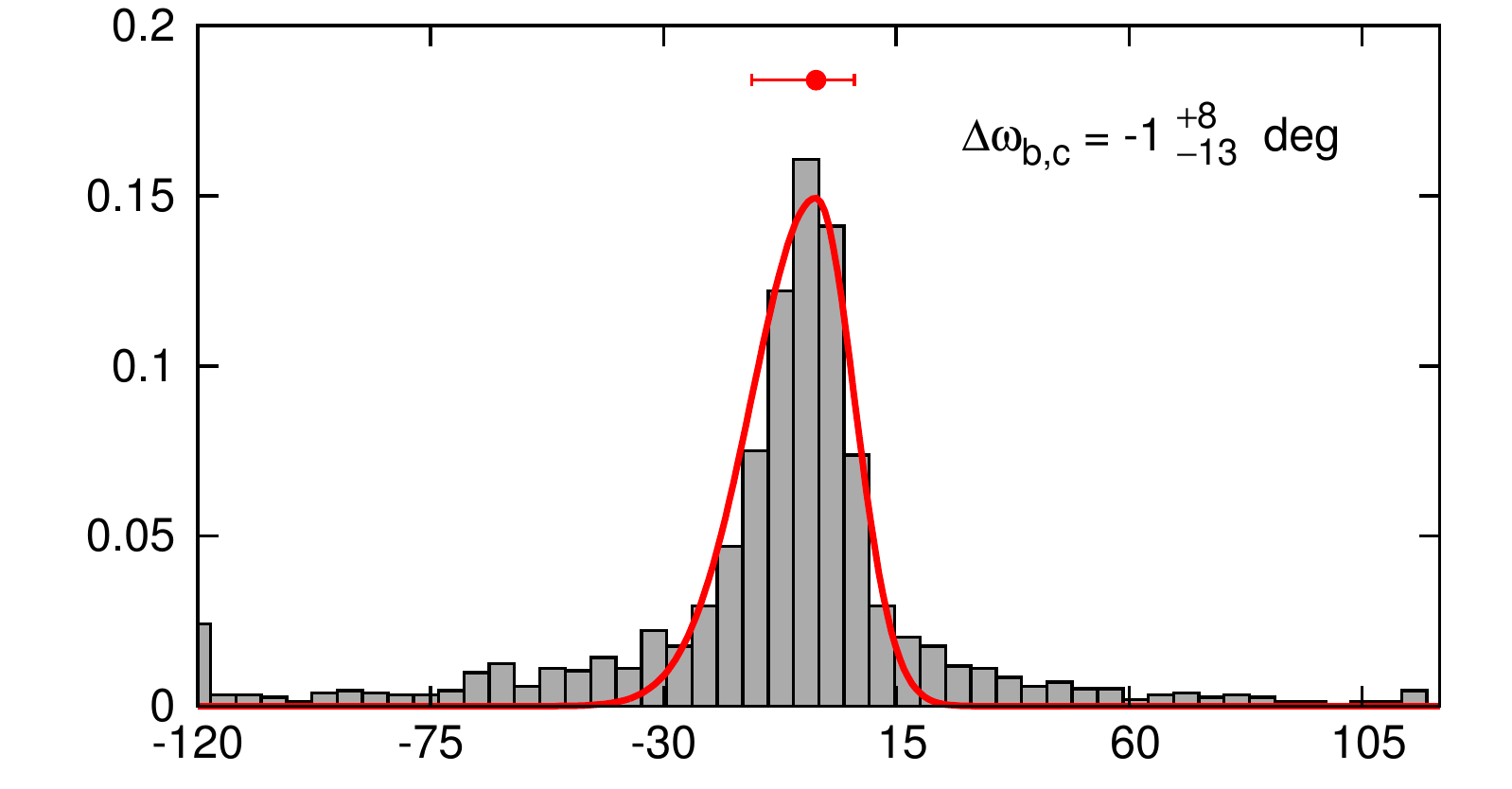}}
\caption{Bootstrap histograms for eccentricities and $\Delta\omega_{\idm{b},\idm{c}}$,
transit model~II.}
\label{fig:eccentricities}
\end{figure*}

\begin{figure*}
\hbox{\includegraphics[width=0.33\textwidth]{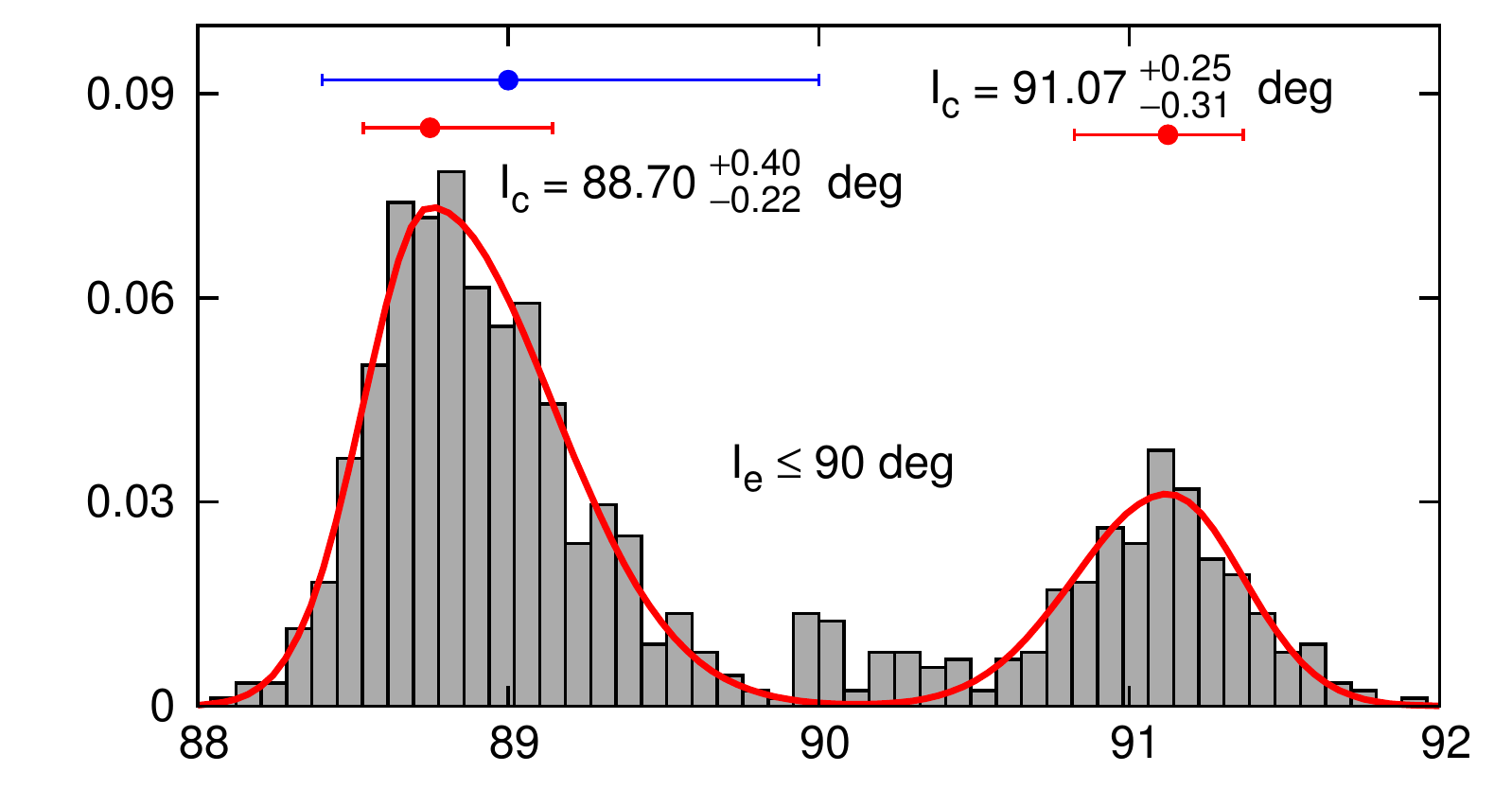}
\includegraphics[width=0.33\textwidth]{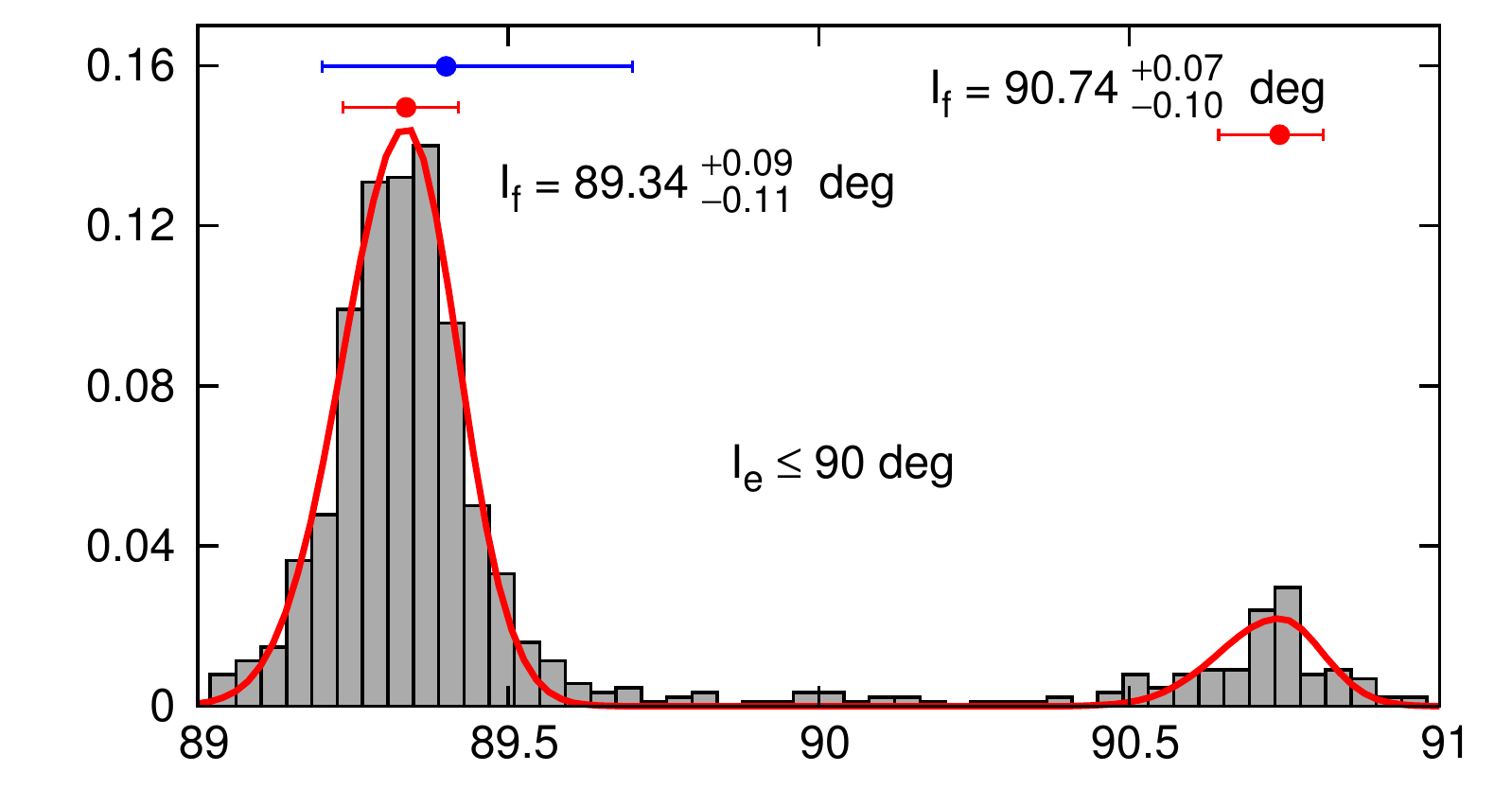}
\includegraphics[width=0.33\textwidth]{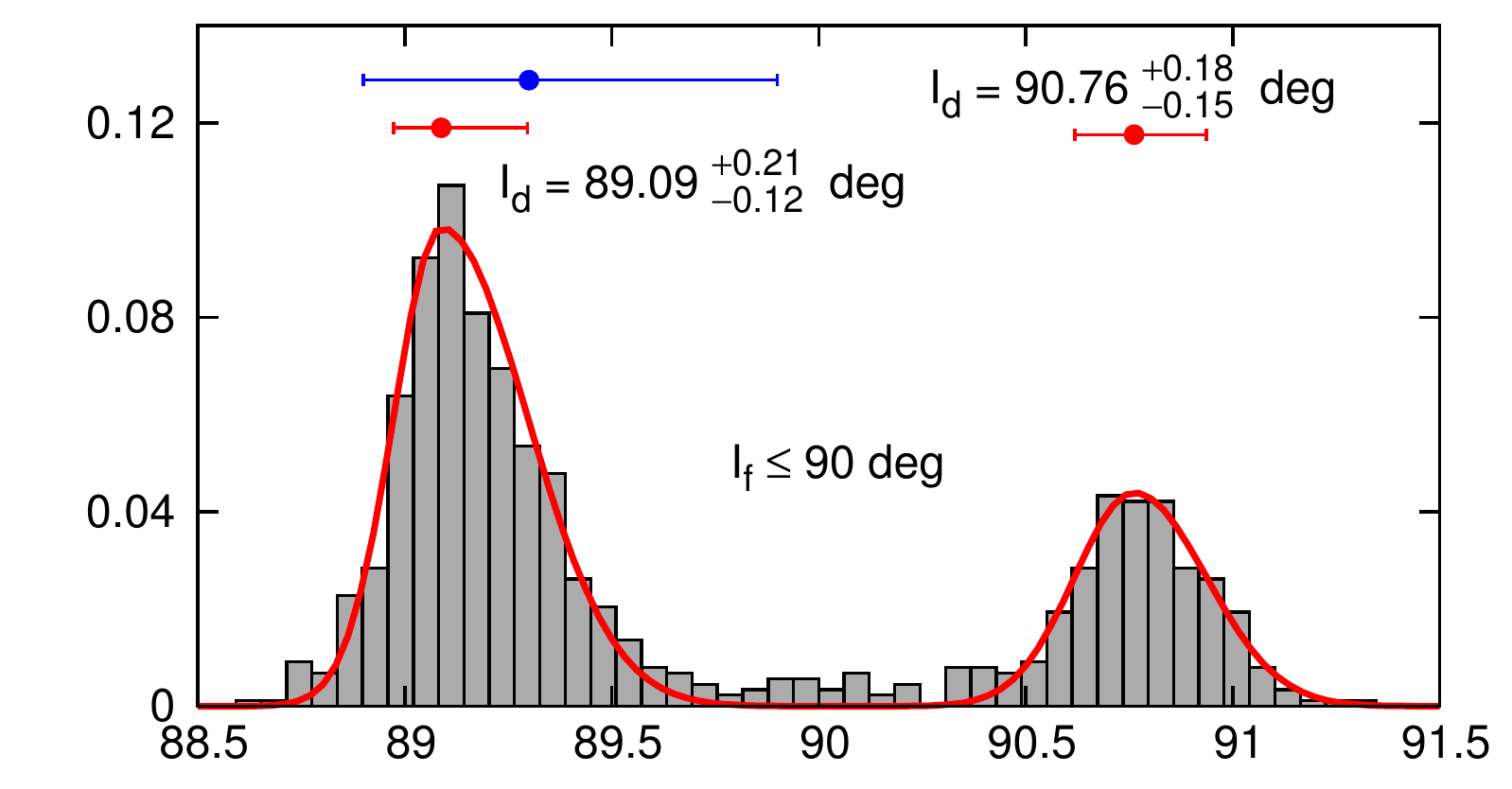}}
\caption{Bootstrap histograms for absolute inclinations, transit model II.}
\label{fig:inclinations}
\end{figure*}

\begin{figure*}
\hbox{\includegraphics[width=0.33\textwidth]{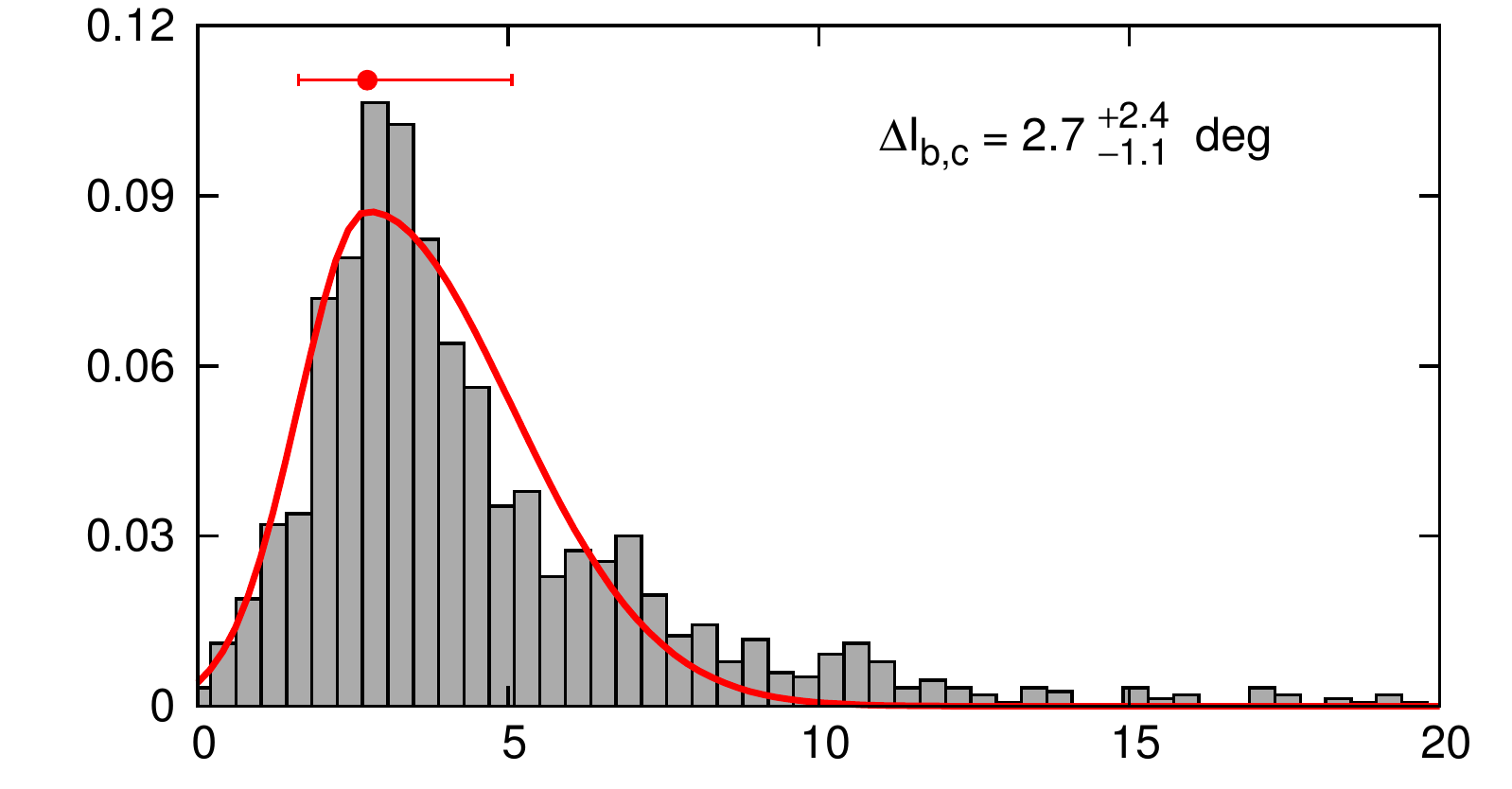}
\includegraphics[width=0.33\textwidth]{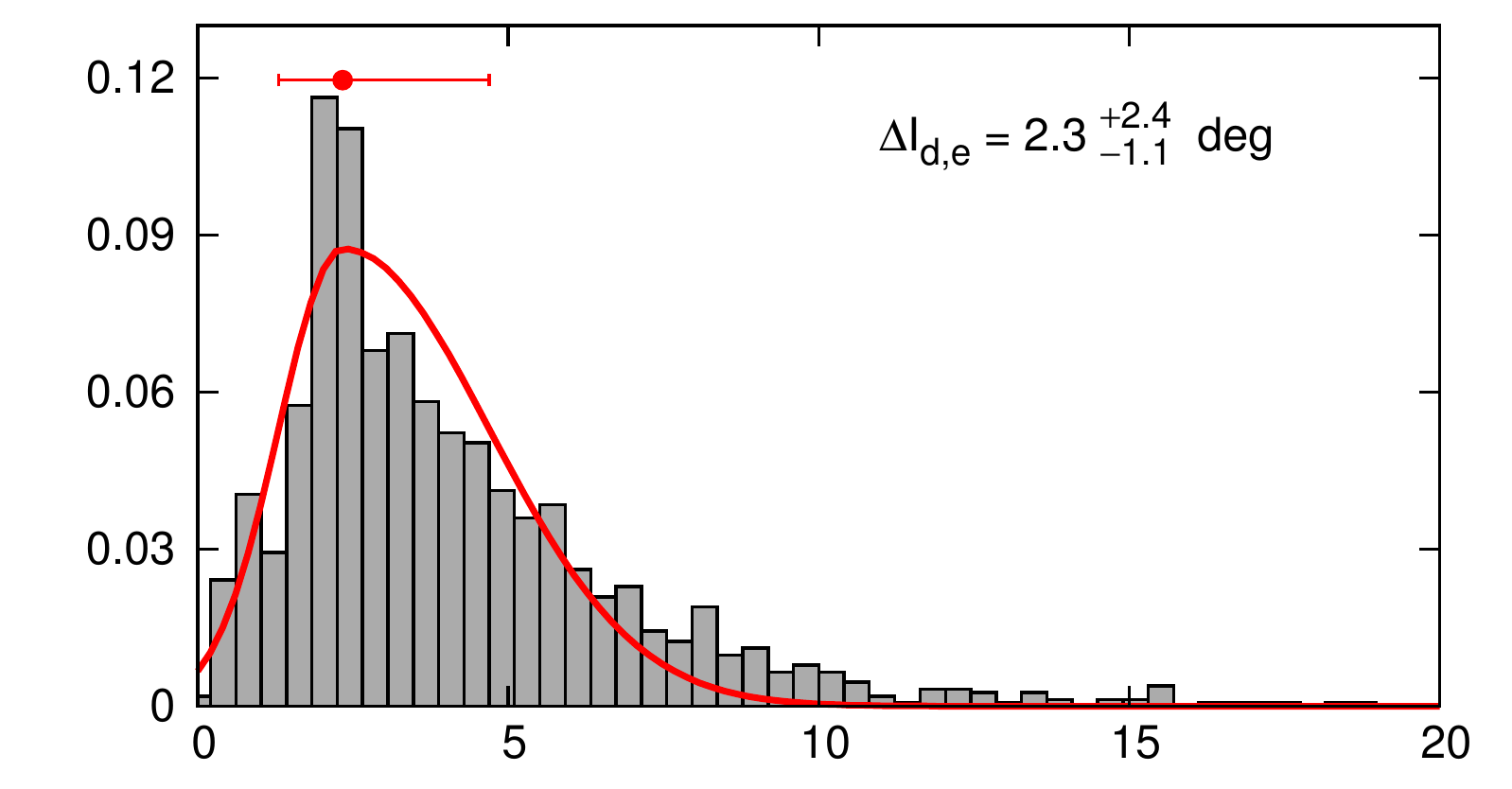}
\includegraphics[width=0.33\textwidth]{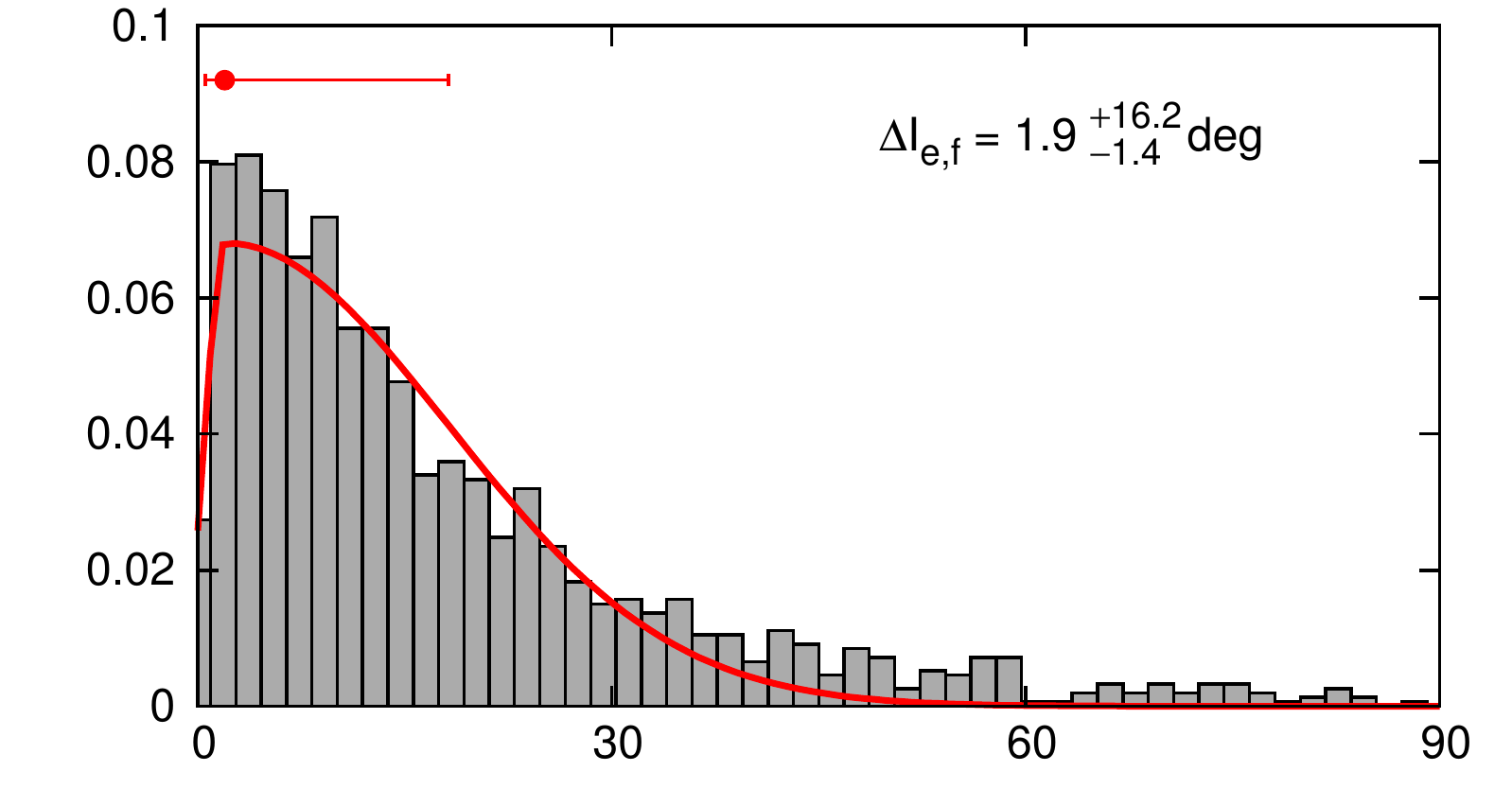}}
\hbox{\includegraphics[width=0.33\textwidth]{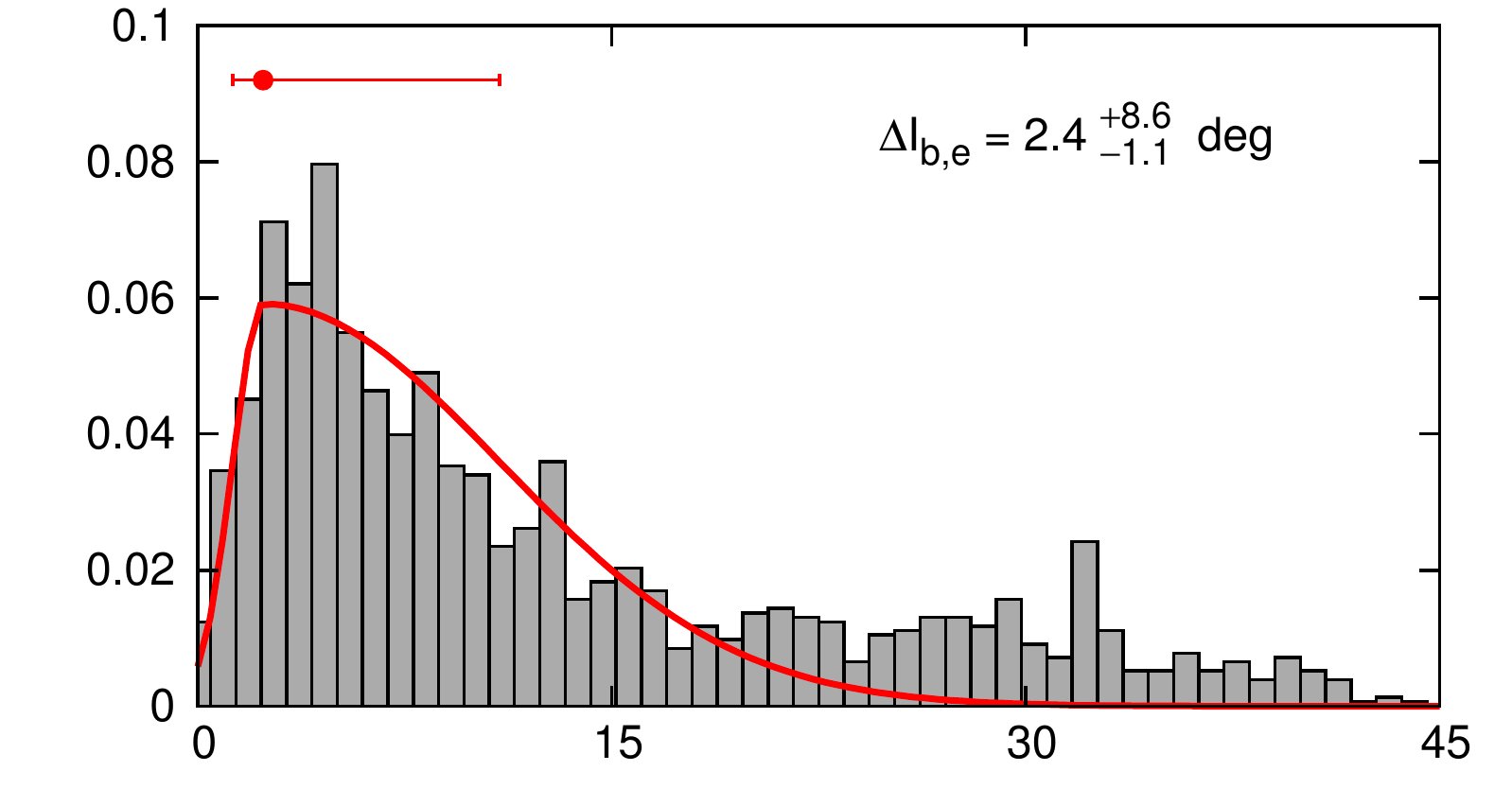}
\includegraphics[width=0.33\textwidth]{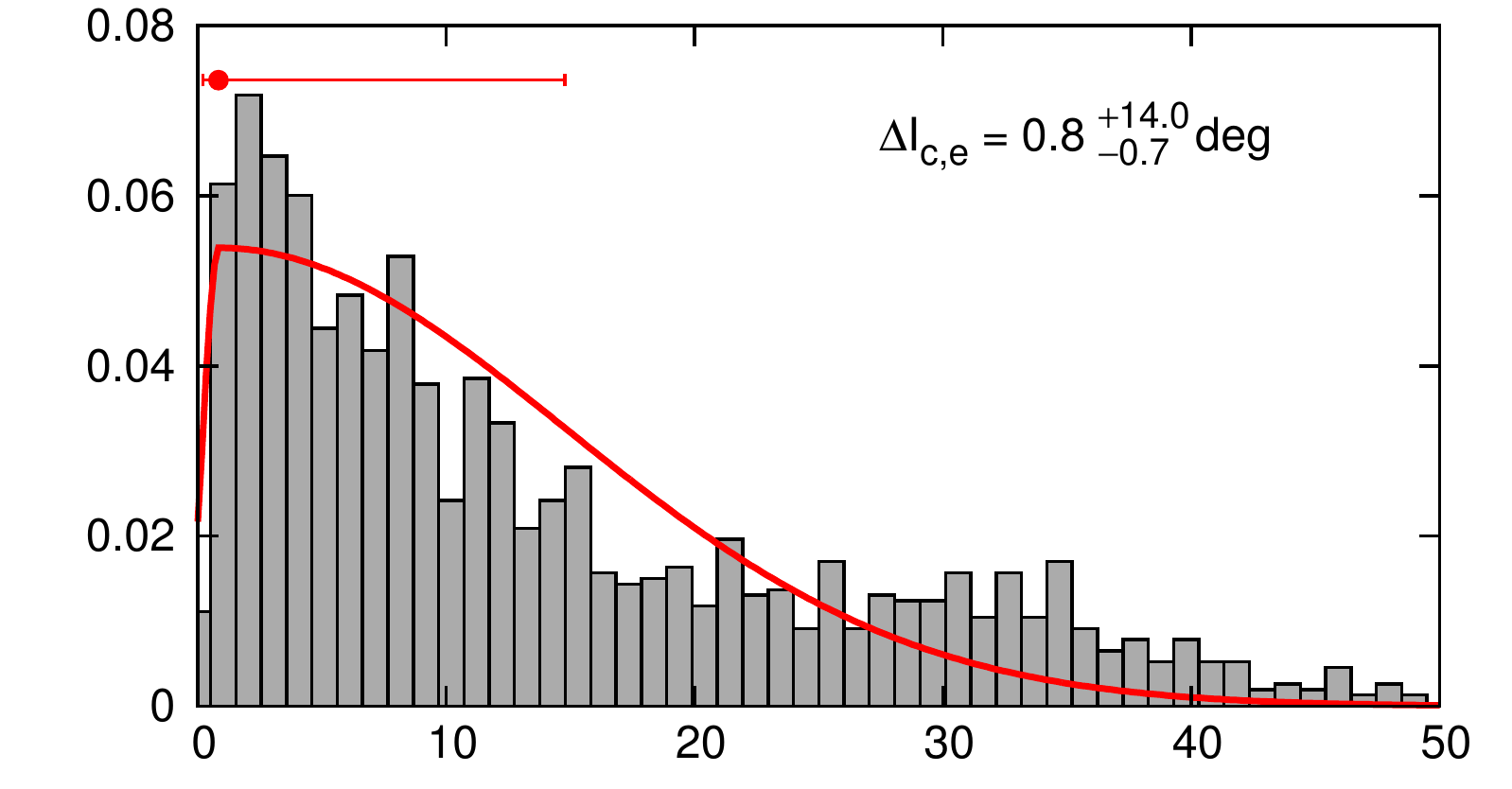}
\includegraphics[width=0.33\textwidth]{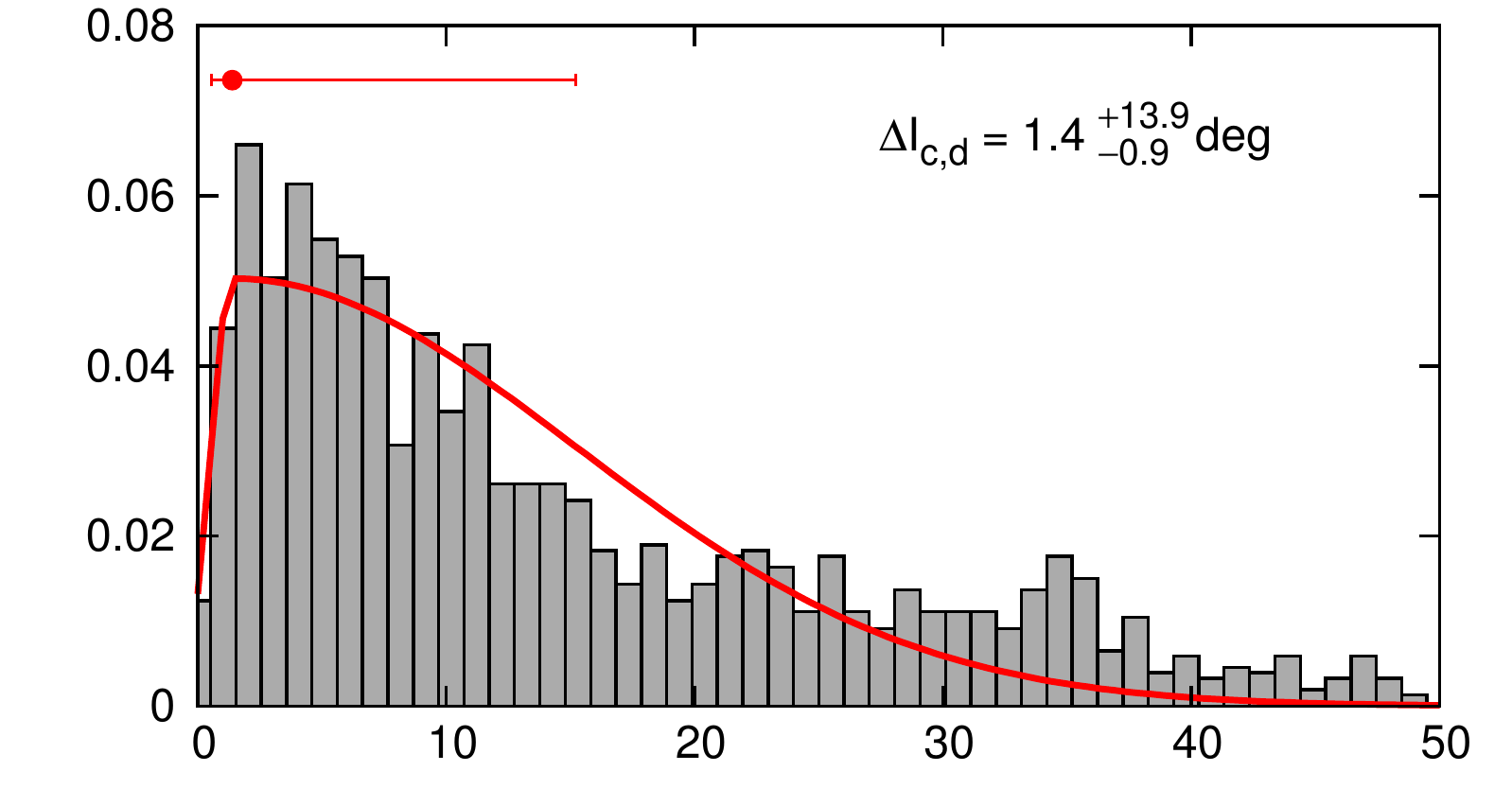}}
\caption{Bootstrap histograms for the mutual inclinations, transit model~II.}
\label{fig:relative_inclination}
\end{figure*}

%
In the next model, we narrow the mostly unconstrained parameters of the  transit
model. We fix the eccentricity $e_g=0$, hence $X_{\idm{g}}$ and  $Y_{\idm{g}}$ are
both equal to $0$. The results of the bootstrap algorithm are  illustrated in
Figs.~\ref{fig:masses}-\ref{fig:relative_inclination}. All  panels in these figures
are constructed in the same manner as Fig.~\ref {fig:modelI}. 
We tested, whether the best--fit parameters encompass at least marginally stable 
solutions with $\langle Y \rangle \approx 2$ after $T=16000$~yr.
Figure~\ref{fig:masses} shows the normalized histograms for masses of  particular
planets expressed in the Earth masses. Besides formal  uncertainties obtained
through the bootstrap (filled  red circles), the best-fit parameters derived in
\citep{Lissauer2011} are  plotted (blue filled circles). Clearly, these estimates
coincide  very well in both cases. There is one exception though, since the mass of 
planet~g is not resolved in \cite{Lissauer2011}. The direct code  helps to resolve
also this mass. It is constrained surprisingly well, in  spite of a narrow
observational window. This result confirms our predictions. Because the orbital
model  is constrained by all measurements, not the TTVs only, the direct algorithm 
makes use of dynamical information contained in the transit depths  and widths.

For a reference, black and green asterisks in Fig.~\ref{fig:masses}   mark masses of
the  Uranus and Neptune, respectively. The masses of planets~b and~f appear
in  a range specific for the super-Earths. They are significantly smaller than the masses of two most distant planets in our Solar system but,  as we will show in the 
next section, their chemical composition has likely much common with the ice  giants
in the Solar system. 
The next Fig.~\ref{fig:radii} shows histograms constructed for planetary  radii
expressed in the unit of the Earth radius. These results confirm data in the 
discovery paper. Similarly to the previous plots, the radii of Uranus and  Neptune
are marked with asterisks. They are also labeled with  $\mbox{R}_{\idm{U}}$ and
$\mbox{R}_{\idm{N}}$, respectively. The derived  radius of planet~g confirms a
hypothesis that it may belong to  the~Uranus/Neptune--class. We note that most of
the planets has radii  smaller than $\mbox{R}_{\idm{U/N}}$, and only planet~e has
its radius larger.
Histograms of the mean densities  are presented in Fig.~ \ref{fig:densities}. The
$x$-axis is for the density expressed w.r.t. the Earth density. Black and green
asterisks mark the values characteristic for Uranus and Neptune,  respectively. The
mean densities of Saturn and water are also marked with  the red and blue symbols,
respectively. According to this plot, the less dense  planet~e has a density of
Saturn. The most dense planet~b may be almost as  dense as the Earth. The densities
of the other planets span a range characteristic  for Saturn and Neptune, from
$\rho_{\idm{S}}$ to $\rho_{\idm{N}}$.

Figure~\ref{fig:semi-major-axes} is for the bootstrap histograms constructed  of the
semi-major axes. These parameters are the best determined among all of the transit
models, with  uncertainties of the order of $10^{-5}\,\au$ only. We do not compare
these  results with data in \citep{Lissauer2011} because they accounted for the 
formal error of the stellar mass. Note that we fixed $m_0 =  0.95\,\msun$, 
\corr{because we found that this parameter is unconstrained by the 
photometric data. Yet it seems that the 
 $\chi_{\nu}^2(m_0)$ function 
monotonically increases in the range of $m_0 \in (0.7,
1.2)\,\msun$.
}

The first five panels of Fig.~\ref{fig:eccentricities} are for the  eccentricities,
and the bottom, right-hand panel is for $\Delta\omega_{\idm{b}, \idm{c}} \equiv 
\omega_{\idm{b}} - \omega_{\idm{c}}$ . These  histograms confirm that the
eccentricities of planets~b to~f are small,  typically less that $0.05$, and the
arguments of pericenters are not well  constrained. The last panel assures us that
$\Delta\omega_{\idm{b},  \idm{c}}$ is determined with an error of only $\sim
10^{\circ}$, recalling  a narrow time--window of the photometric data. The best--fit
parameters  of model~II are given in Tab.~\ref{tab:bootstrapII}. 
\begin{table*}
\caption{
Bootstrap results for model~II (with fixed $e_g = 0$).  Mass of the star is
$0.95\,\msun$ (fixed).  Best fitted stellar parameters are $R_0 =
1.161^{+0.035}_{-0.028}$,  $\corr{\gamma_1 = 0.32^{+0.46}_{-0.30}}$,  $\corr{\gamma_2 =
0.41^{+0.17}_{-0.36}}$,  $\gamma_1 + \gamma_2 = 0.73^{+0.16}_{-0.30}$.  Osculating
Poincar\'e{} elements are given at the epoch of the first observation 
JD~2455964.51128.
}
\begin{tabular}{c c c c c c c c}
\hline
\hline
parameter/planet & b & c & d & e & f & g \\
\hline
$m \, [\mE]$ & $4.2^{+2.8}_{-2.4}$ & $9.2^{+3.6}_{-6.0}$ & $9.2^{+3.5}_{-2.6}$ & $10.5^{+2.4}_{-1.0}$ & $4.4^{+4.7}_{-1.8}$ & $3^{+16}_{-3}$  \\
$R \, [\RE]$ & $2.07^{+0.16}_{-0.13}$ & $3.31^{+0.16}_{-0.12}$ & $3.65^{+0.17}_{-0.14}$ & $4.80^{+0.30}_{-0.19}$ & $2.88^{+0.17}_{-0.15}$ & $3.93^{+0.20}_{-0.13}$ \\
$\bar{\rho} \, [\bar{\rho}_{\earth}]$ & $0.48^{+0.42}_{-0.30}$ & $0.25^{+0.14}_{-0.23}$ & $0.19^{+0.07}_{-0.07}$ & $0.10^{+0.03}_{-0.03}$ & $0.19^{+0.21}_{-0.14}$ & $0.04^{+0.30}_{-0.04}$ \\
$a \, [\au]$ & $0.091087\left(^{+15}_{-7}\right)$ & $0.106521\left(^{+6}_{-11}\right)$ & $0.154233\left(^{+16}_{-10}\right)$ & $0.193926\left(^{+16}_{-17}\right)$ & $0.249511\left(^{+32}_{-23}\right)$ & $0.463924\left(^{+42}_{-37}\right)$  \\
$e \, \cos\omega$ & $0.006^{+0.011}_{-0.022}$ & $0.001^{+0.010}_{-0.017}$ & $-0.013^{+0.007}_{-0.019}$ & $-0.020^{+0.007}_{-0.015}$ & $-0.004^{+0.016}_{-0.013}$ & $0$ (fixed)  \\
$e \, \sin\omega$ & $0.020^{+0.027}_{-0.027}$ & $0.024^{+0.025}_{-0.023}$ & $0.008^{+0.008}_{-0.015}$ & $-0.002^{+0.005}_{-0.014}$ & $-0.007^{+0.020}_{-0.016}$ & $0$ (fixed)  \\
$I^* \,$~[deg] & $88.39^{+0.95}_{-0.24}$ & $91.17^{+0.38}_{-0.22}$ & $89.14^{+0.23}_{-0.13}$ & $88.701^{+0.052}_{-0.076}$ & $89.282^{+0.09}_{-0.11}$ & $90.31^{+0.086}_{-0.055}$  \\
$\Omega \,$~[deg] & $0$ (fixed) & $3.4^{+4.3}_{-2.7}$ & $-23^{+12}_{-12}$ & $-22^{+11}_{-12}$ & $-25^{+24}_{-27}$ & $37^{+63}_{-59}$  \\
$\mathcal{M} + \omega \,$~[deg] & $205.0^{+2.2}_{-2.3}$ & $265.7^{+2.1}_{-1.2}$ & $182.7^{+1.9}_{-0.9}$ & $197.4^{+1.6}_{-1.0}$ & $89.2^{+1.7}_{-1.5}$ & $336.282^{+0.049}_{-0.093}$  \\
$P \, [\mbox{d}]$ & $10.3019\left(^{+23}_{-14}\right)$ & $13.0281\left(^{+14}_{-18}\right)$ & $22.6985\left(^{+34}_{-22}\right)$ & $32.0025\left(^{+45}_{-34}\right)$ & $46.7054\left(^{+89}_{-57}\right)$ & $118.4147\left(^{+90}_{-33}\right)$  \\
$T_0 \, [\mbox{JD}]$ & $471.504\left(^{+21}_{-7}\right)$ & $471.176\left(^{+18}_{-4}\right)$ & $481.454\left(^{+14}_{-6}\right)$ & $487.177\left(^{+19}_{-9}\right)$ & $464.670\left(^{+14}_{-9}\right)$ & $501.916\left(^{+40}_{-11}\right)$  \\
\hline
\hline
\end{tabular}
\label{tab:bootstrapII}
\end{table*}

Inclination $I_{\idm{b}}$ was constrained to the $\leq  90^{\circ}$ range, and due
to the invariance rule implied by the direction of the total angular momentum, the
remaining inclination $I_i$ may  be smaller and larger than $90^{\circ}$. We tested
whether there  is a correlation of the transit events with a given half--disc of
the  star. We found that both cases are equally possible. Because the orbits are 
inclined to the plane of the sky at angles close to $90^{\circ}$,  the relative 
inclinations with the same longitudes of nodes may be $\sim
2^{\circ}$--$3^{\circ}$.  As expected, the indirect parameters $\Omega_i$ are
unconstrained, see  Tab.~\ref{tab:bootstrapII}. Therefore, the main contribution to
the  uncertainties of the relative inclinations comes from ambiguous estimates  of
$\Omega_i$ rather than of $I_i$.

Curiously, there appears a clear correlation between mutual inclinations  in
particular pairs of orbits, namely c~and e, f~and~e, as well as  d~and f. This can
be seen in normalized histograms constructed for the  inclinations,
Fig.~\ref{fig:inclinations}. For a chosen planet, we  transform $I_i$  to $\leq
90^{\circ}$ range (in accord with the inclination invariance rule), and we compute 
the bootstrap histogram for $I_j$. Panels  of Fig.~\ref{fig:inclinations}, from the
left to the right, are for pairs  $(i,j) = (\mbox{e}, \mbox{c}), (\mbox{e},
\mbox{f}), (\mbox{f}, \mbox{d})$. If  $I_{\idm{e}} \leq 90^{\circ}$ then  much more
likely $I_{\idm{c}}, I_{\idm{f}} \leq 90^{\circ}$ than $I_{\idm{c}}, I_{\idm{f}}
\geq 90^{\circ}$. Similarly,  if $I_{\idm{f}} \leq 90^{\circ}$, then $I_{\idm{d}}
\leq 90^{\circ}$  appears more likely than $I_{\idm{f}}  \geq 90^{\circ}$.
\begin{table*}
\caption{
Bootstrap results for model~III ($e_g = 0, \Omega_g = 0$). Mass of  the star is
$0.95\,\msun$ (fixed). Fitted stellar parameters: $R_0 =  1.158^{+0.021}_{-0.038}$,
$\corr{\gamma_1 = 0.32^{+0.44}_{-0.25}}$, $\corr{\gamma_2 =  0.41^{+0.25}_{-0.40}}$, $\gamma_1 +
\gamma_2 = 0.73^{+0.32}_{-0.21}$.  Osculating Poincar\'e{} elements are given at the
epoch of the first observation JD~2455964.51128.
}
\begin{tabular}{c c c c c c c c}
\hline
\hline
parameter/planet & b & c & d & e & f & g \\
\hline
$m \, [\mE]$ & $4.0^{+2.5}_{-3.0}$ & $9.1^{+3.3}_{-6.1}$ & $9.1^{+3.5}_{-2.9}$ & $10.6^{+2.9}_{-2.1}$ & $4.3^{+5.8}_{-2.7}$ & $1^{+28}_{-1}$  \\
$R \, [\RE]$ & $2.07^{+0.13}_{-0.13}$ & $3.30^{+0.13}_{-0.13}$ & $3.64^{+0.14}_{-0.16}$ & $4.79^{+0.19}_{-0.20}$ & $2.88^{+0.17}_{-0.15}$ & $3.92^{+0.12}_{-0.15}$ \\
$\bar{\rho} \, [\bar{\rho}_{\earth}]$ & $0.45^{+0.40}_{-0.33}$ & $0.25^{+0.12}_{-0.17}$ & $0.19^{+0.09}_{-0.07}$ & $0.10^{+0.05}_{-0.02}$ & $0.18^{+0.27}_{-0.09}$ & $0.02^{+0.49}_{-0.04}$ \\
$a \, [\au]$ & $0.091088\left(^{+15}_{-9}\right)$ & $0.106519\left(^{+9}_{-10}\right)$ & $0.154234\left(^{+22}_{-6}\right)$ & $0.193924\left(^{+24}_{-13}\right)$ & $0.249511\left(^{+31}_{-23}\right)$ & $0.463917\left(^{+55}_{-36}\right)$  \\
$e \, \cos\omega$ & $0.015^{+0.013}_{-0.025}$ & $0.008^{+0.015}_{-0.017}$ & $-0.008^{+0.010}_{-0.016}$ & $-0.015^{+0.009}_{-0.011}$ & $-0.002^{+0.012}_{-0.015}$ & $0$ (fixed)  \\
$e \, \sin\omega$ & $0.007^{+0.034}_{-0.019}$ & $0.012^{+0.025}_{-0.022}$ & $0.000^{+0.007}_{-0.013}$ & $-0.010^{+0.005}_{-0.013}$ & $-0.011^{+0.016}_{-0.023}$ & $0$ (fixed)  \\
$I^* \,$~[deg] & $88.35^{+0.83}_{-0.45}$ & $88.80^{+0.38}_{-0.20}$ & $90.88^{+0.20}_{-0.14}$ & $88.701^{+0.071}_{-0.066}$ & $89.26^{+0.10}_{-0.12}$ & $89.699^{+0.077}_{-0.064}$  \\
$\Omega \,$~[deg] & $0$ (fixed) & $-3.5^{+4.2}_{-2.5}$ & $21^{+14}_{-12}$ & $20^{+14}_{-11}$ & $22^{+24}_{-23}$ & $0$ (fixed)  \\
$\mathcal{M} + \omega \,$~[deg] & $204.0^{+2.8}_{-1.7}$ & $264.9^{+1.9}_{-1.8}$ & $182.2^{+1.5}_{-1.2}$ & $196.7^{+1.3}_{-1.2}$ & $89.0^{+1.9}_{-1.6}$ & $336.29^{+0.05}_{-0.13}$  \\
$P \, [\mbox{d}]$ & $10.3020\left(^{+25}_{-15}\right)$ & $13.0278\left(^{+16}_{-18}\right)$ & $22.6986\left(^{+43}_{-18}\right)$ & $32.0019\left(^{+49}_{-34}\right)$ & $46.7056\left(^{+87}_{-55}\right)$ & $118.412\left(^{+26}_{-12}\right)$  \\
$T_0 \, [\mbox{JD}]$ & $471.504\left(^{+20}_{-9}\right)$ & $471.176\left(^{+17}_{-4}\right)$ & $481.454\left(^{+17}_{-6}\right)$ & $487.176\left(^{+21}_{-6}\right)$ & $464.671\left(^{+13}_{-10}\right)$ & $501.915\left(^{+43}_{-14}\right)$  \\
\hline
\hline
\end{tabular}
\label{tab:bootstrapIII}
\end{table*}

For particular pairs of planets, the relative inclinations can be determined 
surprisingly well. Figure~\ref{fig:relative_inclination} shows  the bootstrap
histograms  $\Delta I_{i,j}$ for such pairs which exhibit well constrained values. 
These histograms reveal that orbits of planets~b and~c are almost coplanar. 
Similarly, the pair of planets~d and~e form an almost coplanar sub-system. The  mutual
inclinations of orbits in these pairs are less than $5^{\circ}$,  with most likely
values of $2^{\circ}$--$3^{\circ}$. The remaining panels indicate  that the mutual
inclinations between five inner orbits remain within a~few  degrees range. Their
upper limits are not so small as in the first two  sub-systems. The outermost orbit
of planet~g may by highly inclined to the  rest of the system, see errors of
$\Omega_g$ in Tab.~\ref{tab:bootstrapII}.

These results confirm a hypothesis in the discovery paper. In accord with this work,
planetary orbits in the Kepler--11 system should be mutually inclined by no more
than  a~few degrees. It flows from estimating a probability that for a given
orientation of the orbits, all six planets transit the star. This reasoning assumes
that all inclinations are independent. However, we found that Kepler--11 system is
composed of two or three sub-systems, which exhibit  small mutual inclinations of
orbits. 
\corr{Although a probability that the mutual inclinations between these
sub-systems are significant seems a bit larger than for fully
independent orbits, it still remains very small. We estimate 
that a randomly located observer can detect transits of all 6 planets
with a probability as small as  $\sim 0.05\%$, for both models~I and~II.}

\begin{table*}
\caption{Bootstrap results for model~IV 
($e_g=0, \Omega_i=0, i=b,c,d,e,f,g$). 
Mass of the star is $0.95\,\msun$ (fixed). 
The best--fit stellar parameters: $R_0 = 1.140^{+0.026}_{-0.024}$, 
$\corr{\gamma_1 = 0.30^{+0.32}_{-0.30}}$, 
$\corr{\gamma_2 = 0.42^{+0.18}_{-0.42}}$, 
$\gamma_1 + \gamma_2 = 0.72^{+0.22}_{-0.25}$. 
Osculating Poincar\'e{} elements are given at the epoch of the first observation 
JD~2455964.51128.
}
\begin{tabular}{c c c c c c c c}
\hline
\hline
parameter/planet & b & c & d & e & f & g \\
\hline
$m \, [\mE]$ & $3.3^{+2.4}_{-1.8}$ & $8.8^{+4.0}_{-5.0}$ & $8.9^{+2.3}_{-3.4}$ & $10.3^{+1.9}_{-1.5}$ & $4.1^{+3.8}_{-2.4}$ & $<21$  \\
$R \, [\RE]$ & $2.01^{+0.13}_{-0.13}$ & $3.23^{+0.12}_{-0.12}$ & $3.59^{+0.15}_{-0.13}$ & $4.70^{+0.21}_{-0.15}$ & $2.82^{+0.15}_{-0.15}$ & $3.85^{+0.12}_{-0.12}$ \\
$\bar{\rho} \, [\bar{\rho}_{\earth}]$ & $0.40^{+0.33}_{-0.19}$ & $0.26^{+0.13}_{-0.18}$ & $0.19^{+0.07}_{-0.08}$ & $0.10^{+0.03}_{-0.03}$ & $0.18^{+0.20}_{-0.10}$ & $<0.35$ \\
$a \, [\au]$ & $0.091088\left(^{+13}_{-11}\right)$ & $0.106514\left(^{+6}_{-8}\right)$ & $0.154234\left(^{+18}_{-7}\right)$ & $0.193927\left(^{+19}_{-10}\right)$ & $0.249507\left(^{+35}_{-26}\right)$ & $0.463913\left(^{+57}_{-39}\right)$  \\
$e \, \cos\omega$ & $-0.002^{+0.015}_{-0.021}$ & $-0.006^{+0.015}_{-0.016}$ & $-0.012^{+0.009}_{-0.019}$ & $-0.018^{+0.006}_{-0.017}$ & $-0.007^{+0.015}_{-0.020}$ & $0$ (fixed)  \\
$e \, \sin\omega$ & $0.049^{+0.020}_{-0.050}$ & $0.050^{+0.020}_{-0.044}$ & $-0.010^{+0.008}_{-0.018}$ & $-0.016^{+0.006}_{-0.017}$ & $-0.017^{+0.011}_{-0.021}$ & $0$ (fixed)  \\
$I^* \,$~[deg] & $88.76^{+0.96}_{-0.41}$ & $91.00^{+0.40}_{-0.25}$ & $90.89^{+0.17}_{-0.18}$ & $88.743^{+0.043}_{-0.049}$ & $89.30^{+0.10}_{-0.09}$ & $89.719^{+0.061}_{-0.068}$  \\
$\mathcal{M} + \omega \,$~[deg] & $205.8^{+2.4}_{-2.1}$ & $266.4^{+1.7}_{-1.9}$ & $182.7^{+2.1}_{-1.2}$ & $197.2^{+1.8}_{-0.9}$ & $89.6^{+2.6}_{-1.3}$ & $336.286^{+0.066}_{-0.057}$  \\
$P \, [\mbox{d}]$ & $10.3021\left(^{+24}_{-18}\right)$ & $13.0269\left(^{+10}_{-17}\right)$ & $22.6986\left(^{+43}_{-13}\right)$ & $32.0027\left(^{+41}_{-30}\right)$ & $46.7044\left(^{+88}_{-59}\right)$ & $118.411\left(^{+14}_{-15}\right)$  \\
$T_0 \, [\mbox{JD}]$ & $471.505\left(^{+20}_{-7}\right)$ & $471.177\left(^{+18}_{-3}\right)$ & $481.452\left(^{+17}_{-6}\right)$ & $487.176\left(^{+22}_{-6}\right)$ & $464.670\left(^{+14}_{-9}\right)$ & $501.914\left(^{+26}_{-11}\right)$  \\
\hline
\hline
\end{tabular}
\label{tab:bootstrapIV}
\end{table*}
%
%
\subsection{Models III ($\pmb{e_g = 0, \Omega_g = 0}$) 
and IV ($\pmb{e_g = 0, \Omega_i = 0}$)}
%
%
The results for model~III and model~IV are given in Tabs.~\ref {tab:bootstrapIII}
and \ref{tab:bootstrapIV}. Most of these results are  common for all transit
models~I to~IV. There are some differences regarding  a determination of the mass of
planet~g. In the realm of models III and IV  (note that both have fixed  $e_g=0$ and
$\Omega_g=0$), only an upper limit  of $m_g$ smaller than $20-30$~Earth masses may
be found. The low limits  of $m_g$ in model~I are likely due to weakly constrained
$e_g$ and  $\Omega_g$. 

Let us recall that in the bootstrap set derived for model~II, we found only  two
solutions with $\langle Y \rangle \approx 2$ after $16000$~yr.  However, this
integration time scale is too short to detect weak  instability which actually leads
to catastrophic disruption of these  configurations. 
It was verified by the direct, long--term integrations.
Hence, we did not detect any long--term stable  configuration in the bootstrap set
of model~II. Similarly, stability tests  performed for configurations of model~III
did not reveal any stable models.  As compared to model~II, fixed
$\Omega_{\idm{g}}=0$ seem does not change  the general view of the stability of the
system. 

For model~IV we found many stable configurations which confirm stability  analysis
in \cite{Lissauer2011}. They found some stable solutions assuming  that the
Kepler-11 system is strictly co-planar. We may conclude that a  factor of small
relative inclinations is more important for maintaining the  long term stability
than small eccentricities. This will be discussed in  more detail further in this
work.

\corr
{We examined a probability that a randomly located observer could detect
transits of all planets in the system.  This is basically unlikely for
model~III ($\sim 0.09\%$), while for model~IV a probability of such an event
is larger, and we estimate it $\sim 3.4\%$.}

%
\section{Discussion on the planet interiors}
%
\label{sec:composition}

\begin{figure*}
\hbox{\includegraphics[width=0.33\textwidth]{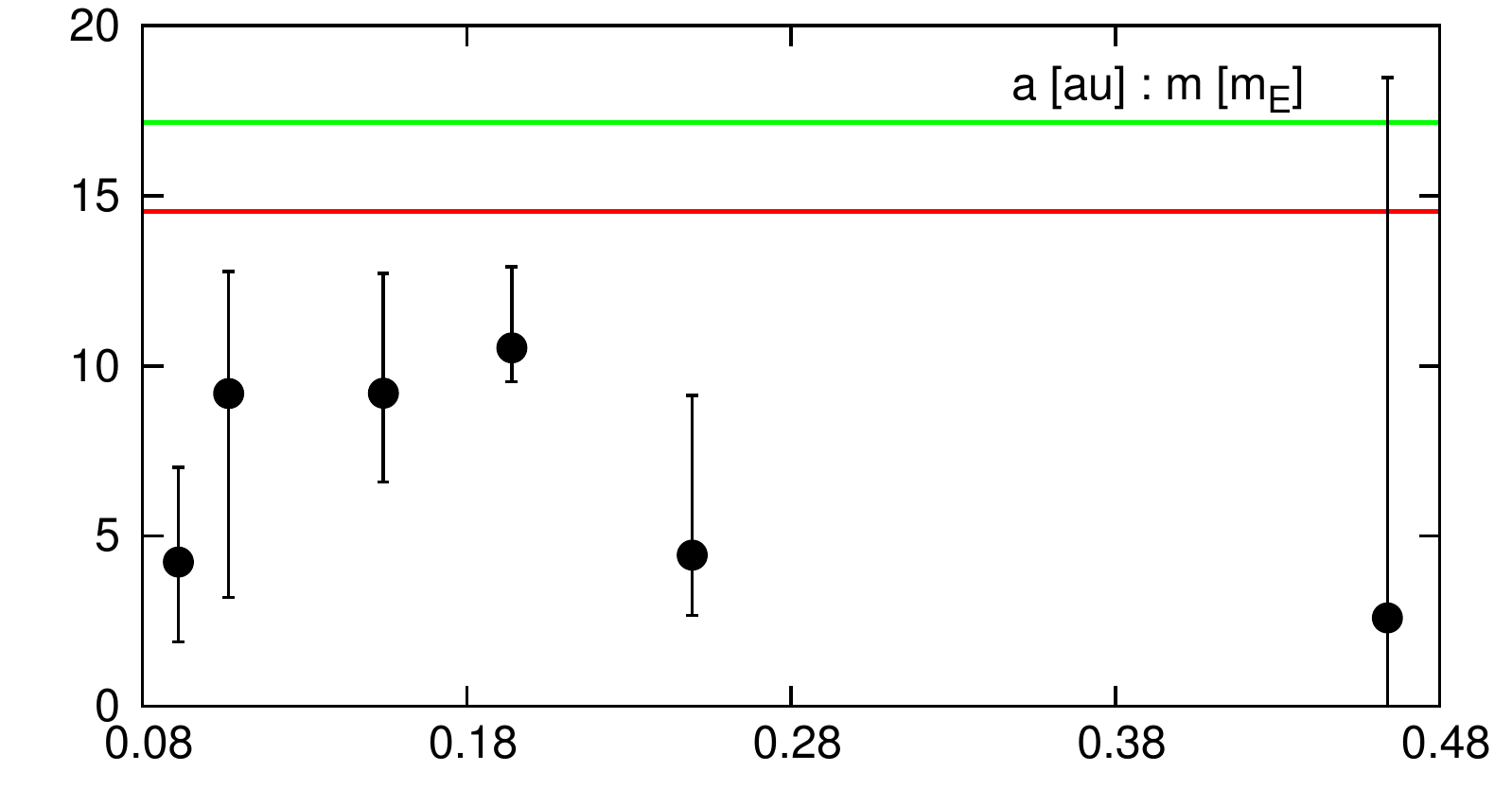}
\includegraphics[width=0.33\textwidth]{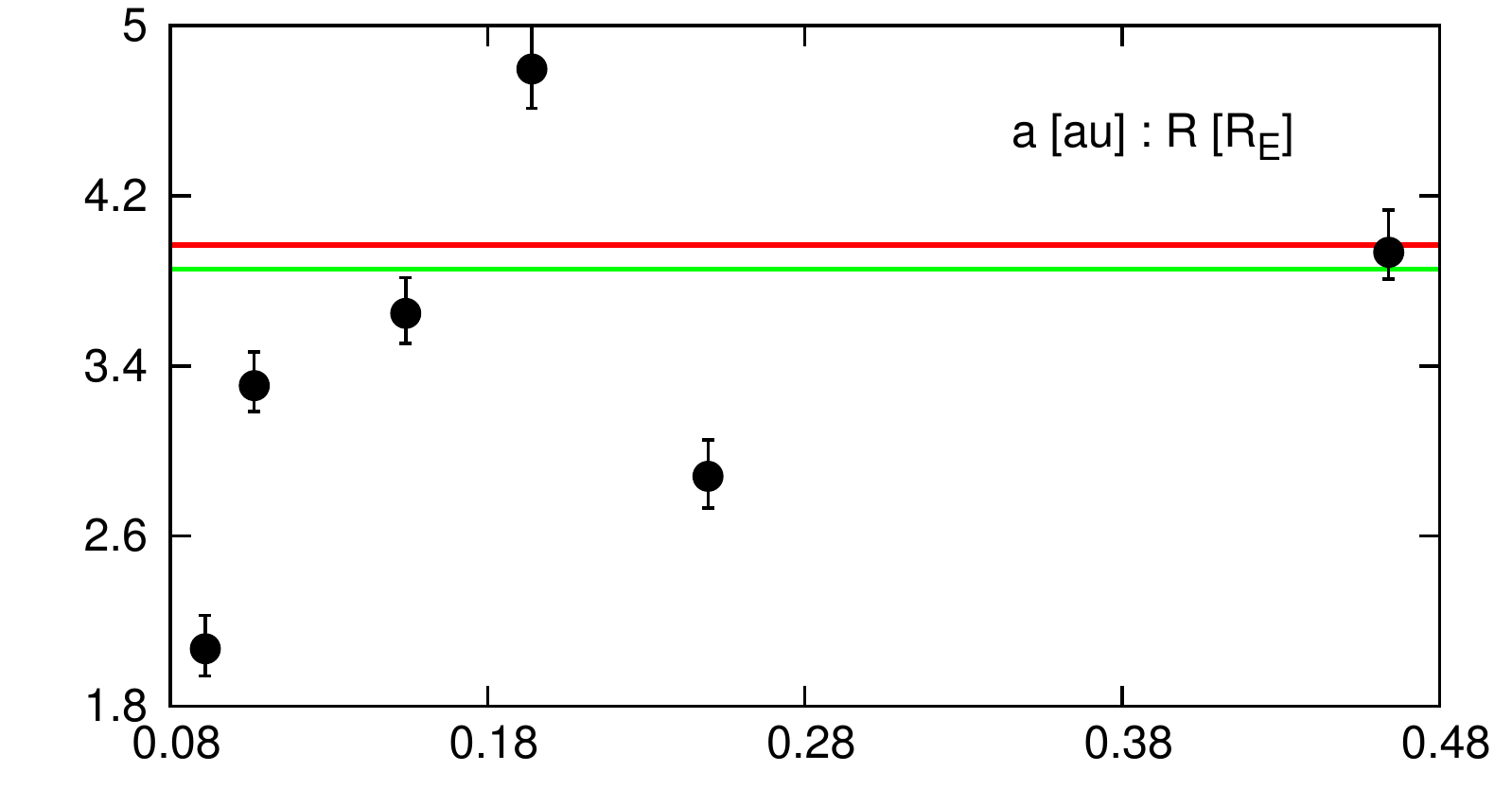}
\includegraphics[width=0.33\textwidth]{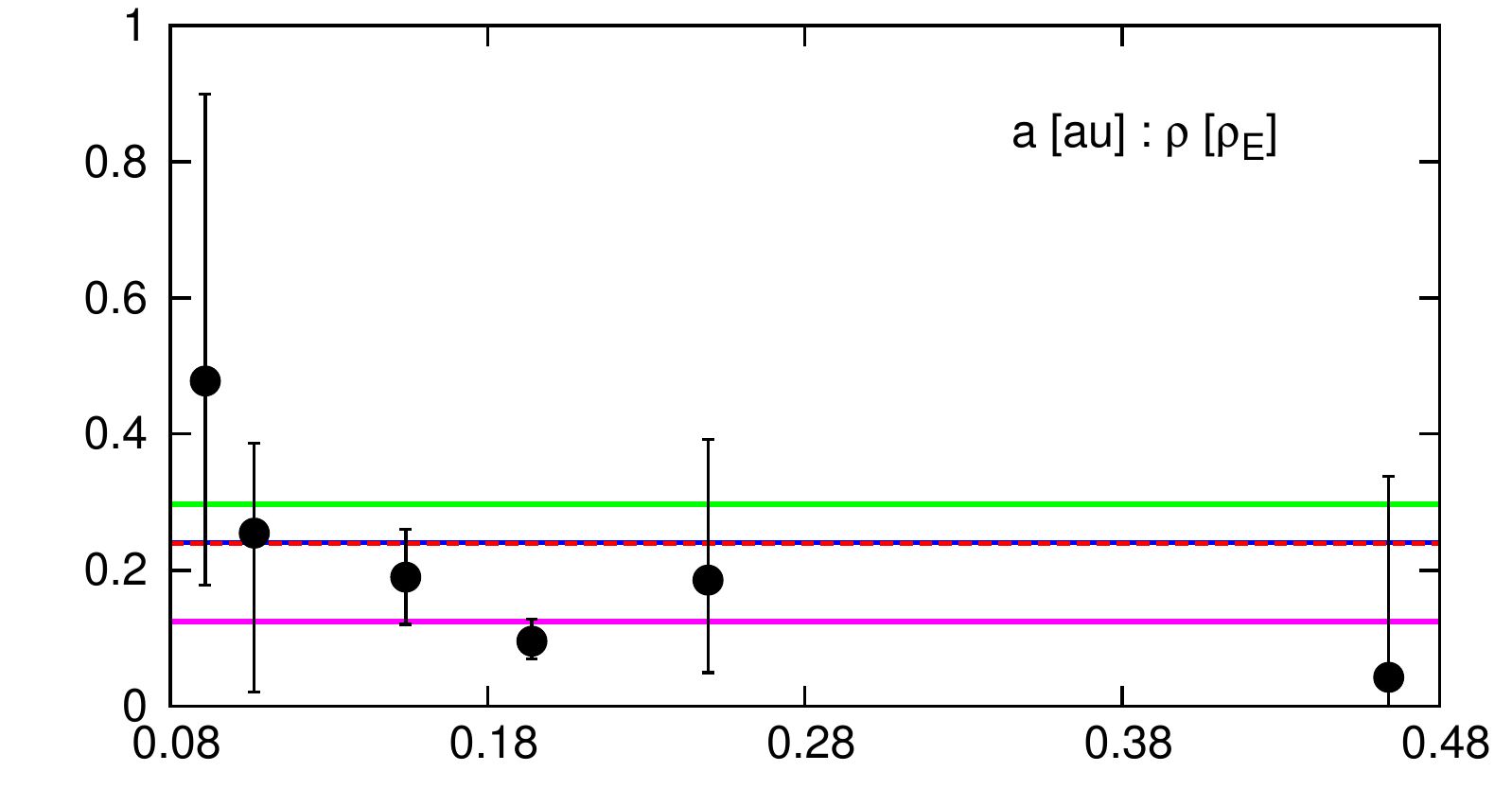}}
\hbox{\includegraphics[width=0.33\textwidth]{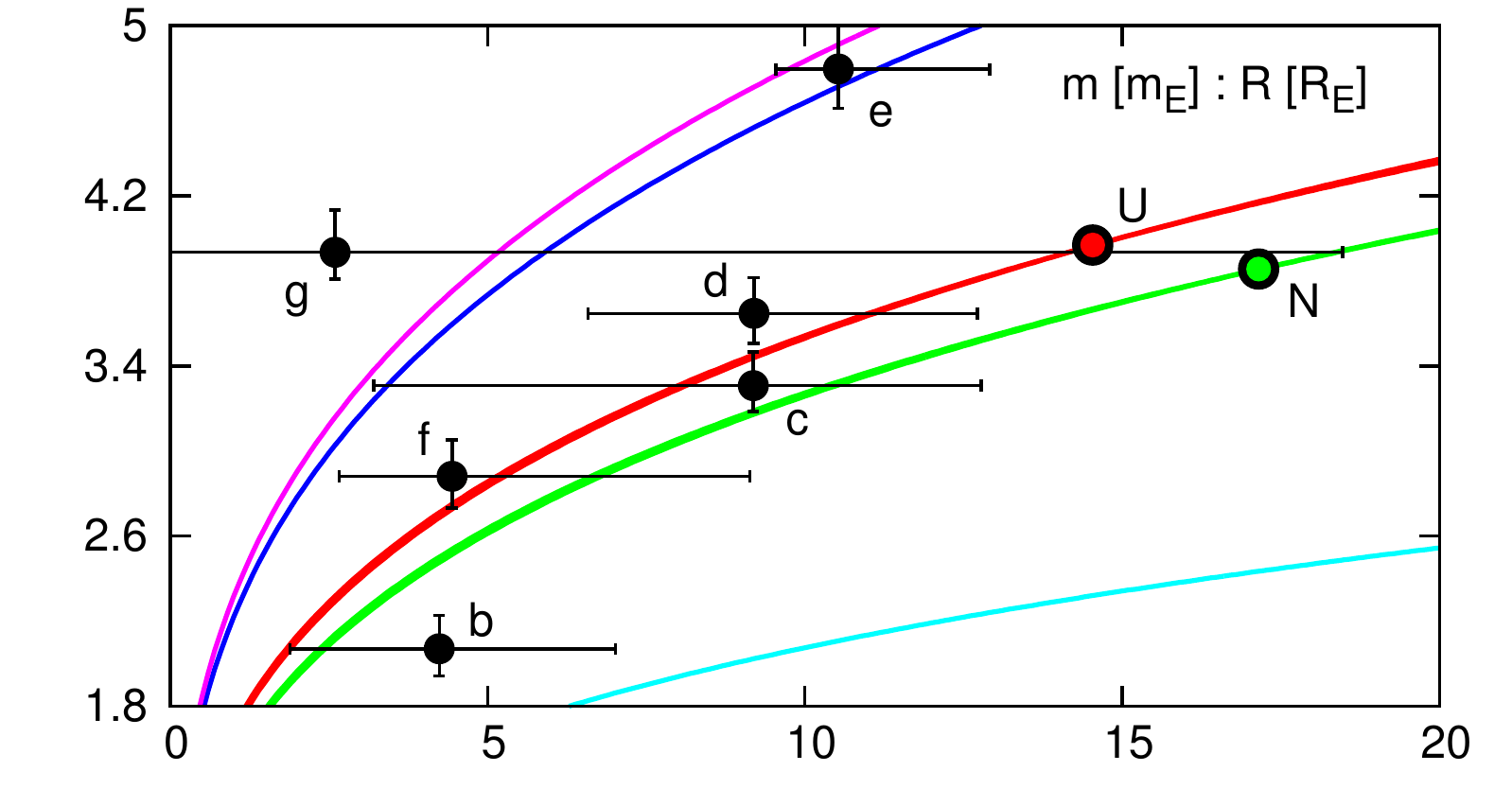}
\includegraphics[width=0.33\textwidth]{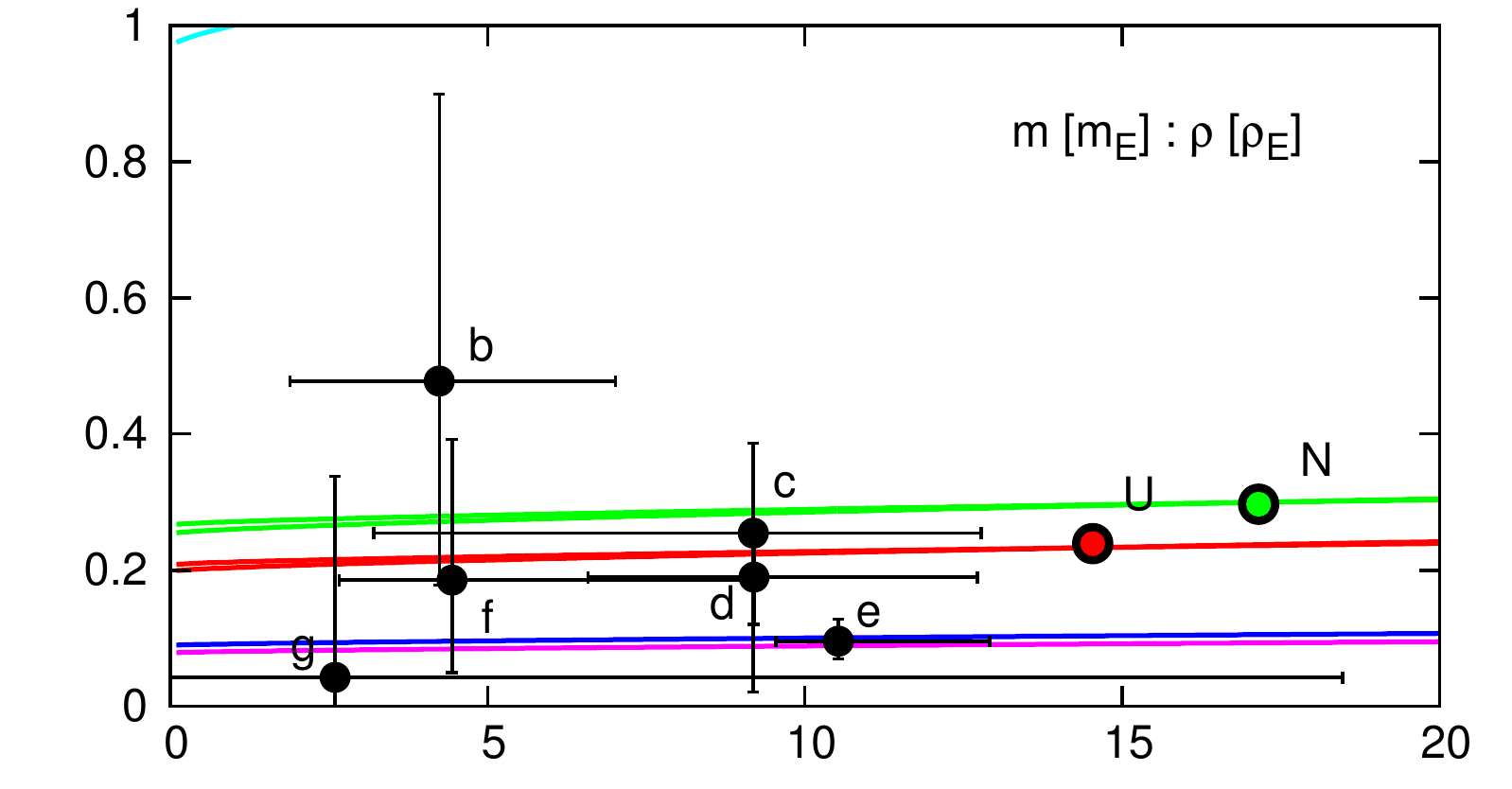}
\includegraphics[width=0.33\textwidth]{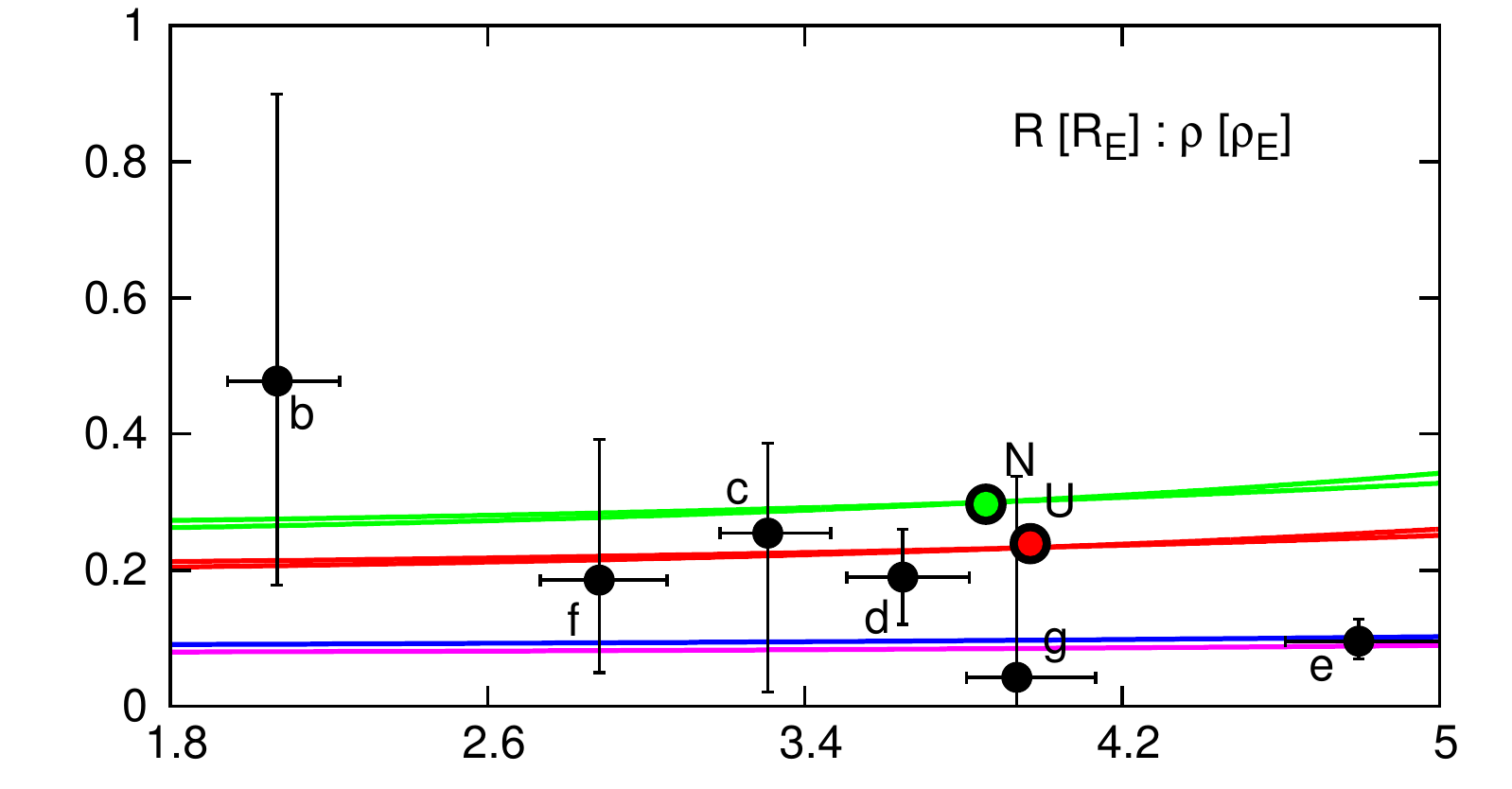}}
\caption{
{\em The top row}: mass, radius and mean density as a function of 
the semi-major axis 
(model~II). Black and green solid curves are for Uranus and Neptune, 
respectively. {\em The bottom row}: mass--radius, mass--mean density and 
radius--mean density relations.
}
\label{fig:masses_radii_densities}
\end{figure*}
Figure~\ref{fig:masses_radii_densities} shows bootstrap diagrams of a few  selected
pairs of parameters. These results are for model~II. The top row is for the 
semi-major axes and the planetary masses, the radii and mean  densities,
respectively. The red and green curves mark the data for Uranus  and Neptune. The
bottom row is for the mass--radius, mass--density and  radius--density relations,
respectively. Similarly, the red and green filled  circles are for Uranus and
Neptune.
As we concluded above, the orbital solutions in set~II are only marginally  stable,
however, it is a matter of unconstrained orbital angles. Note that a discussion in
this Section concerns semi-major axes 
(known with an excellent precision) as well as planetary masses and radii.

This figure reveals that the most inner four planets in the Kepler-11 system exhibit
a clear and curious anti/correlation of masses, radii and densities   with the
semi-major axes. Masses and radii increase with $a_i$, while  densities decrease.
The last panel constructed for $(R, \rho)$ shows a weak  anti-correlation between
the radii and densities, the smaller  radius, the larger density. 
\begin{figure*}
\includegraphics[width=0.99\textwidth]{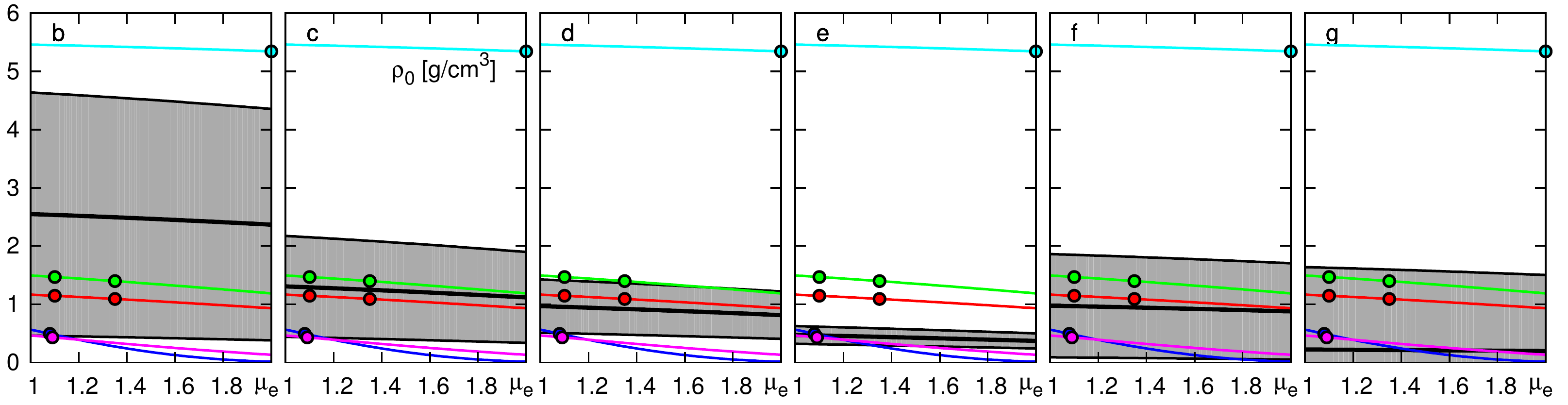}
\caption{
Characteristic density $\rho_0$ of the chemical mixture of planetary  interiors as
functions of the mean number of nucleons per one electron, 
$\mu_e$ (model~II). 
}
\label{fig:rho0}
\end{figure*}
The determined masses and radii of the planets provide some insight into their 
chemical composition. We use a simple analytic relation between the radius and  the
mass of a cold body \citep{Lynden-Bell2001, Lynden-Bell2001b} to estimate  the
characteristic density $\rho_0$ of planetary matter. 
This density can be compared with $\rho_0$ calculated for a given number of nucleons
per  number of electrons of a chemical mixture forming the planet, $\mu_e$.  The
value of $\mu_e$ is a simple mean over the elements in each chemical substance or
component.  For instance the H-He mixture has $\mu_e = 8/7$ for the mass 
proportions $3$ to $1$, and ice or rock has $\mu_e \approx 2$. In this way, we can
obtain some insight into likely chemical composition of the planets.

Our results for Kepler-11 are presented in Fig.~\ref{fig:rho0}. Black curves with
grey  areas are for $\rho_0$ and its uncertainty $\Delta\rho_0$. Each panel is  for
one planet of the Kepler-11 system. Data for planets~b to planet~g are  displayed
from the left to the right, respectively. The colored curves are  for the Solar
system, i.e., for Uranus (red), Neptune (green),  Jupiter (blue), Saturn (violet)
and the Earth (light blue).  The density $\rho_0$ was computed in a wide range of
$\mu_e \in [1,2]$.   These values are known relatively well for the Sun companions. 
Following \cite{Helled2011},  for Uranus and Neptune one finds $\mu_e \approx 1.1$
(for the icy model)  and $\mu_e \approx 1.35$ (for the rocky model). The density
$\rho_0$ in these  particular case is plotted with filled circles. Similarly, for
Jupiter and  Saturn, $\mu_e$ may be also estimated $\approx 1.08-1.09$ \citep
{Guillot1999}. Values of $\rho_0$ for these particular $\mu_e$  are marked with
circles. Let us note that densities $\rho_0$ of Jupiter and Saturn  are almost
identical.

\cite{Lynden-Bell2001b} argue that $\rho_0$ is the zero--pressure density 
$\rho_{0,\idm{p}}$ of the chemical mixture of the planets. Because their  model has
many simplifications, $\rho_0$ is usually $2-5$ times larger then  $\rho_{0,
\idm{p}}$.  Keeping this in mind, the densities $\rho_0$  calculated for Kepler-11
planets can be compared with those of the Solar  system planets. The value of
$\rho_0$ is best determined for planet~e. Its  is very close to the Jupiter/Saturn
(J/S) value $\sim  0.45\,\mbox{g}\,\mbox{cm}^{-3}$. This suggests, that planet~e is
built  mainly of a H/He mixture with mass proportions of the elements close to 
$3/1$ with a~portion of heavier elements contained in ices or rocks. This  makes the
planet classified as a smaller ``cousin'' of Jupiter and Saturn  rather than of
Neptune and Uranus, as suggested in \citep{Lissauer2011}.

The density $\rho_0$ of planet~b is determined worse than for planet~e. It  is
rather unlikely though that it belongs to the same class as planet~e.  Parameter
$\rho_0$ is larger than for Jupiter and Saturn, even taking into  account a large
uncertainty. It is also larger than $\rho_0$ for Uranus and  Neptune (U/N)--like
planets but is smaller than $\rho_0$ for the Earth.  We can conclude that planet~b
is a~small planet containing a large  percentage of heavy elements in its interior, 
which is likely larger than  in the ice giants. It is reasonable to classify this 
planet in the super-Earths family, although, it may be also a small  Neptune--like
planet.

Planet~f has the mass similar to planet~b. However, its composition is likely 
between the Jupiter--Saturn and Uranus--Neptune classes. Planet~d has likely
similar  composition as planet~f.  The best-fit estimate of $\rho_0$ for planet~c is
very close to the Uranus/Neptune value. For planet~g there is only the upper limit
of $\rho_0$. However, it is probably close to the Jupiter/Saturn value or, less
likely, to the Neptune/Uranus value.

These conclusions are reinforced after inspecting the bottom-left panel of 
Fig.~\ref{fig:masses_radii_densities} illustrating the mass--radius diagram  for the
Kepler-11 system. The mass--radius relation computed for $\rho_0$ and  $\mu_e$ of
the Solar system planets are plotted with different colors. Data  are shown for
Uranus (red), Neptune (green), Jupiter (blue), Saturn  (violet) and Earth (light
blue), respectively. Representations of this  relation are plotted in the
mass-density diagram (the middle panel) and the  radius--density diagram (the right
panel).

For small masses, the characteristic density  $\rho_0$ is very close to $\rho$
\citep[see Eq.~34 in][]{Lynden-Bell2001}.  This derived decay of $\rho$ with the
mean distance from the star suggests that the  inner planets may contain larger
amount of heavy elements than more  distant companions. If this correlation can be
confirmed, it may provide an observational constraint for the planet formation
theory.  Allowing for some speculations here, let us note that all Kepler-11 
planets exhibit masses in the same range of a few Earth masses. Hence, they  likely
have formed in a similar way and physical environment \citep {Rogers2011}. Small
eccentricities and small relative inclinations suggest  that the system evolved
orbitally smoothly towards the current state,  conserving the ordering of initial
distances from the star. The observed  relation between $\rho_0$ and $a_i$ may then
indicate the chemical  composition and mass density distribution in the primordial
protoplanetary  disk.

We underline that the results in this section must be considered as  preliminary.
Due to relatively narrow time--window of the photometric data,  masses of the
planets are determined with large uncertainties.
%
%
\section{Results of the dynamical analysis}
%
\label{sec:dynamics}

The best--fit solutions gathered with the help of the bootstrap  algorithm provide
us primary information required to perform  extensive study of the dynamical
stability of the system.  Because the orbits of Kepler-11 super-Earth planets are
confined within the mean distance of Mercury in the Solar system, we could expect
that the dynamics of  this system are extremely complex. Indeed, preliminary
integrations demonstrated that the Kepler-11 system is dynamically packed, in accord
with a definition and the PPS hypothesis in \citep{Barnes2008}.  In spite of
apparently ordered  configurations with quasi--circular, almost coplanar orbits, and
relatively small masses, no  long-term stable model~I solutions were found.  Below,
we try to resolve this paradox and we try to  detect sources of this seemingly odd
and strong instability. To illustrate the structure of the phase space close to the
best--fit configurations, we choose a few representative solutions and we construct
the MEGNO maps in their vicinity.
%
%
\subsection{Quasi-stable solutions in transit model~II}
%
For  model~II,  among $\sim 1500$ initial conditions, we found only 2 
configurations exhibiting MEGNO  close to $2$ after $T=16000$~yr.  Parameters of
these solutions are listed in Tabs.~\ref {tab:solution_II_124} and
\ref{tab:solution_II_937}.
\begin{table*}
\caption{
Orbital parameters of marginally  stable configuration IIa. Mass of the star is
$0.95\,\msun$.  Osculating Poincar\'e{} elements are given at the  initial epoch
JD~2455964.51128.
}
\begin{tabular}{c c c c c c c c}
\hline
\hline
parameter/planet & b & c & d & e & f & g \\
\hline
$m \, [\mE]$ & $4.550$ & $1.542$ & $7.224$ & $15.698$ & $4.340$ & $18.530$  \\
$a \, [\au]$ & $0.091097$ & $0.106515$ & $0.154241$ & $0.193939$ & $0.249532$ & $0.463815$  \\
$e$ & $0.00800$ & $0.00698$ & $0.00918$ & $0.00924$ & $0.01827$ & $0$ (fixed)  \\
$I \,$~[deg] & $89.048$ & $88.938$ & $89.023$ & $88.803$ & $89.339$ & $89.470$  \\
$\Omega \,$~[deg] & $0$ (fixed) & $2.119$ & $15.216$ & $14.804$ & $20.626$ & $52.574$  \\
$\omega \,$~[deg] & $158.534$ & $138.150$ & $61.831$ & $183.853$ & $296.192$ & $0$ (fixed)  \\
$\Mmean \,$~[deg] & $47.28591$ & $128.23248$ & $118.94929$ & $12.27874$ & $151.62481$ & $336.22407$  \\
\hline
\hline
\end{tabular}
\label{tab:solution_II_124}
\end{table*}
The first stable solution (refereed to as IIa from hereafter, see Tab.~\ref
{tab:solution_II_124}), has a relatively low mass of planet~c $\sim 1.5$  Earth
masses.  Two innermost planets b and c have almost coplanar orbits. The next three 
planets, d, e and f also form a nearly coplanar sub-system (d-e-f), which is 
inclined to the first two orbits by large angle $\sim 15^{\circ}$. The  outermost
orbit is inclined even more, by $\sim 50^{\circ}$ to the inner  subsystem of (b-c),
and by $\sim 30^{\circ}$to the triple--planet  subsystem of (d-e-f).
\begin{table*}
\caption{
Orbital parameters of solution IIb. Mass of the star is $0.95\,\msun$.  Osculating
Poincar\'e{} elements are given at the initial epoch JD~2455964.51128.
}
\begin{tabular}{c c c c c c c c}
\hline
\hline
parameter/planet & b & c & d & e & f & g \\
\hline
$m \, [\mE]$ & $6.227$ & $4.422$ & $8.467$ & $11.293$ & $6.866$ & $32.651$  \\
$a \, [\au]$ & $0.091102$ & $0.106525$ & $0.154254$ & $0.193942$ & $0.249501$ & $0.464000$  \\
$e$ & $0.02314$ & $0.01780$ & $0.01148$ & $0.00401$ & $0.01859$ & $0$ (fixed)  \\
$I \,$~[deg] & $88.000$ & $90.849$ & $89.296$ & $91.206$ & $90.677$ & $89.733$  \\
$\Omega \,$~[deg] & $0$ (fixed) & $-1.748$ & $-5.944$ & $-2.902$ & $-2.393$ & $92.907$  \\
$\omega \,$~[deg] & $178.847$ & $175.740$ & $35.793$ & $336.443$ & $350.572$ & $0$ (fixed)  \\
$\Mmean \,$~[deg] & $29.83704$ & $92.04062$ & $144.32568$ & $218.20059$ & $96.06246$ & $336.17121$  \\
\hline
\hline
\end{tabular}
\label{tab:solution_II_937}
\end{table*}
The second stable solution IIb (see Tab.~\ref{tab:solution_II_937}) has all  masses
close to the nominal best--fit values. The mutual  inclinations of five inner orbits
are close to $0^{\circ}$, while the  outermost orbit of planet~g is highly inclined
to the inner orbits  by $\sim 90^{\circ}$, similarly to solution~IIa.
\corr{
Because  the relative inclinations between particular pairs of orbits are
large in these best-fit solutions, such systems might be unlikely observed
by a randomly located observer.  We estimate a probability of such an event 
as $\sim 0.07\%$ and $\sim 0.05\%$ for solutions IIa and IIb, respectively.
}
%
%
\subsection{Triple-planet resonances as the main source of instability}
\begin{figure*}
\centerline{
\vbox{
\hbox{
\includegraphics[ angle=270, width=0.278\textwidth]{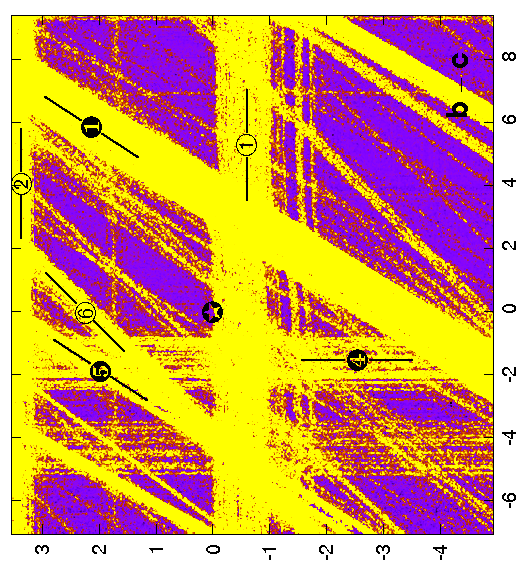}
\includegraphics[ angle=270, width=0.278\textwidth]{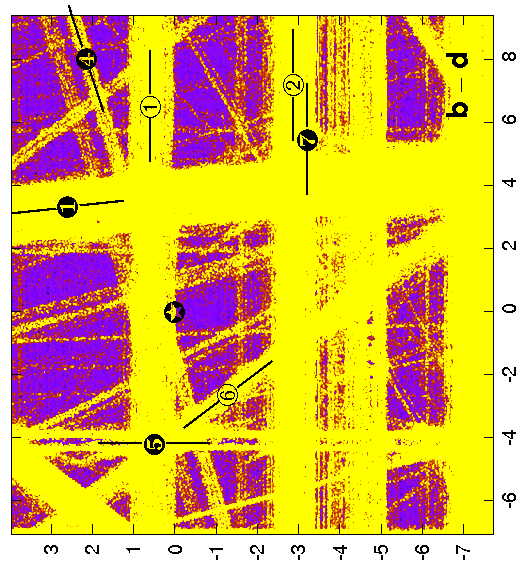}
\includegraphics[ angle=270, width=0.278\textwidth]{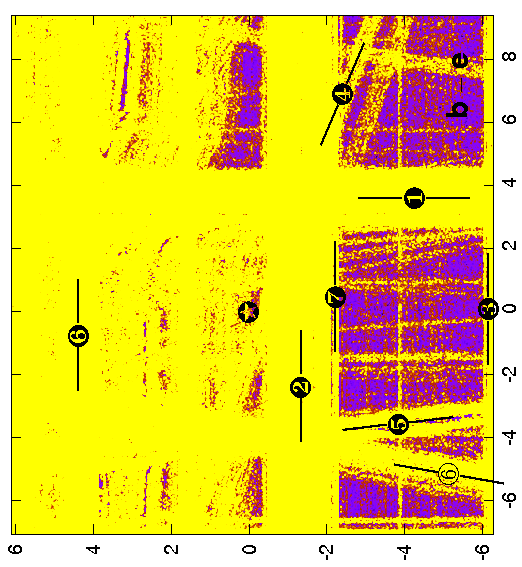}
}
\hbox{
\includegraphics[ angle=270, width=0.278\textwidth]{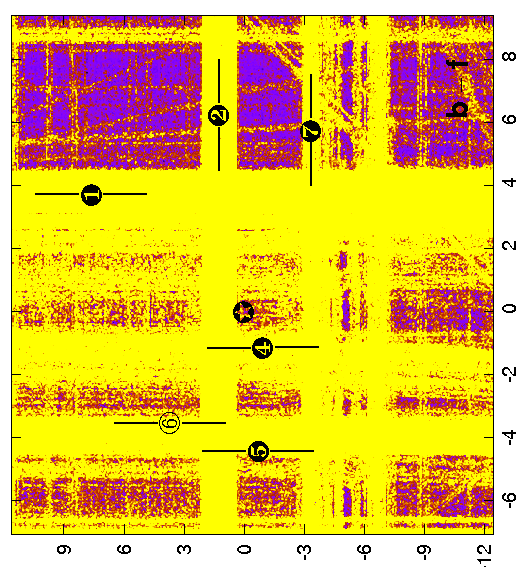}
\includegraphics[ angle=270, width=0.278\textwidth]{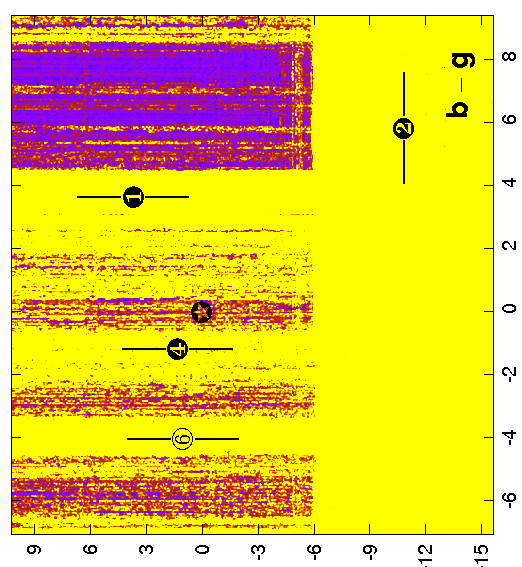}
\includegraphics[ angle=270, width=0.278\textwidth]{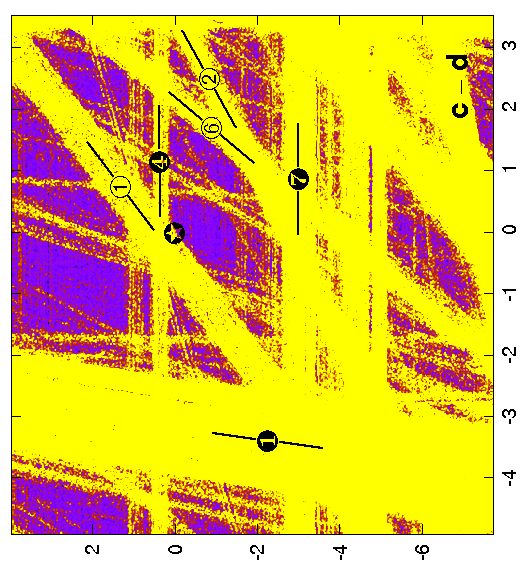}
}
\hbox{
\includegraphics[ angle=270, width=0.278\textwidth]{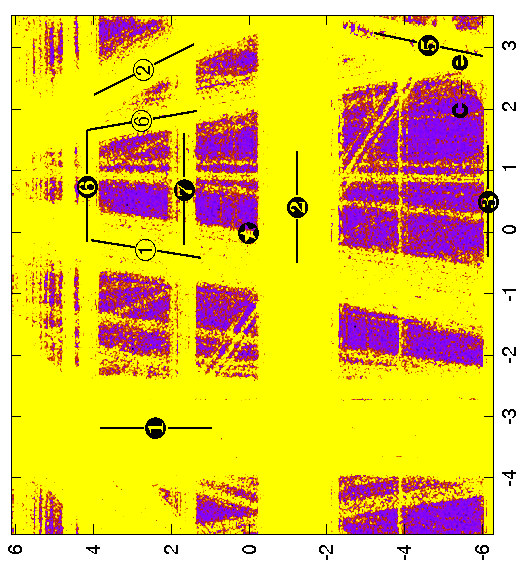}
\includegraphics[ angle=270, width=0.278\textwidth]{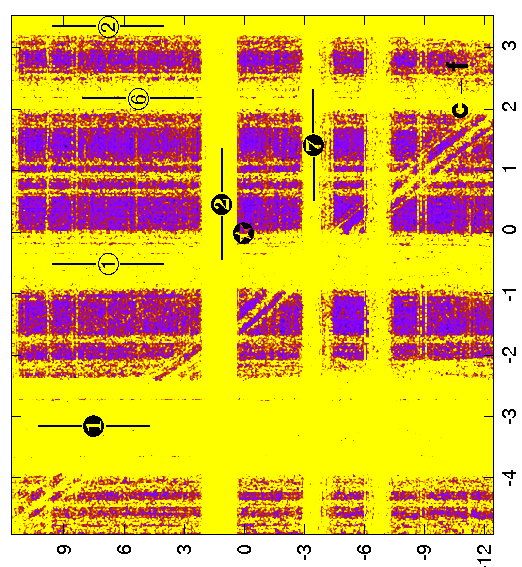}
\includegraphics[ angle=270, width=0.278\textwidth]{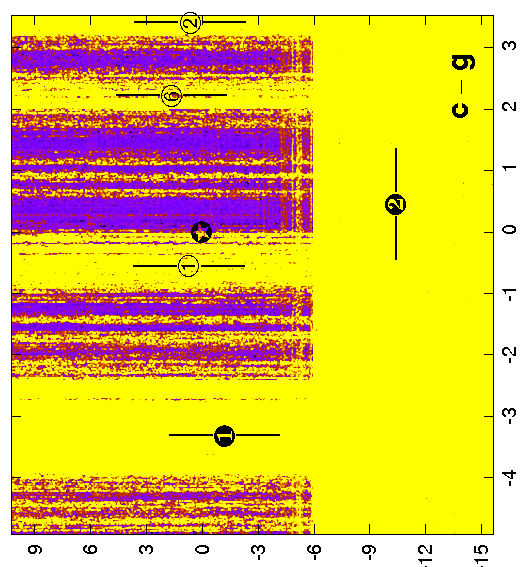}
}
\hbox{
\includegraphics[ angle=270, width=0.278\textwidth]{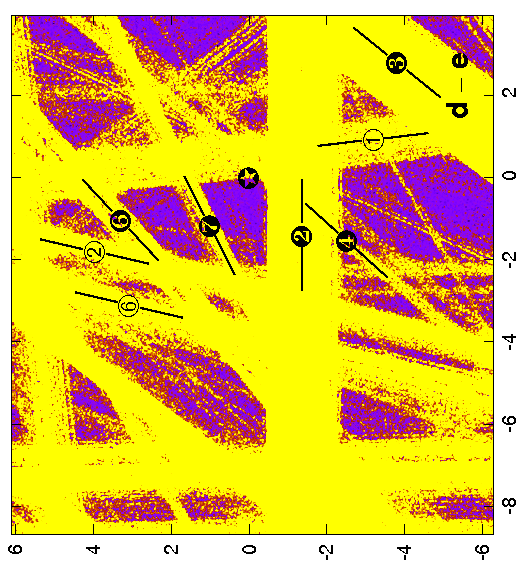}
\includegraphics[ angle=270, width=0.278\textwidth]{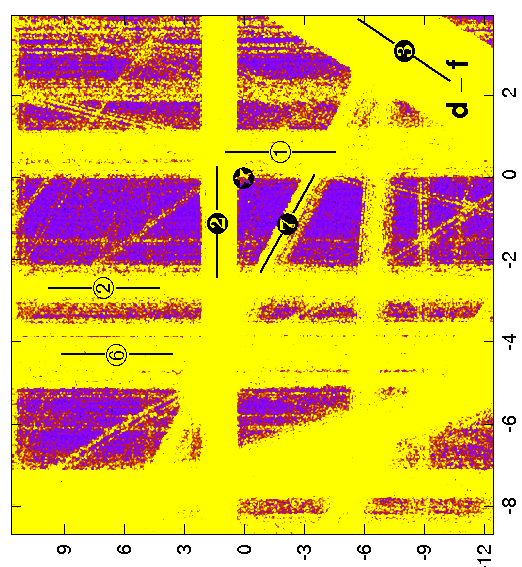}
\includegraphics[ angle=270, width=0.278\textwidth]{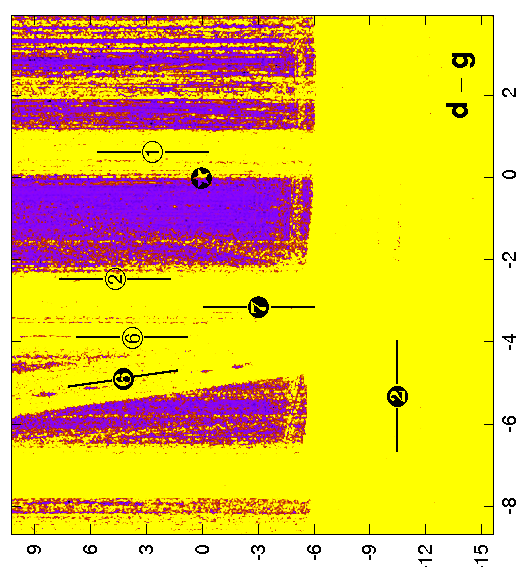}
}
\hbox{
\includegraphics[ angle=270, width=0.278\textwidth]{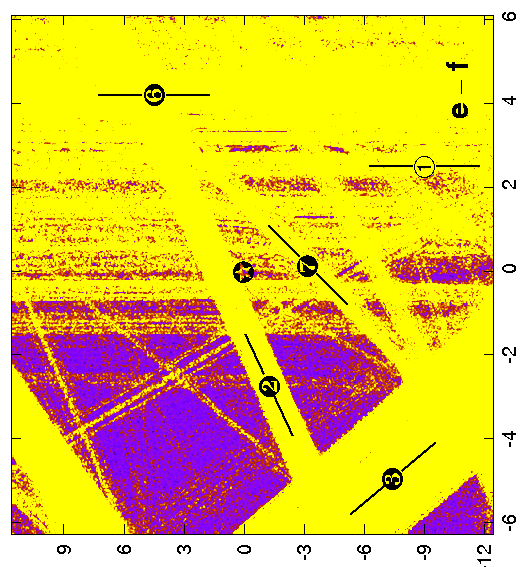}
\includegraphics[ angle=270, width=0.278\textwidth]{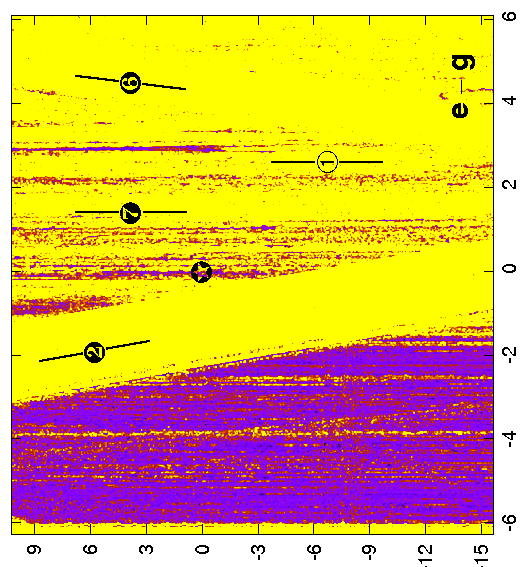}
\includegraphics[ angle=270, width=0.278\textwidth]{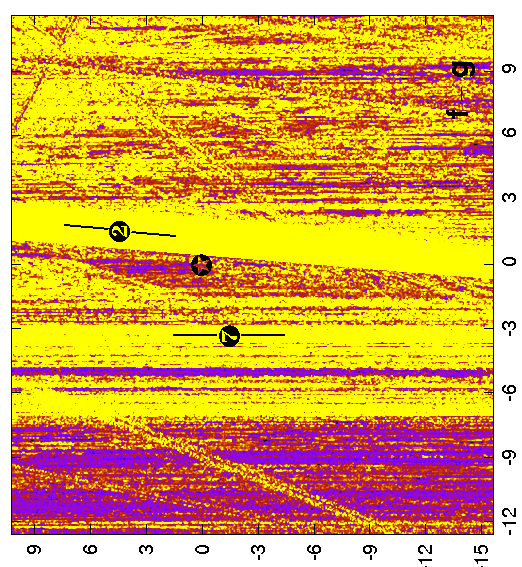}
}
}
}
\caption{
Symplectic MEGNO maps in the $(x_i, x_j)$--plane for solution~IIa (see the text for
details).  Color scale for $\langle Y \rangle$ is  $[1,5]$ (blue means $\langle Y
\rangle \approx 2$ and stable solutions; yellow  is for $\langle Y \rangle \gtrsim
5$ and unstable motions).  Each panel is for different pair of  planets labeled in
its bottom-right corner.
}
\label{fig:124_nn}
\end{figure*}

%
Let us now study the vicinity of these particular solutions through the {\em 
dynamical maps}. For each initial condition of the discrete grid with
$512\times512$~resolution, we compute  $\langle Y \rangle$ over $T=8000$~yr.
Figure~\ref{fig:124_nn} shows  the MEGNO maps for solution~IIa. Each panel is for a
different pair of planets.  The coordinate axes are rescaled mean motions centered
at their  nominal $n_{i,0}$:
\[
x_i \equiv \frac{n_i - n_{i,0}}{n_{i,0}} \times 10^4.
\] 
The $x_i$--axes span $1\sigma$ uncertainties of the semi--major  axes $a_i$, in
accord with Tab.~\ref{tab:bootstrapII}. The semi-major axes are determined very
precisely, hence the $1\sigma$  interval span a range of $10^{-5}$ to $10^{-4}~\au$.
The rescaled $x_i$  are confined to order of $10$. 

The MEGNO is  color-coded in the dynamical maps. Blue regions mean regular solutions
with $\langle  Y \rangle \approx 2$, while yellow color is for chaotic (unstable)
initial  conditions, $\langle Y \rangle \ge 5$. 
The resolution is $512\times512$ pixels, total integration time is $T=8000$~yr  per
pixel, SABA$_4$ integrator step-size is 0.5~days. A single computation of each map
took $\sim 16$~hrs of 1200~CPU cores. The integrations of each pixel were performed
up to the end time $T$, regardless a value of MEGNO.

Still, although the maps cover tiny regions of the phase  space, close to the fixed
initial condition, they reveal a sophisticated  structure. Because we consider the
dynamics in terms of conservative, close to integrable  Hamiltonian system, this
structure is governed by the resonant motions. 
A  relatively short integration time $\sim 10^4$--$10^5$ characteristic periods,
equivalent to the orbital period of the outermost planet makes it possible to detect
unstable MMRs.
They appear as yellow straight bands of different widths and  slopes. Inspecting the
dynamical maps, we can identify particular resonances.

A condition for the mean motion resonance in the $N$-planet system may be  written
in the following form:
\begin{equation}
\sum_{i=1}^{N} p_i \dott{\lambda_i} = \mathcal{O}\left[f_{\omega}, f_{\Omega}\right], \quad
\mbox{or} \quad
\sum_{i=1}^{N} p_i n_i = \mathcal{O}\left[f_{\omega}, f_{\Omega}\right],
\label{eq:resonance_condition}
\end{equation}
where $n_i$ is the mean motion of the $i$-th planet,
$f_{\omega}$ and $f_{\Omega}$ are the fundamental frequencies associated with the
pericenter arguments $\omega_i$ and the longitudes of nodes $\Omega_i$ (for all
orbits), and $p_i$ are relatively prime integers. The linear relations must obey the
d'Alambert rule.

The two-planet MMR takes place when two values of $p_i$ are  non-zero. If three
coefficients in this linear relation are non-zero, it means that the system exhibits
3--body MMR. In the   Kepler-11 system, the fundamental frequencies associated with
$\omega_i$  and $\Omega_i$ are much smaller than $n_i$ (these are the secular
frequencies), hence the right-hand sides of  Eq.~\ref{eq:resonance_condition} are
close to $0$. This makes it possible to skip the secular terms, as the first order
approximation, and to identify the MMRs through approximate resonance  conditions
involving the mean motions only. 

To identify the MMRs in the MEGNO  maps, we apply a simple method described in
\citep{Guzzo2005}. In the $(n_j, n_k)$ --plane\footnote{Let us note that
Fig.~\ref{fig:124_nn} shows the $(x_i,  x_j)$-planes but $(n_i, n_j)$ may be easily
computed from these data.}, the  slope
\[
\alpha_{j,k} \equiv \frac{\mbox{d}\,n_k}{\mbox{d}\,n_j} 
\]
of a particular resonance line determines the ratio of coefficients $p_i$, i.e., 
\begin{equation}
\alpha_{j,k} = -\frac{p_j}{p_k}.
\end{equation}
If $\alpha_{j,k} = 0$ then the MMR forms a horizontal line,  planet $k$ is involved
in the MMR, while planet $j$ is not. If  $\alpha_{j,k}^{-1} = 0$ then the MMR forms
a vertical line, planet $j$ is  involved in the resonance, while planet $k$ is not.
In these cases, other planets may be involved in this particular resonance. If
$\alpha_{j,k}$  is finite and non--zero then both  considered planets are involved
in the MMR. To identify this particular  resonance, one has to compute slopes
corresponding to this resonance  in all $(n_i, n_j)$-planes. It may be not possible,
if the map ranges do  not cover the resonance band. If $\alpha_{j,k}$ is  non-zero
and  finite only for one pair of planets ($j$, $k$), it means that  2--planet MMR is
present. It should be verified whether $p_j n_j + p_k n_k  \approx 0$. Coefficients
$p_j$, $p_k$ of the MMR condition can be  computed from the slopes $\alpha_{j,k}$. 
Similarly, the 3-body MMR takes place, if there exist relatively prime integers $i$,
$j  \neq i$ and $k \neq i,j$, such that $\alpha_{i,j}, \alpha_{i,k}$ and 
$\alpha_{j,k}$ are all finite and non--zero. The 3-body MMR may be  identified by
computing the slope coefficients in  appropriate planes of the mean motions.  An
identification of $4$--body and $N$--body MMRs can be derived as well.

Using this simple MMR identification algorithm, we found most significant MMRs close
to  solution~IIa.   The identified  3--body MMRs were labeled at the  panels, and
listed in Tab.~\ref{tab:resonances_124}. The mean motions of  solution~IIa permit a
few low--order 2--body MMRs in the vicinity of this  solution, e.g., $4 n_{\idm{b}}
- 5 n_{\idm{c}}$, $1 n_{\idm{b}} - 3  n_{\idm{e}}$, $2 n_{\idm{c}} - 5 n_{\idm{e}}$,
$1 n_{\idm{d}} - 2  n_{\idm{f}}$, $2 n_{\idm{e}} - 3 n_{\idm{f}}$. There is no
2-planet MMR in the range of  $n_i$ implied by $1\sigma$ uncertainty. All straight
bands with  finite  and non-zero $\alpha_{i,j}$ have at least two images in the
planes  constructed for other planets. All features seen in the dynamical maps 
correspond then to $3$-- and 4--body MMRs. Possibly, even 
more complex $N$-body resonances may be found. 
\begin{table}
\caption{Three-planet resonances near the best--fit model~IIa. 
See Fig.~\ref{fig:124_nn}.}
\begin{center}
\begin{tabular}{c c c c c c c c}
\hline
\hline
label & $p_{\idm{b}}$ & $p_{\idm{c}}$ & $p_{\idm{d}}$ & 
$p_{\idm{e}}$ & $p_{\idm{f}}$ & $p_{\idm{g}}$ \\ 
\hline
$1$ & $7$ & $-10$ & $2$ & $0$ & $0$ & $0$ \\
$2$ & $0$ & $0$ & $0$ & $7$ & $-11$ & $2$ \\
$3$ & $0$ & $0$ & $5$ & $-5$ & $-3$ & $0$ \\
$4$ & $1$ & $0$ & $-10$ & $11$ & $0$ & $0$ \\
$5$ & $9$ & $-13$ & $0$ & $4$ & $0$ & $0$ \\
$6$ & $0$ & $0$ & $6$ & $-9$ & $0$ & $2$ \\
$7$ & $0$ & $0$ & $6$ & $-16$ & $11$ & $0$ \\
$1^*$ & $0$ & $5$ & $-8$ & $-1$ & $0$ & $0$ \\
$2^*$ & $0$ & $6$ & $-14$ & $5$ & $0$ & $0$ \\
\hline
\hline
\end{tabular}
\end{center}
\label{tab:resonances_124}
\end{table}

Labels in Fig.~\ref{fig:124_nn} corresponds to data in Tab.~\ref
{tab:resonances_124}. Labels with asterisks are written in open circles in the
dynamical maps and denote MMRs in the neighborhood of  solution IIa. Resonances
labeled with numbers without asterisks and written  in filled circles in the
dynamical map are also present in the neighbourhood  of solution~IIb. 

\begin{figure}
\centerline{
\includegraphics[ angle=270, width=0.4\textwidth]{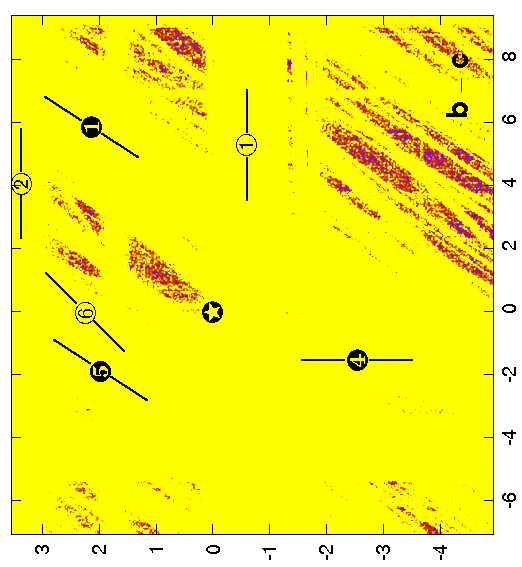}
}
\caption{
Dynamical map of solution IIa but for the integration  time $40000$~yr per pixel.
Mean motion resonances are labeled in accord with Table~7.
}
\label{fig:124_40}
\end{figure}

The MMRs ($7 n_{\idm{e}} - 11 n_{\idm{f}} + 2 n_{\idm{g}}$) labeled as ``2''   and
($5 n_{\idm{c}} - 8 n_{\idm{d}} - 1 n_{\idm{e}}$) labeled as ``$1^*$''  are the
most  close to solution~IIa. Solution IIa passed the $16000$~yr MEGNO test as
stable  solution. However, because it is close to two 3--body MMRs, and is found in
a  a dense web of low--order $3$--body and $4$-body MMRs, its long--term  stability
cannot be guaranteed by this test.  The integration time of $16000$~yr corresponds
to $\sim 50,000$  orbital periods of the outermost planet~g. This time is usually
too short  to detect a chaotic nature of the orbit when the 3--body resonances are
present. Unfortunately, the CPU-overhead does not permit to derive the  dynamical
map with the integration time per single initial condition  which  should be
$10$--$10^{2}$--times longer. Indeed, a test run over $T=40,000$~yr illustrated in 
Fig.~\ref{fig:124_40} reveals that solution IIa is chaotic and unstable. Besides,
almost the whole  $(x_{\idm{b}}, x_{\idm{c}})$-plane corresponds to chaotic motions
with  $\langle Y \rangle > 5$.

\begin{figure*}
\centerline{
\vbox{
\hbox{
\includegraphics[ angle=270, width=0.3\textwidth]{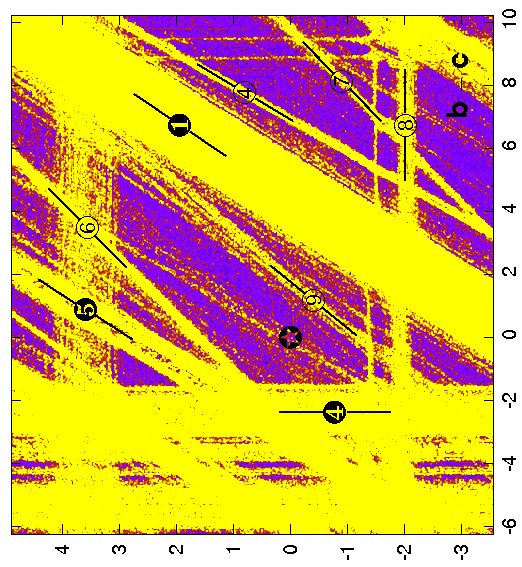}
\includegraphics[ angle=270, width=0.3\textwidth]{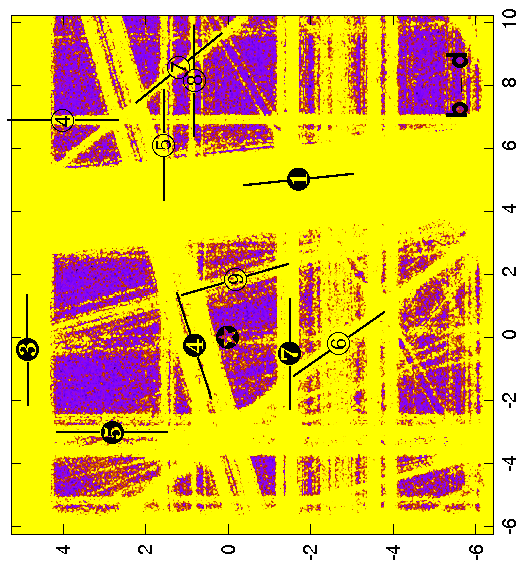}
\includegraphics[ angle=270, width=0.3\textwidth]{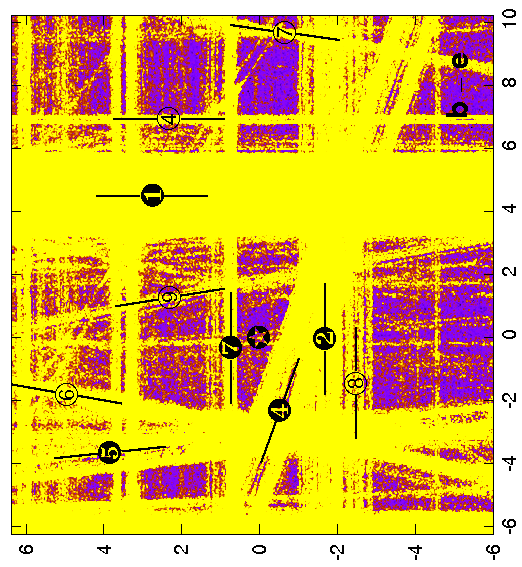}
}
\hbox{
\includegraphics[ angle=270, width=0.3\textwidth]{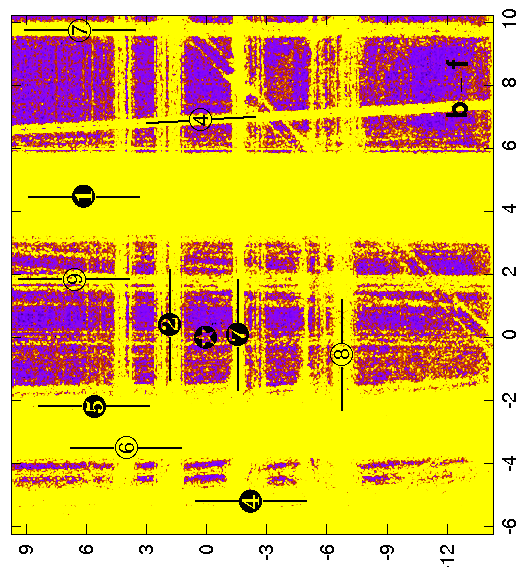}
\includegraphics[ angle=270, width=0.3\textwidth]{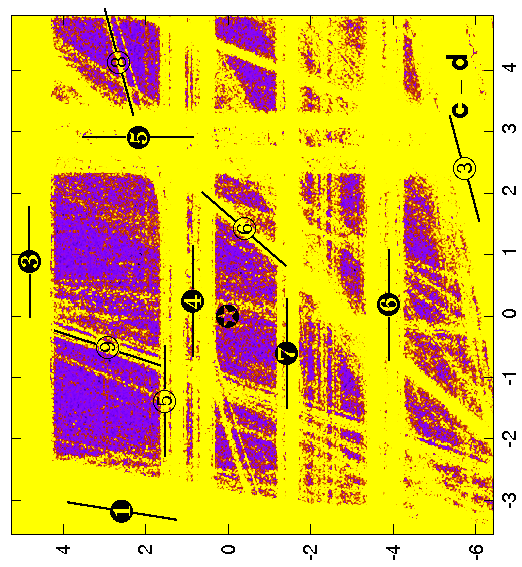}
\includegraphics[ angle=270, width=0.3\textwidth]{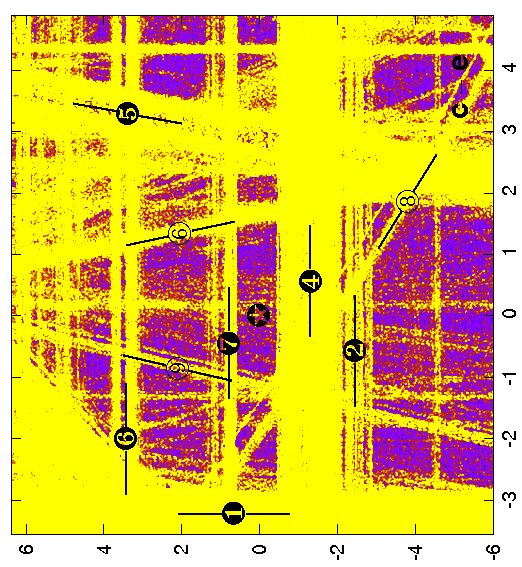}
}
\hbox{
\includegraphics[ angle=270, width=0.3\textwidth]{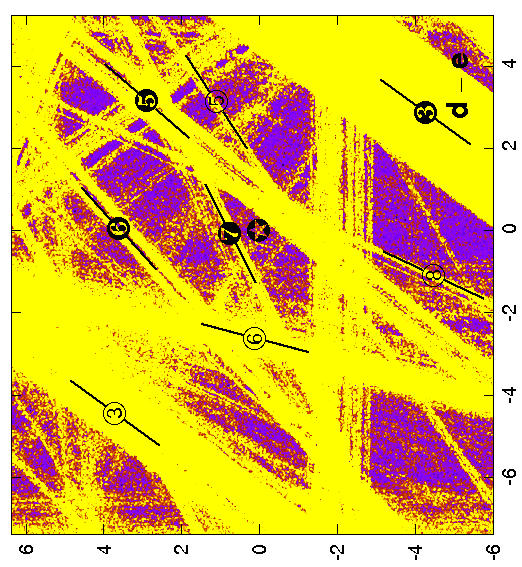}
\includegraphics[ angle=270, width=0.3\textwidth]{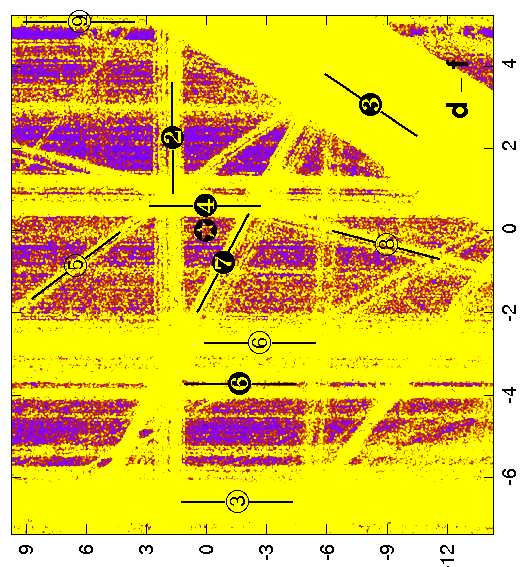}
\includegraphics[ angle=270, width=0.3\textwidth]{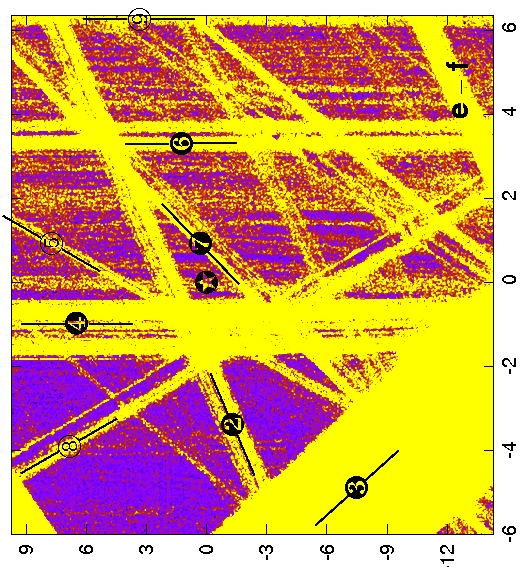}
}
}
}
\caption{
Symplectic MEGNO maps in the $(x_i, x_j)$--plane for solution~IIb.  Blue color means
$\langle Y \rangle \approx 2$ (stable configuration), while yellow is for $\langle Y
\rangle \gtrsim 5$ and unstable systems.  Each panel is for different pair of
planets labeled in the bottom-right corner of each particular panel.  The resolution
is $512\times512$ pixels, total integration time is $8000$~yr per pixel, SABA$_4$
integrator step-size is 0.5~days.
}
\label{fig:937_nn}
\end{figure*}

We analyzed the second solution IIb in the same manner. The MEGNO maps  computed for
$8000$~yr are presented in Fig.~\ref{fig:937_nn}. Also these  maps reveal a dense
net of $3$-body and $4$-body resonances. Some of  these resonances may be identified
as close to solutions~IIa as well as to~IIb.  They are labeled with the same numbers
written within filled circles, as in  Fig.~\ref{fig:124_nn}. We found also a few new
MMRs, labeled  within open circles and labeled by ``3'', ``4'' and ``5''. The
remaining 4--body MMRs which are visible in this figure are not labeled.  All
identified MMRs are listed in Tab.~\ref{tab:resonances_937}.

\begin{table}
\caption{Three-planet resonances near solution~IIb}
\begin{center}
\begin{tabular}{c c c c c c c c}
\hline
\hline
label & $p_{\idm{b}}$ & $p_{\idm{c}}$ & $p_{\idm{d}}$ 
& $p_{\idm{e}}$ & $p_{\idm{f}}$ & $p_{\idm{g}}$  \\ 
\hline
$1$ & $7$ & $-10$ & $2$ & $0$ & $0$ & $0$ \\
$2$ & $0$ & $0$ & $0$ & $7$ & $-11$ & $2$ \\
$3$ & $0$ & $0$ & $5$ & $-5$ & $-3$ & $0$ \\
$4$ & $1$ & $0$ & $-10$ & $11$ & $0$ & $0$ \\
$5$ & $9$ & $-13$ & $0$ & $4$ & $0$ & $0$ \\
$6$ & $0$ & $0$ & $6$ & $-9$ & $0$ & $2$ \\
$7$ & $0$ & $0$ & $6$ & $-16$ & $11$ & $0$ \\
$3^*$ & $0$ & $1$ & $-6$ & $6$ & $0$ & $0$ \\
$4^*$ & $13$ & $-17$ & $0$ & $0$ & $2$ & $0$ \\
$5^*$ & $0$ & $0$ & $11$ & $-21$ & $8$ & $0$ \\
\hline
\hline
\end{tabular}
\end{center}
\label{tab:resonances_937}
\end{table}
Similarly to configuration~IIa, the integration over longer time of  $T=40000$~yr,
reveals that solution IIb is unstable. Almost the whole plane of the  dynamical map
corresponds to $\mmegno > 5$. The MEGNO map is  similar to Fig.~\ref{fig:124_40},
and is not shown here.
%
%
\subsection{Dynamical maps in the $\pmb{(\omega_i, \omega_j)}$-- and 
$\pmb{(e_i, \Omega_i)}$-planes.}
%
%
\begin{figure*}
\centerline{
\vbox{
\hbox{
\includegraphics[ angle=270, width=0.3\textwidth]{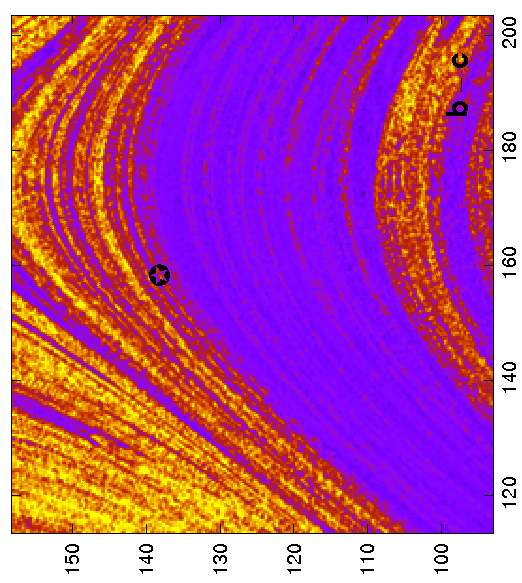}
\includegraphics[ angle=270, width=0.3\textwidth]{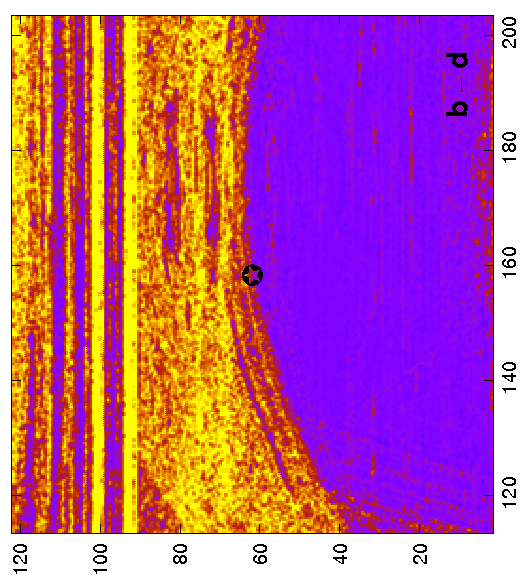}
\includegraphics[ angle=270, width=0.3\textwidth]{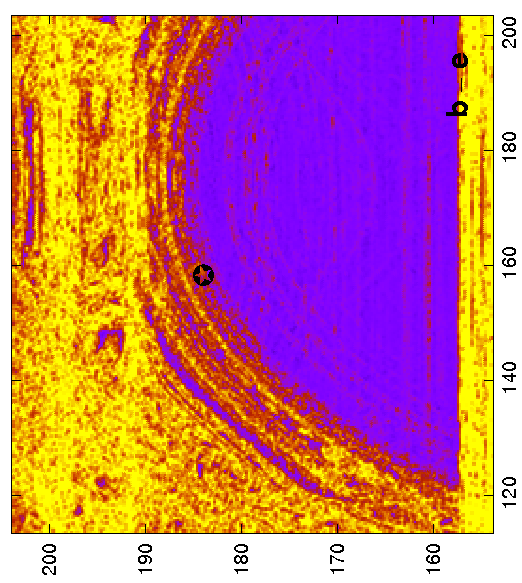}
}
\hbox{
\includegraphics[ angle=270, width=0.3\textwidth]{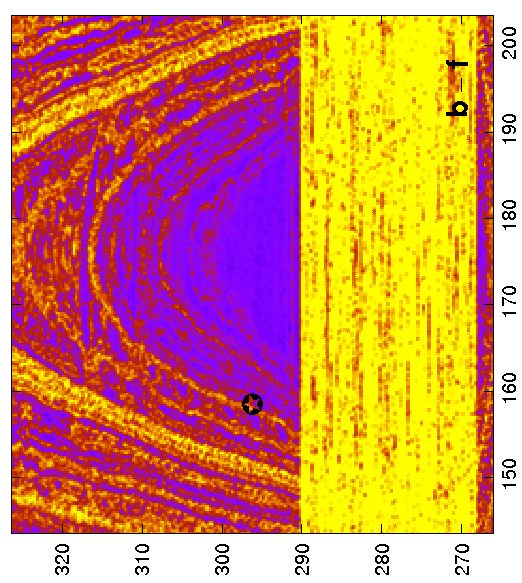}
\includegraphics[ angle=270, width=0.3\textwidth]{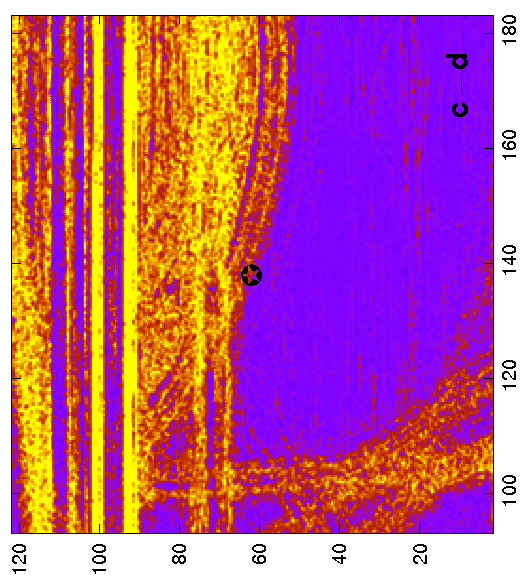}
\includegraphics[ angle=270, width=0.3\textwidth]{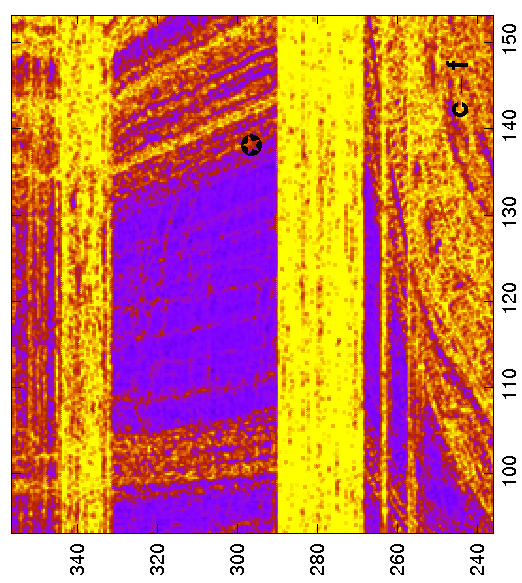}
}
\hbox{
\includegraphics[ angle=270, width=0.3\textwidth]{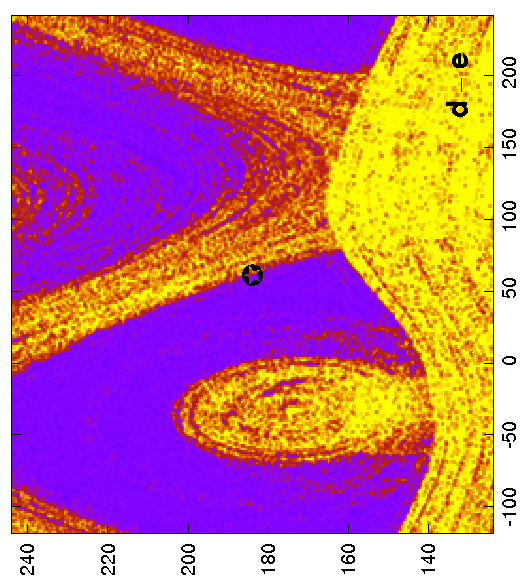}
\includegraphics[ angle=270, width=0.3\textwidth]{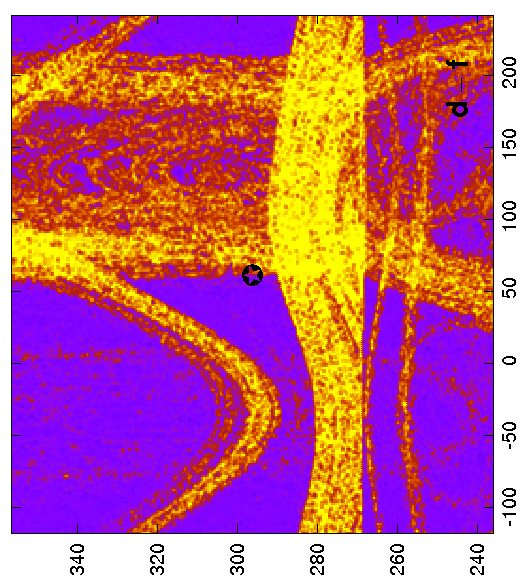}
\includegraphics[ angle=270, width=0.3\textwidth]{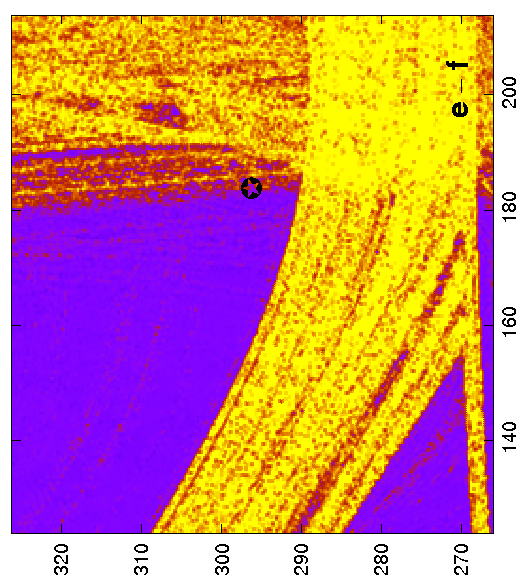}
}
}
}
\caption{
Dynamical maps in the $(\omega_i, \omega_j)$--plane for solution~IIa.  The MEGNO
range is $[1,5]$ and colour--coded: blue is for stable solutions, yellow is for
unstable systems.  Each map has elements of a given pair of planets varied, these
planets are labeled in the bottom-right corner.  The arguments of pericenter are
expressed in degrees.  The nominal solution is marked with the asterisk.
}
\label{fig:124_ww}
\end{figure*}
Figure~\ref{fig:124_ww} illustrates the MEGNO maps in planes of the arguments of 
pericenters. The MEGNO is calculated over $T=8000$~yr. The $(\omega_i,
\omega_j)$--maps are constructed a bit differently  than the mean motions dynamical
maps. Because we intent to analyse  configurations coherent with the observations,
$\omega_i$ can be freely  varied over the grid, if also the mean anomalies are
modified to preserve  the time of the first transit, $T_i$. For instance, for a
circular orbit,  when the argument of pericenter is shifted from the nominal value
by 
$\Delta\omega_i$, the mean anomaly should be shifted by $-\Delta\omega_i$.
For eccentric orbits such a correction flows from the Ist Keplerian law.

The MMRs form even more sophisticated  web in the $(\omega_i, \omega_j)$--planes
than in the mean motion planes. These dynamical maps reveal that  the stability
depends on the initial arguments of pericenters. It is not  obvious \textit{a
priori}, because eccentricities are very small.  The initial phases of the system
are preserved across the maps, and each point corresponds to the same $T_i$. Keeping
in mind that the  photometric data spanning only $\sim 500$~days imply weak
constraints  on angles $\omega_i$, we realize how is difficult to find a stable
initial  conditions in the huge, $50$-dimensional parameter space of the system.
\begin{figure*}
\centerline{
\vbox{
\hbox{
\includegraphics[ angle=270, width=0.3\textwidth]{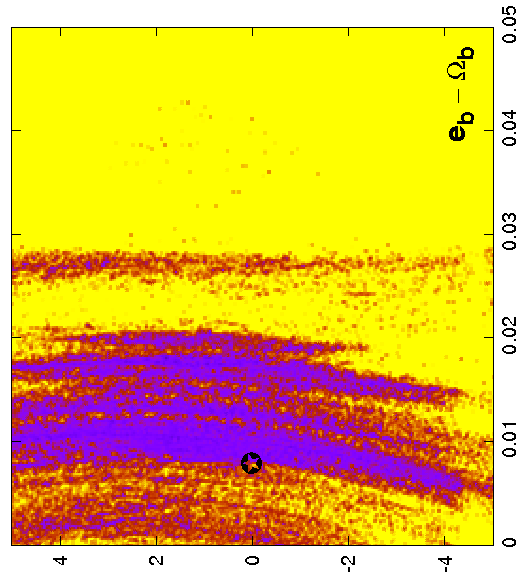}
\includegraphics[ angle=270, width=0.3\textwidth]{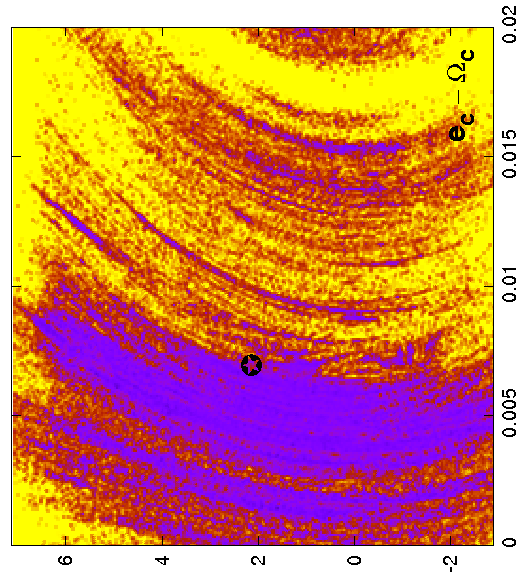}
\includegraphics[ angle=270, width=0.3\textwidth]{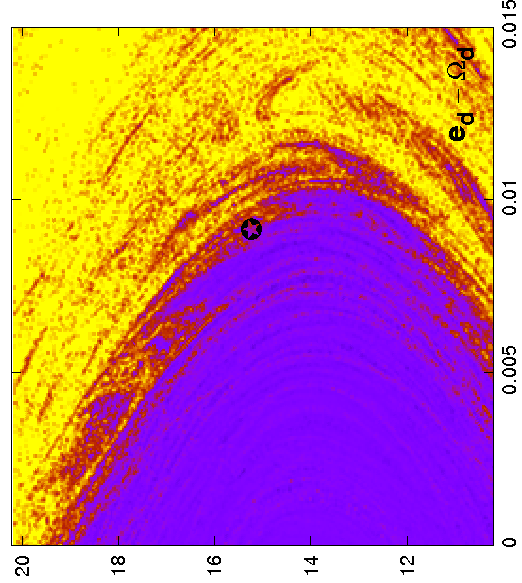}
}
\hbox{
\includegraphics[ angle=270, width=0.3\textwidth]{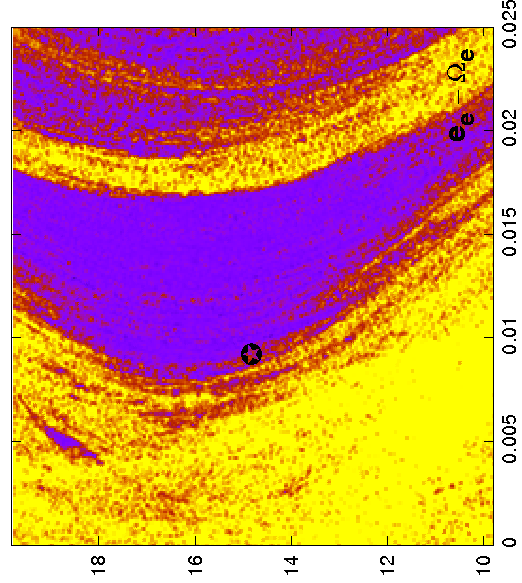}
\includegraphics[ angle=270, width=0.3\textwidth]{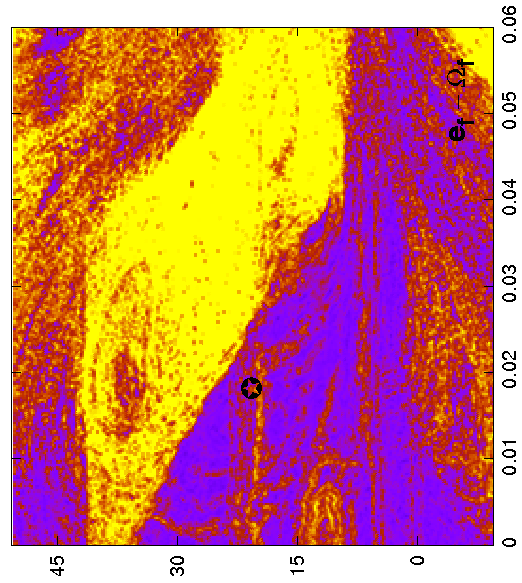}
\includegraphics[ angle=270, width=0.3\textwidth]{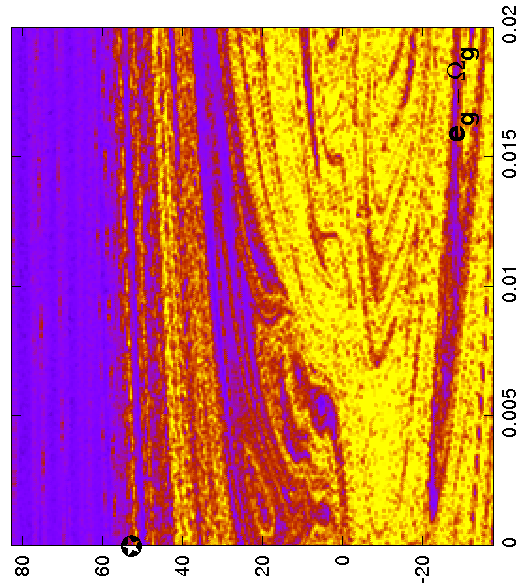}
}
}
}
\caption{
Dynamical maps in the  $(e_i, \Omega_j)$-plane for solution~IIa. The MEGNO range 
$[1,5]$ is colour coded: blue means stable solutions and yellow is for unstable
configurations.  Each map is constructed for a pair of planets labeled in the
bottom-right corner.   The longitudes of ascending nodes are expressed in degrees.  
The nominal solution is marked with an asterisk. See Tab.~\ref{tab:solution_II_124}
for the orbital elements of this best--fit model.
}
\label{fig:124_eW}
\end{figure*}
Figure~\ref{fig:124_eW} illustrates MEGNO maps in the $(e_i, \Omega_i)$ -planes
calculated over $T=8000$~yr. Each panel is for one planet. An  identification of
particular MMRs is much more complex than in the mean  motion planes. It would
require the frequency analysis of these  solutions. Still, regions of stable,
quasi-periodic motions usually form  only  small  islands in the phase space. For
the four innermost planets, the regular motions are  confined to only $\sim
5^{\circ}$ range of $\Omega_i$ around the nominal  value and also to a small range
of eccentricities $\sim 0.01$. For the two  outermost planets~f and planet~g the
maps look like different. A zone of  stable motions of planet g extends towards
large $\Omega_g$. It implies a  large mutual inclination of its orbit to the rest of
the system. Small  $\Omega_g$ provoke unstable motions. The nominal solution is
found at the  very edge between the regular and chaotic zone.
%
%
\subsection{Stable solution of model~IV}
\begin{figure*}
\centerline{
\vbox{
\hbox{
\includegraphics[ angle=270, width=0.3\textwidth]{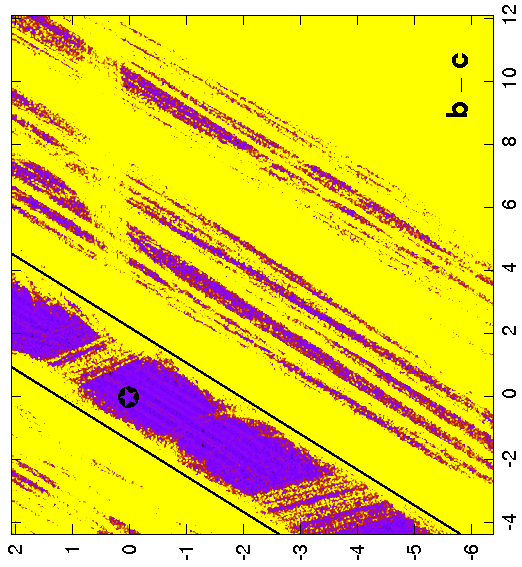}
\includegraphics[ angle=270, width=0.3\textwidth]{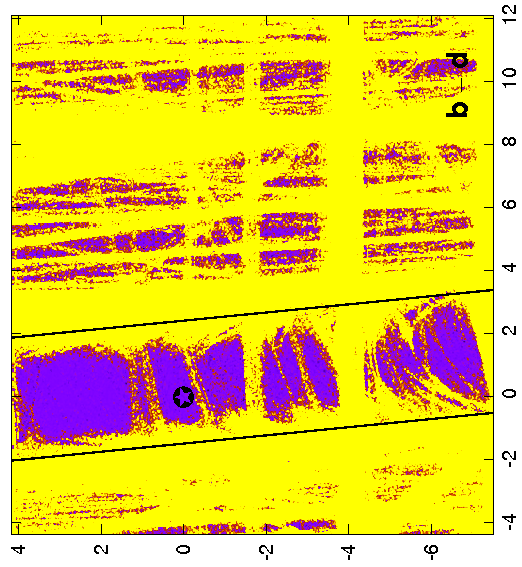}
\includegraphics[ angle=270, width=0.3\textwidth]{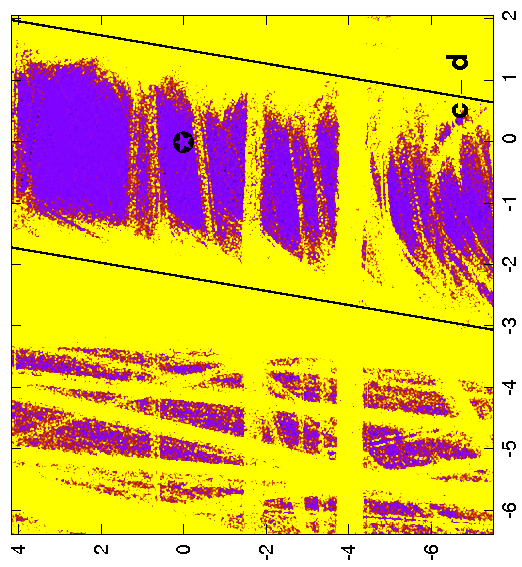}
}
\hbox{
\includegraphics[ angle=270, width=0.3\textwidth]{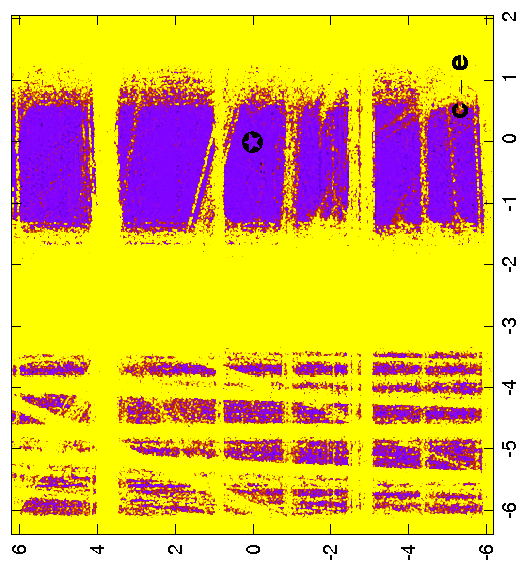}
\includegraphics[ angle=270, width=0.3\textwidth]{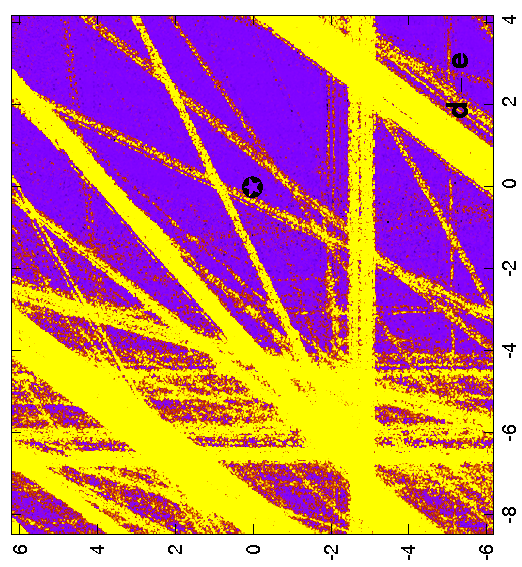}
\includegraphics[ angle=270, width=0.3\textwidth]{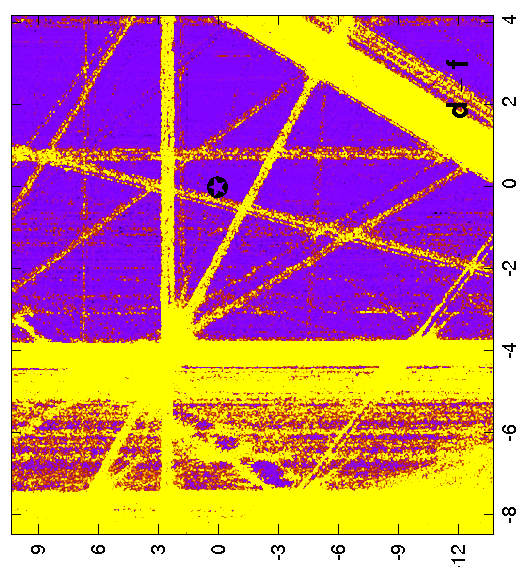}
}
}
}
\caption{
Dynamical maps of solution IVa in the mean motions planes. The  colours and symbols
are the same as in Fig.~\ref{fig:124_nn}.
}
\label{fig:1387_nn}
\end{figure*}

%
%
Let us recall that in transit model~IV all $\Omega_i=0^{\circ}$. Hence, the  mutual
inclinations between all pairs of orbits are close to $0^{\circ}$ but the system
remains non-coplanar because $I_i$ is still  different from $90^{\circ}$ 
\corr{(hence, the transits of all planets in this system can be detected by
a randomly located observer with a significant probability of $\sim
3.4\%$)}.  In this
case we found several solutions with MEGNO converging to $2$  after $T=16000$~yr.
This indicates a possibility of quasi-regular orbits. We  chose one of such
solutions. Its parameters are listed in Tab.~\ref {tab:solution_IV_1387}, and we
compute dynamical maps in its vicinity.  Figure~\ref{fig:1387_nn} shows the results
of this experiment in the mean  motions planes. The integration time is $T=8000$~yr
per pixel.
\begin{table*}
\caption{
Orbital parameters of solution IVa.  Mass of the star is $0.95\,\msun$. Osculating
elements of Poincar\'e{} are given at the epoch JD~2455964.51128.
}
\begin{tabular}{c c c c c c c c}
\hline
\hline
parameter/planet & b & c & d & e & f & g \\
\hline
$m \, [\mE]$ & $2.359$ & $3.386$ & $5.630$ & $10.841$ & $7.524$ & $25.161$  \\
$a \, [\au]$ & $0.091113$ & $0.106505$ & $0.154243$ & $0.193940$ & $0.249511$ & $0.463991$  \\
$e$ & $0.04423$ & $0.01719$ & $0.00633$ & $0.00258$ & $0.01073$ & $0$ (fixed)  \\
$I \,$~[deg] & $89.141$ & $91.215$ & $89.332$ & $88.837$ & $89.394$ & $89.770$  \\
$\omega \,$~[deg] & $20.651$ & $55.728$ & $140.753$ & $236.761$ & $355.845$ & $0$ (fixed)  \\
$\Mmean \,$~[deg] & $178.88174$ & $209.60077$ & $40.79259$ & $318.51831$ & $91.57569$ & $336.26502$  \\
\hline
\hline
\end{tabular}
\label{tab:solution_IV_1387}
\end{table*}
The tested configurations is found in a narrow region of regular motions. The  most
prominent dynamical feature visible in the maps is associated with stable 3-planet 
MMR between planets b, c and d. We identified it as $(7, -10, 2, 0, 0, 0)$  MMR. It
has been also found in dynamical maps associated with solutions~IIa and~IIb  as the
MMR labeled as ``1''. Due to altered parameters of the model, the unstable region of
this resonance is visible in these maps.  The solid black lines plotted across
panels shown in the top row of  Fig.~\ref{fig:1387_nn} have the  slope equal to
${7}/{10}$, $-{7}/{2}$ and $5$, from the left to the  right, respectively. Other
3-body resonances visible in the dynamical maps form  a~dense web, which may be
better seen in the $(x_{\idm{d}}, x_{\idm{e}})$-- and  $(x_{\idm{d}},
x_{\idm{f}})$--planes displayed in the bottom panels.

\begin{figure}
\centerline{
\includegraphics[ angle=270, width=0.4\textwidth]{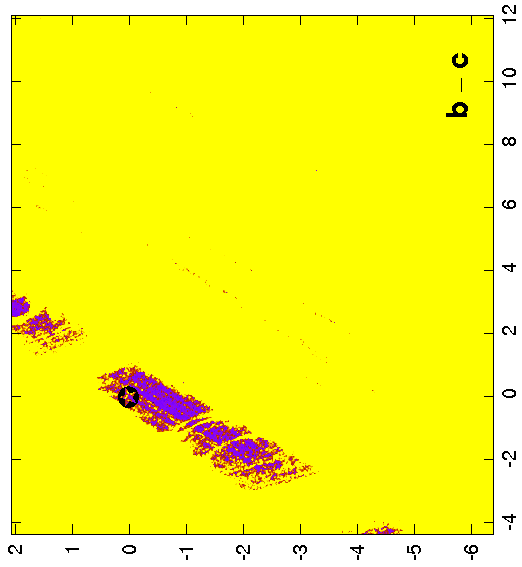}
}
\caption{
Dynamical map of solution IVa in the mean motion plane of planet~b and planet~c but
for longer  integration time of $T=40,000$~yr. See Fig.~\ref{fig:1387_nn} for an
explanation of the symbols and colour coding.
}
\label{fig:1387_40}
\end{figure}
\begin{figure*}
\centerline{
\hbox{\includegraphics[width=0.499\textwidth]{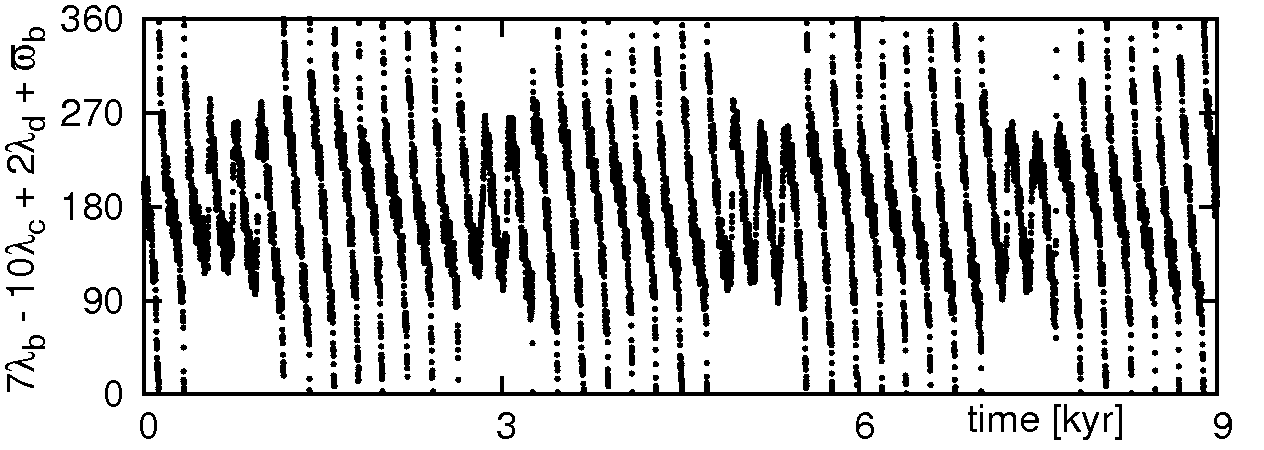}
\includegraphics[width=0.499\textwidth]{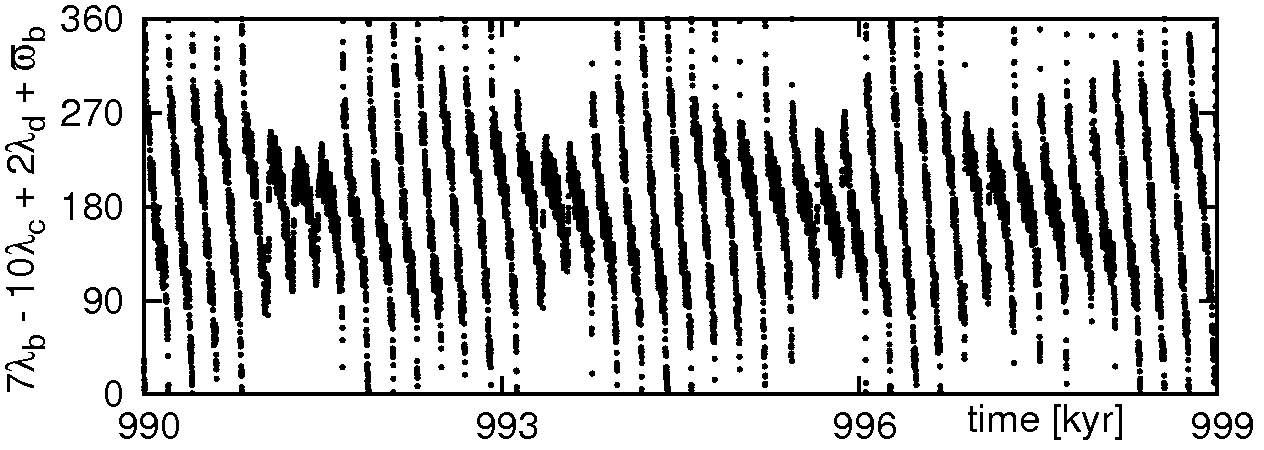}}
}
\caption{
Temporal evolution of the critical argument of the 3-body MMR for the  initial
condition IVa. The left panel is for the beginning  of the integration period, and
the right  panel is for the end of the integration period spanning 1~Myr.
}
\label{fig:critical}
\end{figure*}

To examine the stability of this solution over longer time scale, we  calculated a
single dynamical map in the $(x_{\idm{b}}, x_{\idm{c}})$ --plane for much longer
integration time  $T=40000$~yr. It is shown in  Fig.~\ref{fig:1387_40}. This
solution is located in a tiny region of stable  motions. In this case, the 3--body
MMRs has a protective role for the  system, saving it from a disruption.  Actually,
this solution is chaotic. To demonstrate this, we computed the critical argument of
the 3--body MMR and we plot over two intervals of time, at the beginning of the
1~Myr integration period (the left panel of Fig.~\ref{fig:critical}) and at the end
of this period (the right panel of Fig.~\ref{fig:critical}). The critical arguments
exhibit librations alternated with circulations. Such behavior of the critical
argument indicates a crossing of the separatrix of the resonance, and chaotic
dynamics. In such a case, the configuration may be geometrically  stable over very
long time but it may be suddenly disrupted due to a slow diffusion along the
resonance.
%
%
\subsection{The Arnold web structure in the phase space}
%
The results of  experiments described in the previous sections may be interpreted at
the ground of the dynamical systems theory. The dynamical stability of planetary
orbits in systems with strong  perturbations is influenced by even small errors and
the resulting  diffusion due to resonances overlapping. This dynamical phenomenon
has been   investigated in the Outer Solar system. \cite{Murray2001} identified  
the chaos among the Jovian planets as resulting from the overlap of the  components
of 3--body MMRs among Jupiter, Saturn, and Uranus. In spite of  short Lyapunov time
($10^{7}$ years), they found the dynamical lifetime of Uranus  $\sim 10^{18}$ years.
In this way, the analytic theory of \cite{Murray2001} explained   an apparent
paradox of long--term stable system which is actually chaotic.

The structure of the 3--body and 4--body MMRs in the Outer Solar  system was further
investigated numerically by \cite{Guzzo2005,Guzzo2006}.  Very  recently, it was also
studied by \cite{Quillen2011} for strongly  interacting extrasolar systems. Through
the dynamical maps technique,  \cite{Guzzo2005} found that if the chaotic motions
appear in a regular net, they  may be practically stable over very long times. Such
a state of chaotic  system is called the Nekhoroshev regime. On contrary, if the
chaotic  resonances do not constitute a regular web, and minority of orbits form a 
global chaotic zone, the stability of the system is influenced by strong  chaotic
diffusion. Such regime is related to the resonance overlap, and is  called the
Chirikov regime of the dynamics \citep{Froeschle2000,Guzzo2005}.

These results may be applied to interpret the dynamical maps of the  Kepler-11
system. The dense net of the multiple-body MMRs form the Arnold  web in the
neighborhood of the best fit model configurations. Our  experiments reveal, that
this system may undergo the Chirikov regime rather  than long-term stable
Nekhoroshev regime. We have no strong proof of this  phenomenon, because it would
require non-trivial and intensive computations of the chaotic  diffusion.
Conclusions regarding the real state of the Kepler-11 system require more precise
determination of the initial conditions than obtained in this work.
 
We also note that solutions IIa, IIb investigated in detail are found at the very
border of the chaotic and regular zones. This can be well seen in the ${(\omega_i,
\omega_j)}$-- and ${(e_i, \Omega_i)}$-planes. A change of these angles constrained
by the best--fit errors, could ``move'' the system into larger zones of stable
motions. Because the MEGNO quantifies the stability of the system as a whole, such a
change would imply a shift of the initial condition in all parameter maps
\citep{Gozdziewski2001a}.  The altered initial conditions would be also more
``distant'' from the unstable strips of 3--body and 4--body MMRs in the mean motions
planes. The dynamical maps would be more similar to those computed for solution IV
revealing most extended zones of stable motions. Overall, the dynamical state of the
system is very fragile and depends on subtle changes of the initial conditions.

Still most of the best-fit model configurations obtained with the bootstrap
algorithm, which appear as regular over the  short-term dynamics time--scale  are
self--disrupting. This behavior is similar to  unstable evolution in 3--body MMRs
observed in the HD~37124 system  \citep{Gozdziewski2008a}. In this sense the
dynamical state of the Kepler-11  system is still puzzling. The results of
long--term integrations of the obtained sets of initial conditions indicate that 
the system resides at the edge of dynamical stability.  These conclusions might be
changed, if more data are gathered and analyzed. 
%
%
\section{Conclusions} 
%
%
In this paper, we derived an improved method for the dynamical analysis of
photometric  light curves of stars with multiple transiting planets. Its main
purpose is to  determine planetary masses, as well as a number of {\it indirect} 
parameters affecting the dynamical stability of the system. This algorithm  improved
the well known TTV algorithm. The crucial point of this method is to  model the
whole photometric curve directly with the help of an efficient  symplectic $N$-body
integration. Such {\em the direct approach} make is possible to  account for
multiple transits, as well for the transit depths and their  widths. This in turn
makes it possible to impose dynamical constraints on  parameters which cannot be
determined in terms of the TTV, like the longitude of  nodes and mutual inclinations
of orbits.

With the help of this new method, we re-analyzed available photometric data for 
Kepler-11. The direct algorithm imposes constraints on the mass of the  outermost
planet~g and help us to determine the mutual inclinations between  orbits of
planets~b and planet~c as well as between the (d-e) pair with a good accuracy  of
$2^{\circ}$. These results extend analysis performed in the discovery  paper
\citep{Lissauer2011}. Overall, conclusions in this paper and in our  work coincide
very well, in spite of quite a different transit  models, optimisation algorithms
and uncertainties estimation methods applied.

Thanks to the in--depth analysis of the Kepler-11 light-curves, we investigated  a
possible chemical composition of the planets detected in this intriguing  system.
The most curious finding is a clear anti-correlation of the mean  densities of the
planets with their mean distances from the star. The inner  planets exhibit larger
abundance of heavy elements than the outer  companions. Because all eccentricities
as well as the mutual  inclinations of stable systems remains small,  the  system
unlikely suffered violent scattering processes in the past. It follows that the 
ordering of planets have been preserved since their formation. A dynamical 
relaxation should imply large $e_i$ and $\Delta I_{i,j}$ \citep[see, 
e.g.,][]{Chatterjee2008, Adams2003}. These factors indicate that the  primordial
protoplanetary disk might have a significant gradient of  metallicity and the
present Kepler-11 system is a real fossil of this disk.  If this suggestion is
confirmed, it can constrain the planet formation theories, in particular concerning
multiple systems of super--Earth planets.

This conclusion is reinforced by the dynamical analysis of the system. We  found, in
accord with the discovery paper, that the system is basically  free from 2-planet
MMRs. However, its global dynamics is governed by  $3$-body and $4$--body MMRs.
Overlapping of these resonances near the best-fit  solutions imply an extended zone of
dynamical chaos and very unstable  configurations. Particular multi-body resonances
may stabilize the system.  We identified such a 3--body resonance. However, the
observation window spanning only  $\sim 500$~days does not make it possible to pick
up this MMR as the only possible.  Besides,  the MMRs form the structure of the
Arnold web characteristic for the  Chirikov regime. In this dynamical state, the
phase-space orbits are strongly unstable  due to overlapping of the MMRs. In such a
case, the chaotic diffusion in the  actions (semi--major axes) space is significant
and easily leads to strong  geometric changes of orbits. Actually, our numerical
experiments indicate this in the  Kepler-11 system. It remains an open question
though, how such an apparently  ordered configuration of planets could survive the
formation phase in extremely  complex and fragile dynamical environment. 

The discovery team argue that the system is non-resonant.  This factor would 
prevent a scenario of trapping the planets into MMRs during the inward  migration at
the early stages of the evolution. As we showed here, the  system is in fact
extremely resonant, but in quite a different sense.  Combining this fact with small
values of the eccentricities and  anti--correlation  of the chemical composition
with the distance to the star, we may conclude  quite an opposite: the
migration/trapping scenario could be the only way  of preserving the primary
architecture in the present form. However, these  suggestions might be verified only
if more photometric data are gathered  and are available. 

Most likely, many other Kepler--discovered multiple  extrasolar systems with
transiting planets exhibit qualitatively similar behaviours to that one  we found in
the Kepler-11. Hence, the approach in this paper is general and may be applied in
the studies of other compact systems composed of low-massive, super--Earth or
Neptune-like planets.

\section*{Acknowledgments}
%
We would like to thank the anonymous referee for a review and comments
that improved the text.
This work is supported by the Polish Ministry of Science and Higher  Education grant
No. N/N203/402739. CM is a recipient of  the  Foundation for Polish Science
Fellowship (programme START, editions 2010 and 2011).  This research was carried out
with the support of the ``HPC Infrastructure  for Grand Challenges of Science and
Engineering'' project (POWIEW), co-financed by  the European Regional Development
Fund under the Innovative Economy  Operational Programme.
\bibliographystyle{mn2e}
\bibliography{ms}
\label{lastpage}
\end{document}